\documentclass{aa}  

\usepackage{graphicx}
\usepackage{txfonts}

\usepackage{subfigure}
\usepackage{siunitx}
\usepackage{stfloats}

\usepackage{color}
\let\ACMmaketitle=\maketitle
\renewcommand{\maketitle}{\begingroup\let\footnote=\thanks \ACMmaketitle\endgroup}

\begin{document}

\title{Study of the thermal and nonthermal emission components in M31: the Sardinia Radio Telescope view at 6.6\,GHz\footnote{Table \ref{Tab:SRT_Sources_catalogue} and the 6.6 GHz SRT map are available in electronic form
at the CDS via anonymous ftp to cdsarc.u-strasbg.fr (130.79.128.5)
or via http://cdsweb.u-strasbg.fr/cgi-bin/qcat?J/A+A/ }}

\titlerunning{M31 map at 6.6\,GHz}        

\author{S. Fatigoni\inst{1,2}, F. Radiconi\inst{1,3}, E.S. Battistelli\inst{1,3,4},  M. Murgia\inst{4},  E. Carretti\inst{5}, P. Castangia\inst{4}, R. Concu\inst{4}, P. de Bernardis\inst{1,3}, J. Fritz\inst{6}, R. Genova-Santos\inst{7,8}, F. Govoni\inst{4}, F. Guidi\inst{7,8}, L. Lamagna\inst{1,3}, S. Masi\inst{1,3}, A. Melis\inst{4}, R. Paladini\inst{9}, F.M. Perez-Toledo\inst{8}, F. Piacentini\inst{1,3}, S. Poppi\inst{4},  R. Rebolo\inst{7,8}, J.A. Rubino-Martin\inst{7,8}, G. Surcis\inst{4}, A. Tarchi\inst{4}, V. Vacca\inst{4}}

\institute{Sapienza University of Rome, Physics Department, Piazzale Aldo Moro 5, 00185 Rome, Italy
 \and
University of British Columbia, Department of Physics and Astronomy, Vancouver, BC, V6T 1Z1, Canada
 \and
 INFN - Sezione di Roma, Piazzale Aldo Moro 5 - I-00185, Rome, Italy
 \and 
 INAF - Osservatorio Astronomico di Cagliari, Via della Scienza 5 - I-09047 Selargius (CA), Italy
 \and 
 INAF - Istituto di Radioastronomia - Via P. Gobetti, 101 - I-40129 Bologna, Italy
 \and 
 Instituto de Radioastronomia y Astrofisica, UNAM, Campus Morelia, A.P. 3-72, C.P. 58089, Mexico
 \and 
 Instituto de Astrofisica de Canarias, C/Via Lactea s/n, E-38205 La Laguna, Tenerife, Spain
 \and 
 Departamento de Astrofsica, Universidad de La Laguna (ULL), E-38206 La Laguna, Tenerife, Spain
 \and 
 Infrared Processing Analysis Center, California Institute of Technology, Pasadena, CA 91125, USA
 \\ \email{sfatigoni@phas.ubc.ca, federico.radiconi@roma1.infn.it, elia.battistelli@roma1.infn.it, matteo.murgia@inaf.it}
} 
             

\authorrunning{Fatigoni, Radiconi, Battistelli, Murgia et al.}

\date{Received 27 November 2020 / Accepted 20 April 2021}

\abstract
{The Andromeda galaxy is the best-known large galaxy besides our own Milky Way. Several images and studies exist at all wavelengths from radio to hard X-ray. Nevertheless, only a few observations are available in the microwave range where its average radio emission reaches the minimum.}
{In this paper, we want to study the radio morphology of the galaxy, decouple thermal from nonthermal emission, and extract the star formation rate. We also aim to derive a complete catalog of radio sources for the mapped patch of sky.}
{We observed the Andromeda galaxy with the Sardinia Radio Telescope at 6.6 GHz with very high sensitivity and angular resolution, and an unprecedented sky coverage.}
{Using new 6.6\,GHz data and Effelsberg radio telescope ancillary data, we confirm that, globally, the spectral index is $\sim 0.7-0.8$, while in the star forming regions it decreases to $\sim 0.5$. By disentangling (gas) thermal and nonthermal emission, we find that at 6.6 GHz, thermal emission follows the distribution of HII regions around the ring. Nonthermal emission within the ring appears smoother and more uniform than thermal emission because of diffusion of the cosmic ray electrons away from their birthplaces. This causes the magnetic fields to appear almost constant in intensity. Furthermore, we calculated a map of the star formation rate based on the map of thermal emission. Integrating within a radius of {$R_{max}=15$ kpc}, we obtained a total star formation rate of $0.19 \pm 0.01$ M$\ensuremath{_\odot}$/yr in agreement with previous results in the literature. Finally, we correlated our radio data with infrared images of the Andromeda galaxy. We find an unexpectedly high correlation between nonthermal and mid-infrared data in the central region, with a correlation parameter $r=0.93$.
Finally, by computing the logarithmic $24_{\mu m}/21_{cm}$ ratio $q_{24 \mu m}$, we find a decreasing trend with increasing galactocentric distance and an increasing dispersion. The logarithmic far-infrared-to-radio ratio is found to be $2.41 \pm 0.04$.}
{}


\maketitle

\section{Introduction}
\label{sec:1}

The Andromeda galaxy (Messier 31, M31) is the largest galaxy of the Local Group, which includes the Milky Way, the Triangulum galaxy (M33), and several other smaller galaxies including the satellite Messier 32 (M32) and Messier 110 (M110). Messier 31 is a large disk galaxy at a distance of 780 kpc from Earth \citep{stan98}, and has a proper redshift of z=-0.001 \citep{deV91}, where the negative value shows that it is moving towards the Milky Way.

The galaxy presents a large central bulge. The bulge is defined as the central region that has an extension of $5^{\prime}$ in the optical band, corresponding to a proper size
of $1.1 \,$kpc \citep{mort77}.
The bulge contains dilute ionized gas and a few dense star clusters. V- and I-band Hubble Space Telescope (HST) Planetary Camera images show that the central region seems to have two separate concentrations of light that are about 0.5" apart \citep{hst93}. One of these probably contains a dense central object, presumably a black hole of mass $M_{BH} \simeq 10^{6}$  M$_\odot$, while the other might be a star cluster. There is a {ring c}ircling the bulge at a radius of about 10 kpc, and most of the young disk stars lie in this ring or just outside it \citep{haa98,gor06,for13}. Going to greater distances from the center we find the HI \citep{braun09} layer that flares out to become thicker. The M31 galaxy is the brightest of the Messier objects, with an apparent magnitude (V) of 3.4 \citep{M31magn07} and an apparent extension of approximately $1\times3$ deg$^{2}$ \citep{Nil73} due to the high inclination with respect to the line of sight, which is assumed to be of $\ang{75}$ \citep{gor06} considering that $\ang{90}$ means an edge-on orientation.

Observations by the infrared {\it Spitzer} Space Telescope revealed that M31 contains about $\simeq 10^{12}$ stars \citep{StarN14}, which is at least twice the number of stars contained in the Milky Way, and the total mass of the galaxy is estimated to be $M_{M31} = 1.5 \times 10^{12}$ M$_\odot$ \citep{bar06}. High-resolution (FWHM=$9"$) Westerbork Synthesis Radio Telescope (WSRT) observations of the 21 cm line by \cite{braun09} were used to calculate the total neutral gas mass, which turned out to be in the range $4.4$ to $5.5 \times 10^9$ M$_\odot$ \citep{fri12}. CO observations by \cite{nieten06} within a field of $0.5^\circ \times 1.8^\circ$ led to an estimate of the mass of molecular gas of the order of $2.63\times 10^8$ M$_\odot$ \citep{fri12}. While the former observations do not include the whole galaxy disk, they most certainly encompass the majority of the molecular gas (about 95\% of the total molecular mass, according to \citealt{nieten06}). 

Because of its size and distance in the sky, M31 is an excellent laboratory with which to further study and perhaps confirm  the physical mechanisms and properties that have so far only been studied in detail in our own Galaxy. It also provides information about the general features of a typical disk galaxy. The M31 galaxy has been studied at several wavelengths: Fermi LAT was used to detect gamma-ray emission in the energy range 200 MeV - 20 GeV \citep{abd10}. NuSTAR permitted the study of hard X-ray emission allowing  X-ray binaries and neutron stars/black holes to be resolved \citep{wik16}. XMM, ROSAT, and CHANDRA were used to study its X-ray emission \citep{sti09}. Ultraviolet (UV) measurements were used to study interstellar extinction \citep{cla14} as well as the circumgalactic medium, finding a very extended ionized medium which may overlap with a similar one originating from our Milky Way \citep{leh20}. Optical and near-infrared (NIR) observations were used to investigate the stellar composition \citep{sik14}. The IR and submillimeter (submm) emission of M31 were studied by \cite{fri12} and the HELGA collaboration. These latter authors used data obtained from the {\it Herschel} satellite to characterize the dust emission of M31, and to construct maps of the dust surface density, the dust-to-gas ratio, the starlight heating intensity, and the  abundance of polycyclic aromatic hydrocarbons (PAHs).

The radio and microwave emission of M31 have been observed at high angular resolution (FWHM $< 5'$) in the last three decades with the Effelsberg radio telescope \citep{beck20,ber03,ber83}, the Westerbork Synthesis Radio Telescope \citep[WSRT,][]{braun09}, and the Very Large Array \citep[VLA,][]{con98}.  Combined Effelsberg and VLA observations at 1.46\,GHz and Effelsberg data at 4.85\,GHz indicate a global nonthermal spectral index of $\alpha _{s} = 1.0 \pm 0.1$ \citep{ber03}, which is consistent with the value found by \cite{bat19}, who studied the overall flux density emission from radio to IR wavelengths. \cite{ber03}, assuming a nonthermal spectral index equal to 1, derived (gas-) thermal and nonthermal maps at 4.85\,GHz, finding that the thermal emission is stronger in the northern part of M31, whereas the nonthermal emission is nearly homogeneous over the whole ring. This homogeneity suggests that recent star formation does not lead to a local increase in the number of relativistic electrons and/or magnetic field strength. Strong thermal emission was also found within the central region of the galaxy. \cite{beck20} used three new deep radio surveys of M31 made with the Effelsberg radio telescope at $2.645 \, GHz$, $4.85 \, GHz,$ and $8.35 \, GHz$ to perform a detailed study of the spectral index maps as well as thermal and nonthermal emission scale lengths. \cite{beck20}, using nine frequencies, found a total spectral index of $0.71 \pm 0.02$ and a synchrotron spectral index of $0.81 \pm 0.03$; they also  found that the synchrotron emission ring is wider than the thermal emission one.

In \cite{bat19} we showed that the total integrated emission at 6.6\,GHz from M31, integrated over an elliptical region of $91.5^{\prime}$ and $59.5^{\prime}$ semi-axes, is at the level of $1.20\pm 0.06$ Jy. Comparing different ancillary data allowed us to calculate the emission budget; we found strong and highly significant evidence for anomalous microwave emission (AME) at the level of 1.45$^{+0.17}_{-0.19}$ Jy at the peaking frequency of $\simeq$25\,GHz. By decomposing the spectrum into known emission mechanisms, such as synchrotron, free-free, thermal dust, and AME, we found that the overall emission from M31, at frequencies below 10\,GHz, is dominated by synchrotron emission with a spectral index of 1.10$^{+0.10}_{-0.08}$, with subdominant free-free emission. At frequencies of $\sim$10\,GHz, AME has a similar intensity to that of synchrotron and free-free emission, overtaking these components at frequencies between 20\,GHz and 50\,GHz, whereas thermal dust emission dominates the emission budget at frequencies above 60\,GHz, as expected. For the purpose of this paper, we refer to the free-free emission as thermal emission, and to the synchrotron emission as nonthermal emission, assuming any other emission mechanism to be negligible at 6.6\,GHz.
K-band observations with improved angular resolution (e.g., $0.9'$ with the Sardinia Radio Telescope at 22\, GHz) will be key to disentangling AME models, and to studying AME over an entire galaxy outside of our own with high angular resolution (\cite{bat12}, Radiconi et al. in preparation). 

This paper is organized as follows. The characteristics of the observations, the scanning strategy, and the data analysis pipeline of C-band data are described in Sect. 2. In Sect. 3 we describe all the ancillary maps used in this paper and the recipes that were adopted to make their geometry and angular resolution uniform. The removal of compact sources is presented in Sect. 4. In the same section we also compare the normalized differential count of compact sources with theoretical predictions. In Sect. 5 we carry out an analysis of the morphology of  M31 by investigating the galaxy spectral index map. Thermal and nonthermal maps at 6.6\,GHz are presented in Sect. 6, while in Sect. 7, starting from the thermal map, we derive a map of the star formation rate (SFR), and compare this to what was previously found in the literature. In Sects. 8 and 9 we analyze the correlation between radio-continuum and IR data, studying how thermal and nonthermal emissions are related to IR emission. We also compute the parameter $q_{FIR}$ and compare the result with the expected value.
Finally, in Sect. 10 we summarize our results.

\section{Sardinia Radio Telescope observations of M31}
\label{sec:2}

\subsection{Sardinia Radio Telescope}
\label{sec:2.1}

The Sardinia Radio Telescope (SRT \footnote{http://www.srt.inaf.it/}) is an Italian facility for radio astronomy run by the Istituto Nazionale di Astrofisica, INAF, which was formally inaugurated in 2013 after completion of technical commissioning \citep{bol15}. The scientific commissioning of the telescope was carried out in the period 2012-2015 \citep{mur16,pra17}. At the beginning of 2016 the first call for single-dish early science programs (ESPs) was issued, and the observations started on February 1, 2016. The SRT observations of M31 at 6.6\,GHz presented in this work are part of these ESPs.
The SRT is placed 35 km north of Cagliari (Lat: $\ang{39}$ 29'34''N - Long: $\ang{9}$ 14'42''E) on the island of Sardinia at 600 m above sea level.
The optical system is based on a quasi-Gregorian dish antenna, with a primary mirror of 64 m in diameter, and a secondary mirror of 9 m, which is shaped in such a way as to minimize the standing wave bouncing between the two reflectors. Three additional mirrors increase the number of focal positions.
A key feature of the SRT is its active surface, composed of a total of 1116 electromechanical actuators able to correct deformations induced by gravity on the primary surface. The SRT has an angular resolution of 2.9\arcmin (FWHM) at a frequency of 6.6\,GHz. 

The suite of backends currently available on site includes SARDARA \citep[SArdinia Roach2-based Digital Architecture for Radio Astronomy;][]{mel18}. The SARDARA backend is a wide-band digital backend based on ROACH2\footnote{https://casper.ssl.berkeley.edu/wiki/ROACH2} technology, and can divide the signal in the given bandwidth into up to 16384 channels for spectropolarimetric observations. We used this backend for our observations. 

\subsection{C-band observation}
\label{sec:2.2}

The main goal of the observations presented in this paper is to map the whole M31 galaxy in the C-Band down to a sensitivity of 0.4 mJy/beam in order to create a high-sensitivity/high-angular-resolution map of M31 microwave emission covering its full extension, limited by the confusion noise at this frequency and angular resolution \citep{bon03,con74}. We set the Local Oscillator at 5.9\,GHz and we use the SARDARA configuration with 1.5\,GHz bandwidth sampled with 16384 channels of 91.6\,kHz in width. The data cube will be analyzed in
greater detail in a search for spectral line emission from various transitions (Tarchi et al., in prep.).
We set the Focus Selector filter to select the frequency range from 6.0 to 7.2\,GHz, which defines our observing bandwidth. The central frequency of our observations is therefore 6.6\,GHz. Traditionally, C-band is centered at about 5 GHz, while the band around 6.6 GHz is known as the C-high band. For simplicity, in this paper we refer to the 6.6 GHz data as C-band data.

We decided to cover the galaxy with a rectangle of 7.4~deg$^2$ centered at (RA;~Dec) = (0h~42m~48s; +41$^{\circ}$~16$'$ ~48$''$). To the best of our knowledge, this is the biggest map ever produced at this frequency and angular resolution with a single-dish radio telescope.

The map size was determined by the minimum contour that includes the galaxy, to which we added a \textit{Map Edge} of 0.25 deg in each of the two scanning directions. The minimum contour was estimated by comparing the two maps provided by {\it Herschel}/SPIRE at 250 $\mu$m \citep{fri12} and {\it Planck} at 857\,GHz \citep{pla14}; which are supposed to trace the galaxy in its maximum extension. \emph{Planck} and \emph{Herchel} $5 \sigma$ contours are presented in the right panel of Fig. \ref{fig:example fit}.
The aim of adding an extra padding is to have a sky region that is completely free from M31 emission and that we can use to estimate the background and decouple this from the signal coming from the galaxy. 
We adopted an {on-the-fly} map-scanning strategy, which means orthogonal subscans along right ascension (RA) and declination (Dec) spaced by 54 arcsec (FWHM$/3$). We used a scanning speed of 6\arcmin/s. Covering the 7.4 deg$^{2}$ of our map required 209 RA subscans and 161 Dec subscans. Table \ref{tab:1} reports the main details of the observations.

\begin{table}
\begin{tabular}{ll}
\hline
\hline
Telescope & SRT (64m) \\
Receiver & C-band (5.7-7.7\,GHz)\\
Backend & SARDARA \\
Spectroscopy & 16384 channels  \\
Backend bandwidth & 1.5\,GHz\\
Channel spacing & 91.6\,kHz\\
\hline
Observed frequency range &  6.0-7.25\,MHz \\
Beam size &  2.9\arcmin @ 6.6\,GHz\\
Polarization & Full Stokes \\
Scanning strategy & Orthogonal raster subscans\\
Scanning speed & 6\arcmin/s \\
Sample rate & 25 spectra/s \\
$<T_{sys}>$ &  37 K \\
\hline
Map size & $2.4 \,$ deg $\times \, 3.1 \,$ deg $\simeq 7.4 \, $deg$^2$ \\
Total number of scans  & 22 RA + 22 Dec   \\
Total observation time & 64h \\
Reached sensitivity & 0.43 mJy/beam \\
\hline
\hline
\end{tabular}
\caption{Main characteristics of the C-band SRT observations on M31.}
\label{tab:1}
\end{table}

Before starting the observations, a set of daily preliminary operations were performed. These include the configuration of the backend, setting of the amplifier attenuation factors (to ensure linearity), secondary mirror positioning (by focusing a bright radio source), and spectroscopic system acquisition and pointing checks. The calibrators were observed during each session, when M31 was over the antenna critical elevation of 85~deg, the SRT being equipped with an Alt-azimuthal mount. The observations were completed in $64$~h, divided in 6 days, during which we carried out $22$ complete scans of the galaxy.

\subsection{Data acquisition}
\label{sec:2.3}

Once collected, the data were analyzed using the \textit{Single-dish Spectral-polarimetry Software} (SCUBE, see \citealt{mur16}), which was created specifically for treating data acquired with the SRT. SCUBE is a package dedicated to the reduction and analysis of single-dish data and its combination with interferometric observations. SCUBE is written in C++ and is developed and maintained by M. Murgia and F. Govoni; it allowed us to import the data, extract, analyze, and flag the spectra, do the baseline subtraction, perform calibration, and generate the final maps with different methods.

As a first step in the data reduction process we removed the backend "birdies" (dead channels occurring in SARDARA data) and the well-known persistent radio-frequency interference (RFI) that affect specific spectral windows at the SRT. These omnipresent signals were removed a priori by applying the same flag table to all observing days. All the remaining sporadic RFI signals 
(e.g. satellites or other transient ambience disturbances) were removed later in the analysis with automated flagging methods.
Overall, the amount of data erased due to RFI accounts for about 30\% of the total.

The baseline subtraction was carried out differently for the calibrators and for the M31 scans. To estimate the baseline for calibrators we made a linear fit over $10\%$ at the beginning and at the end of each subscan, and then removed this linear trend from the data.

To create a baseline model for the M31 scans we used a mask to cover the part of the map that we knew could possibly contain either galactic emission from M31 itself or any bright background point source. 
The remaining part of the field of view was considered as "cold sky" and was used to perform a linear fit to the baseline subscan by subscan (see Sect. \ref{sec:2.5}).

\subsection{Calibration}
\label{sec:2.4}

Data calibration included bandpass and flux density calibration. These were carried out at the beginning of the map extraction process and performed channel by channel. The bandpass calibration was necessary to set the instrument responsivity for every channel. For  this  purpose  we  used  specific  calibrators, like J0542+4951 (3C147),  with  stable and known  spectra.

We applied the gain-elevation curve correction to account for the gain variation with elevation due to the gravitational stress changes in the telescope structure. We do not compensate for the atmosphere opacity however, considering that the correction is negligible at these frequencies.

The counts-to-Jansky conversion factor was derived by means of a two-dimensional Gaussian fit to the Stokes L and R images of the standard flux density calibrators 3C147, J0137+3309 (3C48), J1331+3030 (3C286), and J1411+5212 (3C295).  We adopted the flux density scale of  \cite{per13}. The systematic uncertainty is assumed to be $5\%$ of the calibrator flux density.

All calibrator observations were made during the interruption of the M31 scans when the galaxy was too highly elevated on the sky or at the end of the scans.
Data were acquired during  four cross scans on each of the calibration sources. The cross scans were made at a velocity of $1$\arcmin/s and each subscan was $20\arcmin$ long.

\subsection{Baseline subtraction}
\label{sec:2.5}

Before stacking the subscans to form the maps, we performed a subscan baseline removal in order to set all the emission coming from blank fields  to zero. To this purpose we applied a mask. Our mask map was created starting from three different maps of the M31 region: the NRAO VLA Sky Survey (NVSS) map at 1.4\,GHz \citep{con98} which contains point sources, the {\it Spitzer} MIPS map at 12.5~THz \citep{gor06}, and the { \it Planck} 850\,GHz map \citep{pla18} that traces the extension of the galaxy with thermal dust emission. We renormalized and combined the set of maps creating a mask from which the baseline is calculated. The mask, combined with the 0.25 deg $MapEdge$, was finally used to estimate the background level by  fitting a linear model to the emission in the nonmasked region.

At this point, in order to generate the final map, we performed a new flagging of the data to remove isolated spikes or short-time RFIs in an automated fashion. Following this approach, we were able to flag all the data that deviate by more than $3 \sigma$, where $\sigma$ is a combination between the single subscan rms and the sky model noise according to:\begin{equation}
    \sigma ^{2} = \left( rms_{sub-scan} ^{2} + \left( mod\_ errsys\cdot model  \right)^{2}   \right)
.\end{equation}
The parameter $mod\_ errsys$ was chosen to be 0.5. This method was iterated until convergence was reached: three iterations for each set of data were needed. At this point, it was possible to create a new set of images where, as expected, the $rms$ level was lower with respect to the previous images.

\subsection{Map making}
\label{sec:2.6}

In the creation of our final total intensity map, we combined the scans by taking an average of all of them in order  to increase the
signal-to-noise ratio. A direct stacking would be heavily affected by the scanning noise present in the individual \textit{on-the-fly} maps, which takes the form of the typical pattern of "stripes" oriented along the scanning direction.
These stripes can be due to imperfect RFI removal or short-term fluctuations in the atmosphere or in the receiver gain.
Sometimes these features are not totally removed during the flagging and the baseline subtraction process, and so they persist in the final map. These features are easier to identify if we analyze the map in the perpendicular direction with respect to the direction of the corrupted subscan \citep{eme88}. 
To this purpose, SCUBE implements both Fourier and wavelet map-making methods.
In the Fourier transform plane, the scanning noise terms from any single scan appear only on a narrow band passing through the origin, with a width inversely proportional to the characteristic scale lengths of the baseline drifts in the original scan.
The final map is created by incorporating a second scan in an orthogonal direction so that once the second scan is also transformed in the Fourier plane, the second map only intersects the error terms of the first map at the origin. The knowledge of the approximate scale length of the baseline drift of each map, and so the width of each band error in the Fourier plane, enables an optimally weighted combination of Fourier terms of each data set to be made \citep{eme88}. In the wavelet space, the logic is the same, except that the unwanted scanning noise is isolated and filtered out using a convenient merging of the stationary wavelet transform (SWT) coefficients \citep{mur16}. One advantage of the SWT method compared to the Fourier method is that there is no need to determine an optimal width for the weights of the noise bands. While the noise bands partially overlap at the center of the Fourier plane, they are completely separated in the wavelet domain. Indeed, the horizontal and vertical detail coefficients of the two orthogonal scans can be opportunely mixed while the diagonal detail coefficients and the approximation coefficients are averaged.

We verified that both the Fourier and SWT de-stripping techniques give consistent results. However, we noted that the SWT performed slightly better than the Fourier method and therefore we opted the former.

\begin{figure*}[!h]
    \centering
    \begin{tabular}{cc}
        \includegraphics[scale=0.38]{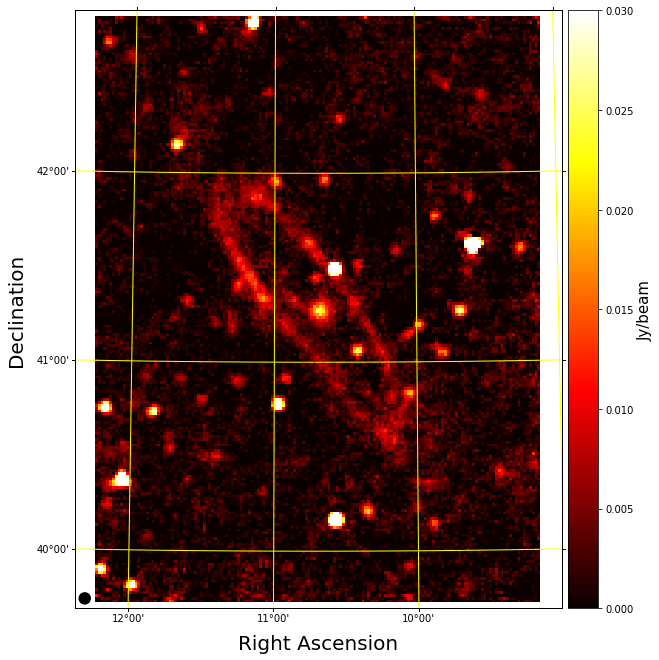} &
        \includegraphics[scale=0.38]{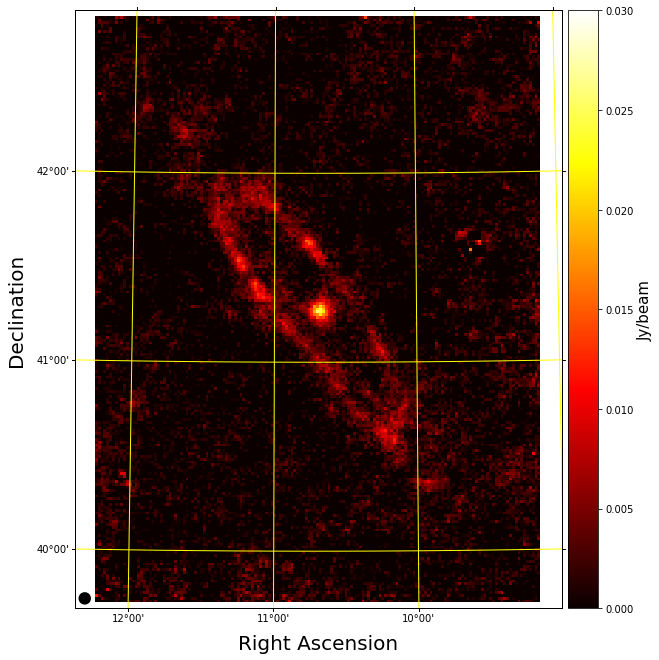}\\
    \end{tabular}
    \caption{Left: Final M31 images after averaging over the whole bandwidth at 6.6 GHz. Right: Final  M31 image after subtraction of point sources. Both the maps are characterized by an angular resolution of $2.9 ^{\prime}$. The typical beam size is indicated with the black circle in the bottom-left corner of each panel.}
    \label{Fig:M31_Letter}
\end{figure*}

As a last step in the map-making process we corrected the final map for the residual base level. To this purpose, we masked the emission from M31 and applied the Papoulis-Gerchberg Algorithm \citep[PGA,][]{pap75} to model and interpolate the surrounding base level over the excluded data.
SCUBE provides an application of the PGA that allows the user to reconstruct the missing part of an image by imposing a low pass filter in the spectral domain. The signal extrapolation is carried out by iterating alternately between time (spatial) and spectral domains until convergence is reached.
To this end, we masked out an elliptical region of 91$'$.5 and 59$'$.5 semi-major and semi-minor axis, respectively (PA=-52$\deg$), centered on the innermost regions, and we reconstructed the missing signal by imposing that the reconstructed map retains no power below a spatial scale corresponding to the ellipse major axis. This is achieved by a convolution with a Gaussian tapering in the spectral domain.
The reconstructed base-level image is then subtracted from the final map, which allows the removal of residual large-scale features left over after baseline removal and  map-making, and large-scale features in foreground and background emission, bringing the map back to a zero-level outside the galaxy, as already shown in \cite{bat19}. Figure \ref{Fig:M31_Letter} shows our final map of M31 at 6.6 GHz.

\subsection{Map statistics}
\label{sec:2.7}

In order to provide an estimate of the sensitivity level reached in our map, we masked the emission from point-like radio sources. Details on how compact sources are removed from the map are reported in Sect. \ref{sec:4}.  In our source-subtracted map, considering our pixel size of $0.9\arcmin$, and the SRT beam in the C-band, we derived the $rms$ fluctuations from two $0.4 \, $deg$^{2}$ uncontaminated regions.  We repeated the same analysis on the SWT map before and after applying the flagging.
We found consistent (although slightly different) results in the two maps: we obtained a rms mean value of 0.44 mJy/beam before flagging and 0.43 mJy/beam after flagging. Therefore, from this first analysis, we have a final value for the rms of 0.43 mJy/beam. This value is fully compatible with the expected value from  confusion noise, assuming a spectral index of 0.7: $confnoise$ = 0.44 mJy/beam \citep{con74}.

\section{Ancillary data}
\label{sec:ancillaryData}

In this section we describe the ancillary maps used throughout this work. The ancillary maps are described in two different sections, one relative to radio and microwave wavelengths and the other to IR observations.

\subsection{Radio data}
For radio wavelengths, we used three maps acquired with the Effelsberg Radio Telescope\footnote{https://www.mpifr-bonn.mpg.de/en/effelsberg}$^{,}$\footnote{ https://www.mpifr-bonn.mpg.de/3265873/m31}, at 1.46\,GHz, 2.7\,GHz, and 4.85\,GHz. With its 100-metre primary mirror diameter, Effelsberg reaches an angular resolution of $ 5.0\arcmin$ \citep{bec82} and $2.8\arcmin$ \citep{ber03} at 2.7\,GHz and 4.85\,GHz, respectively. The 1.46\,GHz map angular resolution of $0.75 \arcmin$ \citep{bec98} was achieved combining Effelsberg data with VLA data.
Moreover, we used the SRT map at 1.385 GHz \citep{mel18}, with an  angular resolution of $13.9^{\prime}$, to correlate radio and IR data.

\subsection{Infrared data}

Infrared data were used to characterize stellar and dust emission and explore the presence of possible correlations with the radio continuum emission (see Sect. \ref{sec:8}). For this purpose, we used the {\it Spitzer} Space Telescope\footnote{http://www.spitzer.caltech.edu/} maps at 24\,$\mu$m and 3.6\,$\mu$m. These two maps were chosen for correlation purposes because the 24\,$\mu$m \citep{gor06} map is a SFR tracer, while the 3.6\,$\mu$m \citep{bar06} is a tracer of older stars. The angular resolution of the {\it Spitzer} 24 $\mu$m map is $0.1\arcmin$, while the 3.6\,$\mu$m map is characterized by a resolution of  $0.04\arcmin$.

In order to estimate the radio/IR ratio, $q_{FIR}$, following the convention by \cite{helou1985}, we also used the IRAS 60 $\mu m$ map \citep{xu96} and the {\it Herschel} 100 $\mu$m map \citep{fri12}. The angular resolutions of the two maps are $4.0^{\prime}$ and $0.21^{\prime}$, respectively.

\subsection{Ancillary data treatment}

Each one of the aforementioned maps is characterized by a given angular resolution and geometry (i.e., pixelization on the sky). In order to compare different maps, it was necessary to bring all of them to the same coordinate system and angular resolution.

We used the $Montage$ $6.0$ software\footnote{http://montage.ipac.caltech.edu/} to modify the coordinate system of the  maps and convert it to the SRT C-band pixelization. Once all the maps were in the same World Coordinate System \citep[WCS,][]{wcs} reference frame, they were convolved to the same angular resolution.

Depending on the purpose, three different final angular resolutions were used: $14^{\prime}$, $5.0^{\prime}$, and $3^{\prime}$. \begin{itemize}
    \item For the sole purpose of estimating the $q_{FIR}$ parameter, the {\it Herschel} 100\,$\mu$m map and the IRAS 60\,$\mu$m map were convolved to the common resolution of $14^{\prime}$.
    \item Convolution of all the maps to the resolution of the Effelsberg 2.7\,GHz map ($5.0^{\prime}$) was carried out to disentangle thermal and nonthermal emission (see Sect. \ref{sec:6}) and to correlate these two emission mechanisms to IR data (see Sect. \ref{sec:8}).
    \item We convolved the two {\it Spitzer} maps, the 1.4\,GHz Effelsberg, and our SRT C-band image to $3'$ to extract a spectral index map (see Sect. \ref{sec:5}).
\end{itemize}

\section{Compact sources}
\label{sec:4}

A large number of sources can be seen in the $2.4 \times 3.1 \space$ deg$^2$ C-band map shown in the left panel of Fig. \ref{Fig:M31_Letter}. In order to calculate a proper estimate of the intrinsic brightness of M31, it is necessary to identify and remove all the signals coming from these sources. To this end, we generated a point-source catalog of M31 at 6.6\,GHz. Aside from point-source subtraction, we used this catalog to verify the theoretical prediction on the number of sources we expect to find with a certain flux density in a given area, according to a specific model.

\begin{figure}[h!]
    \centering
    \includegraphics[width=0.45\textwidth,trim=50 380 40 0]{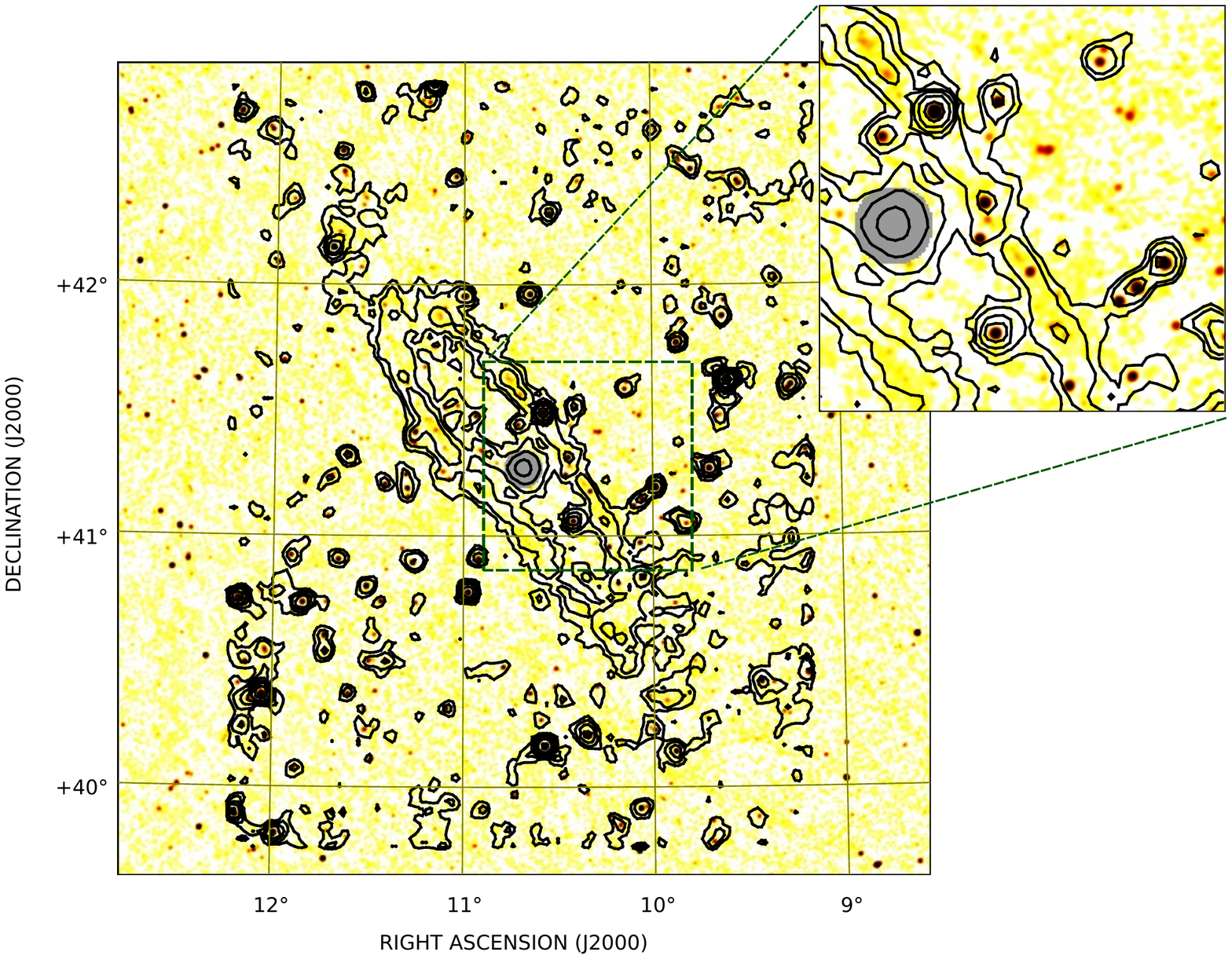}
    \caption{ NVSS image at 1.4\,GHz with the SRT contours at 6.6\,GHz. Levels start at 2\,mJy/beam and increase by a factor of two. The inset shows a detail of the M31 ring. The extended emission of the central region (grey area) has been excluded from the point-source search. }
    \label{fig:NVSS}
\end{figure}

We started by extracting a point-source catalog from the NRAO VLA Sky Survey at $1.4$\,GHz \citep{con98} which has an angular resolution of 45\arcsec~and a rms of 0.5\,mJy/beam.
We relied on the relatively high angular resolution and sensitivity of the NVSS 
to obtain a good prior for the coordinates of the points sources in the direction of the M31 field. This is particularly helpful in the subtraction of those point sources
seen in projection against (or embedded in) the extended emission of the galaxy, as we show in Figure \ref{fig:NVSS} where the SRT C-band contours are added to the NVSS image. Indeed, the finer NVSS resolution also allowed us to resolve nearby sources not easily discernible from our single-dish resolution. We note that we excluded the M31 central region extended emission from the NVSS image in the point-source subtraction process.

We select in the NVSS map all the point sources with a peak flux density of $ F > 5$ mJy at 1.4\,GHz. By fitting these point sources with a 2D Gaussian, we compiled a catalog that contains the peak flux densities and the coordinates for a large number of sources both inside and in the immediate surroundings of the M31 field.

In order to account for nonzero offset over which the source may lie, we added to the Gaussian model a plane with an offset and a tilt as free parameters. This made the fit more accurate for sources with a significant local base level, in particular those seen in projection over the M31 emission.
As a further step, the NVSS catalog coordinates were used as input parameters in the CATS\footnote{www.sao.ru/cats/} \citep{ver97} database in order to extract the radio flux density information available in the literature about these sources at other frequencies in the radio band. We then extracted the flux densities of the point sources present in the 6.6\,GHz SRT image using as a prior the coordinates of the NVSS
catalog. By fitting the same 2D Gaussian + base level model over each one of the  NVSS coordinates pairs we obtained the peak flux density at 6.6\,GHz. The fit was performed only if the SRT peak in the NVSS position is $> 3 \sigma$, where $\sigma$ is the noise of the map, or 1.6\,mJy/beam/pixel for the SRT C-band map.  Out of the all the sources present in the initial NVSS catalog, we detected 93 counterparts at 6.6\,GHz inside the field of the C-band SRT map broadly delimited by the black contours in Fig.\,\ref{fig:NVSS}.

Finally, we merged the CATS catalog with the SRT C-band catalog leading to a final  list of point sources containing both the source positions and all the spectral information available in the literature for these objects.
The radio spectra of all these compact sources were modeled with a modified power: \begin{equation}
    log (F) = -\alpha \cdot log \left( \frac{\nu}{\nu _{ref}} \right) + log (A) + k \cdot e^{ -log \left(\frac{\nu}{\nu _{ref}} \right) }
    \label{eq:modif-power-law}
,\end{equation}
where $k$ is a curvature term, $A$ is the amplitude at the reference frequency $\nu _{ref}$, and $\alpha$ is the spectral index of the unmodified power law. The model converges to the usual power law with index $\alpha$ in the limit $k\rightarrow 0$. We adopted for reference frequency the fixed value $\nu_{ref}=1$\,GHz. The observed radio spectra were fitted with the modified power law model when at least three data points were available, otherwise we fixed the curvature to $k=0$, resorting to an ordinary power law. 

\begin{figure*}[!h]
    \centering
    \begin{tabular}{cc}
    \includegraphics[scale=0.45]{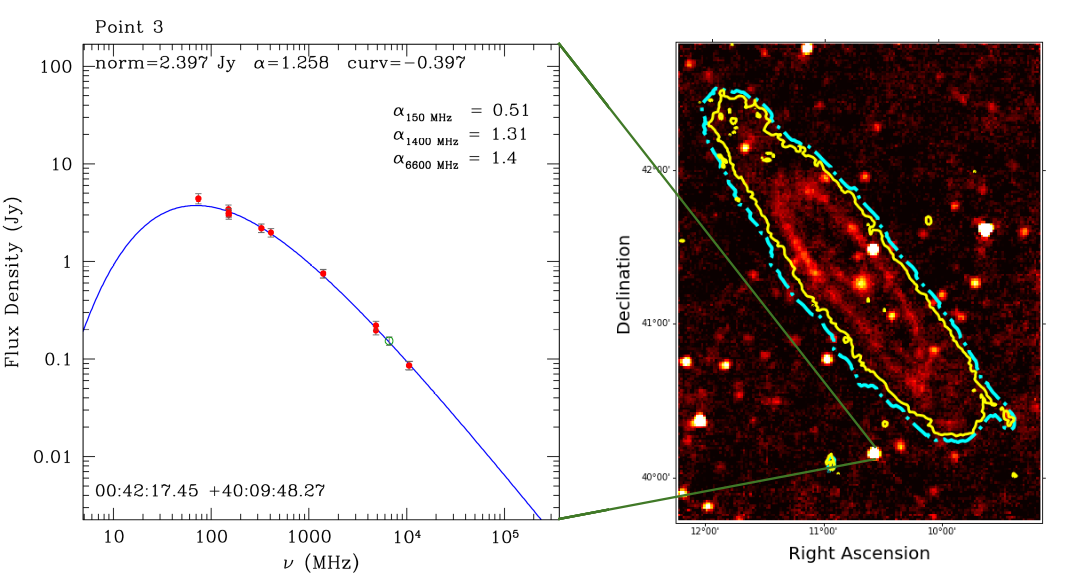}
    \end{tabular}
    \caption{Fit of the spectrum of the source 4C\,+39.03 detected in the SRT map. The green open point represents the SRT flux density measurement at 6.6 GHz. The filled red dots represent the measurements taken from the literature. The values of the "local" spectral indexes at 0.15, 1.4, and 6.6\, GHz are also reported. In the right panel we also outline the { \it Planck} 857 GHz and the { \it Herschel} SPIRE $250 \, \mu$m $ 5 \sigma $ contours in dot-dashed cyan and yellow, respectively.}
    \label{fig:example fit}
\end{figure*}

For significantly curved spectra, only the spectral index between a very close couple of frequencies is meaningful. Nonetheless, a "local" spectral index value at a given frequency, $\alpha _{\nu}$, can be easily computed differentiating Eq. \ref{eq:modif-power-law}  with respect to $log \left( \frac{\nu}{\nu _{ref}} \right )$:

\begin{equation}
    \alpha_{\nu} = \alpha - k \cdot log\left(  \frac{\nu}{\nu _{ref}} \right) e^{-log\left(\frac{\nu}{\nu _{ref}}\right)}
.\end{equation}
This can be viewed as the spectral index between an infinitesimally close couple 
of frequencies at $\nu$.

An example of a spectral fitting of one of the sources is shown in Fig. \ref{fig:example fit}. In general, we observe very good agreement between the SRT flux densities and the measurements from the literature. Moreover, in Fig.\,\ref{fig:example fit} we show that in the case of this bright source there is also 
evidence that the SRT {\it in-band} spectral index measured at four distinct frequencies between 6.0 and 7.2\,GHz is consistent with the trend of both the model fit and the measurements at nearby frequencies taken from the literature.

\begin{figure}[!h]
    \centering
    {\includegraphics[scale=0.4]{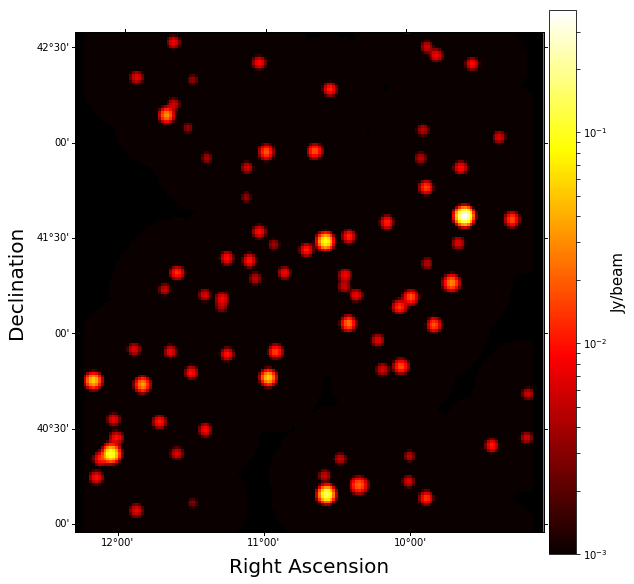} }
    \caption{The SRT C-band point-source model.}
    \label{fig:SRT+SOURCES and subtracted}
\end{figure}

The image of the SRT point-source model at 6.6\,GHz is shown in Fig.\,\ref{fig:SRT+SOURCES and subtracted} at an angular resolution of 2.9\arcmin.  We recall that the SRT at C-band point-source model is a subset of the larger initial catalog based on the NVSS which comprises a larger field in the sky.  Indeed, by using the NVSS-based coordinates and the best-fit parameters of the modified power law in Eq. \ref{eq:modif-power-law}, it is in principle possible to generate a point-source model image at any desired frequency and angular resolution.
Moreover, this procedure was followed to remove the point-source contribution from the global M31 spectral energy distribution (SED) in \citet{bat19}. 

A catalog of all the compact radio sources in the SRT map and the best-fit parameters are reported in Table \ref{Tab:SRT_Sources_catalogue}, while in Fig. \ref{fig:all_fit} we show all the 93 fits of the spectra. Table \ref{Tab:SRT_Sources_catalogue} also reports the spectral indexes measured between 1.4 and 6.6\,GHz, $\alpha _{1.4GHz} ^{6.6\,GHz}$. The data at 1.4\,GHz refer to the NVSS catalog. 

\subsection{Spectral index distribution}

\begin{figure}
    \centering
    {\includegraphics[scale=0.65]{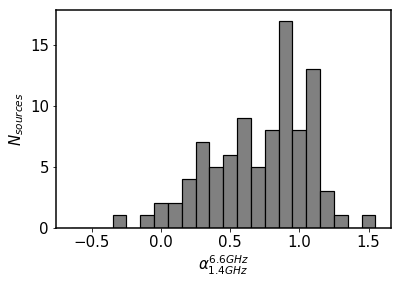} }
    \caption{Spectral index distribution between 1.4 and 6.6\,GHz for the 93 detected point sources in the field of M31.}
    \label{fig:S_index_Hist}
\end{figure}

After obtaining the spectral index $\alpha _{1.4\,GHz} ^{6.6\,GHz}$ for each source detected in C-band, it is possible to study the spectral index  distribution, which is shown in Fig. \ref{fig:S_index_Hist}. The average spectral index value is 0.72, while the median value is 0.81. The spectral index bin with most occurrences is $0.85-0.95$.

As expected, in most cases, the spectral index exhibits a positive value. Only 2 out of the 93 sources exhibit a negative spectral index, or equivalently an inverted spectrum. Approximately 74 \% of the sources exhibit a spectral index of $\alpha > 0.5, $ indicating that their emission is most likely optically thin nonthermal synchrotron emission. This population is typically composed of background sources and/or supernova remnants (SNRs) in the M31 field, as has been shown by \cite{gal14} and \cite{lee14}.

\subsection{Differential source counts}

Given the C-band flux densities for each of the 93 sources, the differential source counts were computed by binning the 0.007-0.45 Jy flux density range according to the values reported in Table \ref{Tab. SourceCount}. It is possible to compare the source counts with a theoretical model: here we use the latest version of the 4.8\,GHz model\footnote{http://w1.ira.inaf.it/rstools/srccnt/} by \cite{bonato}. As our measurements  were performed at $6.6\,GHz$, we extrapolated them to 4.8\,GHz flux density for each source, according to:\begin{equation}
    F_{4.8\,GHz} = F_{6.6\,GHz} \left( \frac{6.6}{4.8} \right) ^{\alpha _{1.4\,GHz} ^{6.6\,GHz}}
\label{eq:correct_to_4,8GHz}
,\end{equation}
where $\alpha _{1.4\,GHz} ^{6.6\,GHz}$ is the source spectral index; see Table \ref{Tab:SRT_Sources_catalogue}. 

The choice of 0.007 Jy as the lower value to study the differential source counts can be justified assuming a Gaussian distribution, an rms value of $\sim 1.6$\,mJy/pixel within the C-band map extrapolated to 4.8\,GHz, and a 3$\sigma$ peak cut-off. We find that the sources with flux density of 0.007 Jy are detected with a probability of greater than $95 \%$. It is important to stress that a non-negligible fraction ($\sim 19 \% $) of the detected sources have extrapolated flux densities at 4.8\,GHz lower than 0.007 Jy; we do not make use of these. 

Before comparing the measurements extrapolated to 4.8\,GHz with the source-count model, the source counts $n(F)$ derived from the SRT flux densities were fitted with a power law $n(F) = A \left(  \frac{F}{Jy} \right)^{-\gamma}$, where both the number density of sources per flux density interval at 1\,Jy $A$ and the spectral index $\gamma$ are free parameters. The  result of the fit in logarithmic scale is:
\begin{equation}
    \frac{dN}{dF} = \left( 300 \pm 80 \right) \left( \frac{F}{Jy} \right)^{-1.98 \pm 0.07} Jy^{-1} sr^{-1}
.\end{equation}

\begin{figure}
    \centering
    {\includegraphics[scale=0.6]{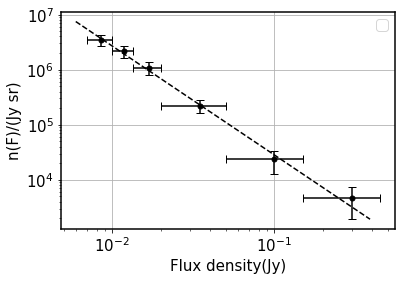} }
    \caption{Point-source differential source counts and the fitted power law in double log scale. The error bars parallel to the flux density axis indicate the bin widths.}
    \label{fig:SourceCounts}
\end{figure}

The differential source counts and the best-fit curve are shown in Fig. \ref{fig:SourceCounts}. The horizontal error bars represent the bin widths, while the vertical ones correspond to the Poisson error associated with the measurements. 

\begin{center}
\begin{table}
\begin{tabular} { c c c c }
\hline
Range (Jy) & Count & $Log_{10}$(Diff. Count) & Theor. count\\
\hline\hline
0.007 - 0.01 & 22 & 6.54 & 21 \\
\hline
0.01 - 0.0135 & 16 & 6.33 & 14\\
\hline
0.0135 - 0.02 & 15 & 6.03 & 14\\
\hline
0.02 - 0.05 & 14 & 5.34 & 18\\
\hline
0.05 - 0.15 & 5 & 4.37 & 7\\
\hline
0.015 - 0.45 & 3 & 3.67 & 2\\
\hline
\end{tabular}
\caption{Number of sources detected in C-band map in a given flux density bin. In the last column we show the number of expected sources in each bin from the model we are considering.}
\label{Tab. SourceCount}
\end{table}
\end{center}

\begin{figure}[!h]
    \centering
    {\includegraphics[scale=0.6]{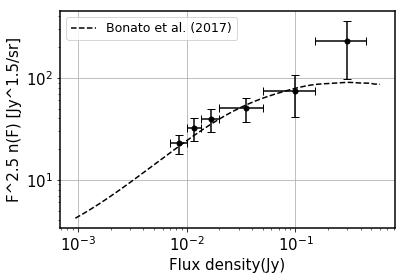} }
    \caption{Normalized differential point-source counts compared with the Bonato et al. (2017) model at 4.8\,GHz.}
    \label{fig:Source+DeZotti}
\end{figure}

At this point we can compare our source sample with the model from \cite{bonato}. This theoretical model returns the predicted Euclidean normalized differential radio source counts per steradian for a discrete series of flux densities. The Euclidean normalized differential counts is defined as $F^{2.5} \frac{dN}{dF}$, where $F$ is in Jy. In Fig. \ref{fig:Source+DeZotti}, our source sample is compared with the 4.8\,GHz model taking into account the $\sim 7$ deg$^2$ observed area. The data are in good agreement with the model for the low-, intermediate-, and high-density values. The tension in the last point in Fig. \ref{fig:Source+DeZotti} can be solved by removing from the counts the BLAZAR $B3-0035+413$ that at 4.8\,GHz has an extrapolated flux density of $\sim 0.44 Jy$.  

Integrating the model from 0.007 to 0.45 Jy, we find that the expected total number of sources is $\sim 76$, while in the same range we detect 75 sources. This appears to confirm the agreement between the detected and expected point sources.

\subsection{The SRT C-band sample of supernova remnants}

\begin{figure}[!h]
    \centering
    \hspace*{-0.2cm}\includegraphics[scale=0.45]{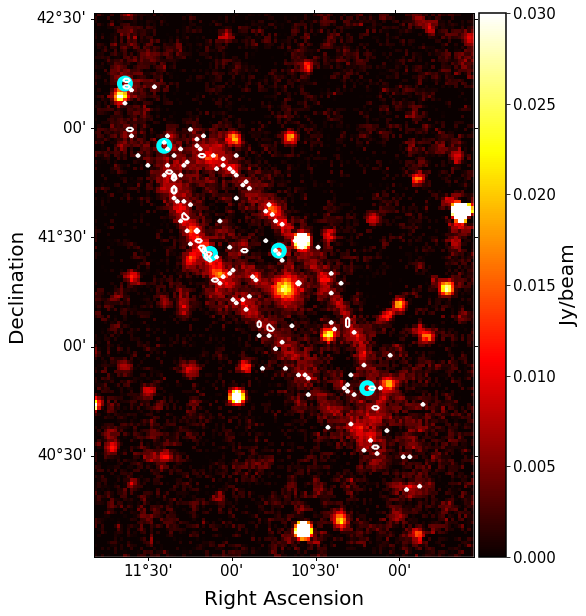} 
    \caption{The SRT C-band map with compact sources. The 156 SNRs and
SNR candidates from \citep{lee14} are overplotted with white contours. The SRT sources that have a corresponding source within a radius of 1.45’  are shown with cyan circles.
}
    \label{fig:SNR}
\end{figure}

With the previously compiled catalog of compact sources detected in our map, we can perform a correlation between SNRs and SNR candidates in the field of M31. For this purpose, we used the \cite{lee14} SNR and SNR candidate catalog, which includes 156 sources, obtained with H$\alpha$ and {\sc [Sii]} images of M31. Using a $1.45^\prime$ search radius we cross-matched the SNRs and SNR candidates catalog with our sample of point sources. The choice of a search radius of $1.45^\prime$ is motivated by the size of the FWHM$_{SRT}=2.9^\prime$. We found then that source numbers 241, 359, 622, 626, and 645 have at least one corresponding source in the SNR catalog. In particular, for the SRT source number 645 we have four corresponding sources within $1.45^\prime$, three corresponding sources for the SRT source number 622, and 2 for number 241. The other two SRT sources have only one corresponding source each. In Fig. \ref{fig:SNR} we  show all the 156 SNRs and SNR candidates (white contours) and the sources that have a corresponding source within a radius of $1.45^\prime$ (cyan circles).

In Table \ref{Tab:SNR_candidates} we report the list of sources with a match in the \cite{lee14} catalog, together with the ID in that same catalog, the distance from the center, and the spectral index between 1.4 and 6.6\,GHz.

\begin{table}[]
\centering
\begin{tabular} { c c c c }
\hline
SRT source n. & Lee SNR n. & Distance($ ^{\prime}$) & $\alpha _{1.4\,GHz} ^{6.6\,GHz}$ \\
\hline
241 & 50 & 0.87 & 0.42\\
\hline
359 & 17 & 0.93 & 0.49\\
\hline
622 & 156 & 0.96 & 0.16\\
\hline
626 & 144 & 0.78 & 0.31\\
\hline
645 & 97 & 0.09 & 0.01\\
\hline
\end{tabular}
\caption{The SRT C-band sources with a correspondence in the Lee et al. (2014) catalogue.}
\label{Tab:SNR_candidates}
\end{table}

We note that the sources with a corresponding source in the SNR catalog are mostly characterized by a relatively flat spectrum. This is a bias most likely introduced by the achieved sensitivity at 6.6\,GHz with SRT: if we assume two sources with the same 1\,GHz amplitude, it is much more likely to detect the source with a flat spectrum than the source with a steeper one.

\section{M31 morphology}
\label{sec:5}

\begin{figure}
\hspace*{-0.2cm}  
    \centering
    {\includegraphics[scale=0.42]{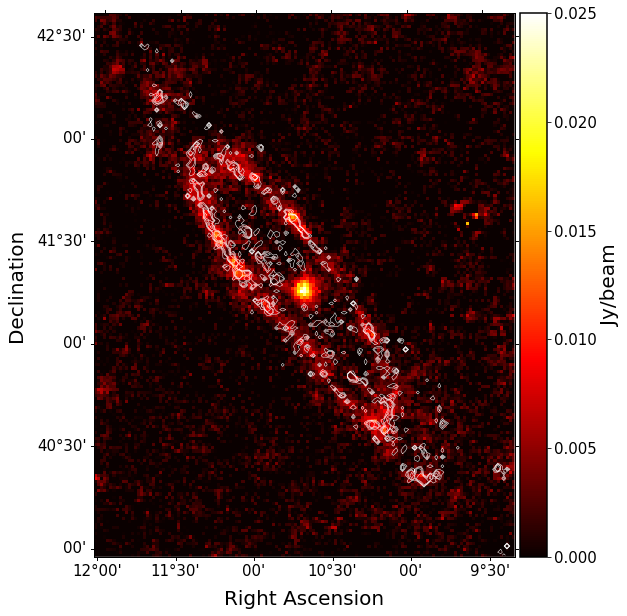} }
    \caption{The SRT C-band map of M31 with selected {\sc Hii} region contours. We note the remarkable agreement between the distribution of the {\sc Hii} regions and the location of the 10 kpc ring and the 15 kpc structure. }
    \label{fig:SRT+HII contour}
\end{figure}

Studies of M31 have revealed a normal disk galaxy with spiral arms wound up clockwise. The spiral arms are separated from each other by at least 4 kpc. The spiral pattern is somewhat distorted by the interactions of the galaxy with M32 and M110 \citep{gor06}. The {\it Spitzer Space Telescope} (SST)  revealed two spiral arms emerging from a central bar. The arms have a segmented structure and continue beyond the galaxy’s ring mentioned above. The ring shape of M31 was discovered in the 1970s \citep{poo69,ber74}. Comparing the radio emission at 6.6\,GHz and the location of the 3691 {\sc Hii} regions presented in \cite{azi11} we find a strong correlation in the ring, as one can see in Fig. \ref{fig:SRT+HII contour}. The {\sc Hii} level contours denote 1, 3, or 6 {\sc Hii} regions inside each SRT C-band pixel, and so they are not flux density contours but are simply an indication of  the positions of the {\sc Hii} regions. 

The ring structure is  believed to be the result of the interaction with M32 over 200 million years ago. The M32 galaxy likely passed through the disk of M31, which left M32 stripped of more than half of its mass \citep{sou18}. The  nucleus of  M31 is home to a dense, compact star cluster. A total of 35 black holes have been detected in M31, of which 7 are within 0.3 kpc of its center \citep{bar13}. These black holes formed by gravitational collapse of massive stars and have a mass of between approximately five and ten times that of the Sun.

Radio and microwave observations revealed emission from the central region and only from the external ring, while the central bar is not detected. The central bar, which is detected in all the IR images, is also not visible in the ancillary Effelsberg radio maps. As one can clearly see in Fig. \ref{fig:SRT+HII contour}, the 10 kpc ring and the 15 kpc structure, noted in IR maps, \citep{haa98, gor06} are well correlated with the location of the {\sc Hii} regions in the SRT map. The 15 kpc structure is unfortunately at the edge of the Effelsberg maps, and  in the following section, when these maps are used, it is not possible to disentangle thermal and nonthermal emission within these external regions.

The M32 and NGC205 galaxies, which fall inside the SRT coverage, are not detected because of their low radio emission. \cite{bro11} found that, at 1.4\,GHz, M32 and NGC205 have respectively a flux density of $ 0.7 \pm 0.5$ mJy and $0.1 \pm 0.5$ mJy. Also, if a flat radio spectral index is assumed for these galaxies (most favorable scenario), their flux densities are compatible with the SRT confusion noise and therefore cannot be detected. \newline

In the following, we investigate the M31 radio features by creating a spectral index map, and by studying the spectral index trend within the central region and the entire galaxy.

\subsection{Spectral index map}
\label{sec:5.1}
We start by building a spectral index map across the whole M31. In general, a flat spectral index is expected in regions whose emission is dominated by a thermal component, while a steeper spectrum is an indication of a substantial presence of nonthermal emission.

The spectral index value has been calculated on a pixel-by-pixel basis by taking into account only the pixels with flux densities greater than $3 \sigma$ in all considered maps, where $\sigma$ is the map  sensitivity expressed in Jy beam$^{-1}$. Indeed, pixels below this threshold are those where the spectral index becomes very uncertain because of the presence of faint radio emission and radio noise.
The spectral index map was created using Effelsberg maps at $1.46$\,GHz and $4.85\,$GHz and the SRT map at 6.6\,GHz. We used the maps projected and then convolved to $3^{\prime}$ as explained in Sect. 3.\newline

\begin{figure*}[!h]
    \centering
    \begin{tabular}{cc}
        \includegraphics[scale=0.4]{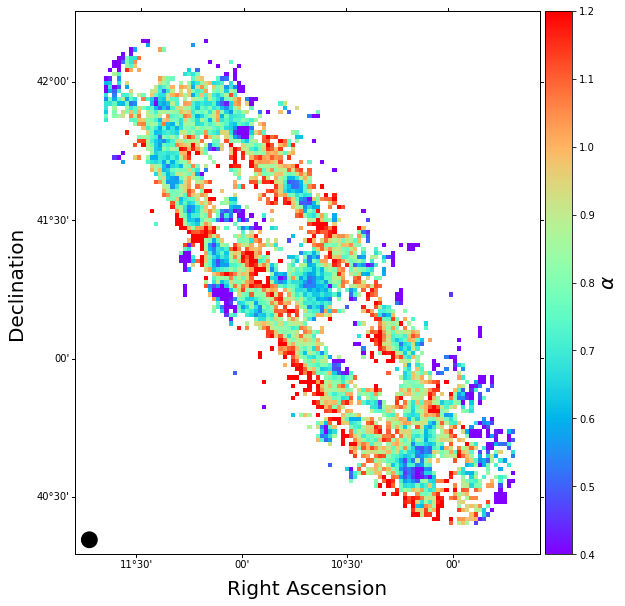} &
        \includegraphics[scale=0.4]{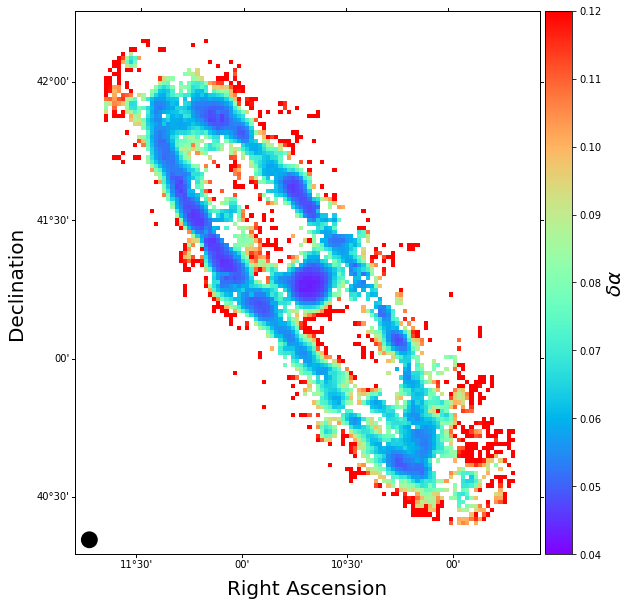}\\
    \end{tabular}
    \caption{Left panel: Spectral index map extracted from 1.46 and 4.85 GHz Effelsberg maps and our SRT 6.6 GHz map. Right panel: Spectral index 1$\sigma$ noise map. The typical beam size (FWHM$=2.9^{\prime}$) is indicated with the black circle in the bottom-left corner of each panel.}
    \label{fig:S_indexMap_Eff+SRT}
\end{figure*}

The spectral index value and its $1 \sigma$ uncertainty were computed pixel by pixel by fitting the relation: 
\begin{equation}
    f(\nu) = A \left(\frac{\nu}{1\,GHz} \right) ^{-\alpha} 
    \label{eq:power_law}
,\end{equation}
where both $A$ and $\alpha$ are free parameters. To perform the fitting, we used a Python implementation of Goodman and Wear's Markov chain Monte Carlo (MCMC) Ensemble sampler \citep{emcee1}. The same MCMC was used in the following sections, to fit the emission in the central region and that in the ring, and to disentangle thermal and nonthermal emission maps. The spectral index map and its $1 \sigma$ uncertainty are shown in Fig. \ref{fig:S_indexMap_Eff+SRT}.

Globally, the spectral index map shows values of $\alpha \simeq 0.7$ (light blue and light green pixels). In the northern and southern ring and in the central region, we can also find small patches (violet and blue pixels) not located at the edge of the map where the SRT and Effelsberg S/N is lower, with a spectral index around $\alpha=0.4-0.5$. These are the regions where thermal free-free emission is expected to be dominant and are likely to be {\sc Hii} regions, which are mainly located in these parts of the galaxy, as shown in \cite{azi11}. In some inner and outer ring regions we find that the spectral index is larger than 0.9. This means that these regions are characterized mostly by synchrotron emission, with the steep slope indicating that an old population of electrons is responsible for the nonthermal emission.

Spectral index maps of M31 have been built before \citep{ber03, beck20}. We find excellent agreement between our spectral index map shown in Fig. \ref{fig:S_indexMap_Eff+SRT} and the one extracted by \cite{ber03} , which is not surprising given the fact that, with the exception of the SRT C-band map, to extract the spectral index we used the same set of maps, at the same angular resolution. The spectral index values found by \cite{beck20} using Effelsberg maps at 1.465 and 8.350 GHz are generally smaller than ours by 0.2-0.3.
We can ascribe this discrepancy to the difference in total flux density between the SRT 6.6 GHz map and the new Effelsberg 8.35 GHz map. More specifically, the latter shows a flux density that is $\sim 15$\% higher than the former. We identified some factors that could account for this discrepancy:\begin{itemize}
    \item The point sources in the two maps have been subtracted with different methods. Residual point sources in one map that have been removed from the other can affect the flux difference in both directions.
    \item Maps have been made with different telescopes, the data have been analyzed independently, and different software has been used.
    \item The total area covered by the two maps is different; this can affect the background estimate and the baseline subtraction.
\end{itemize}

Moreover,  \cite{bat19}   showed that AME is the dominant emission mechanism in M31 in the frequency range 10-60 GHz. This means that, at 8.35 GHz, AME must make up $10-15$\% of the total integrated flux density, causing the spectral trend to drift away from a simple power law. Therefore, AME can account for part of the difference between the two spectral index maps.

In the following sections, we explain how we used our spectral index map  to choose regions in which to perform a detailed study of the decomposition between thermal and nonthermal emissions.

\subsection{Spectral index trend}
\label{sec:5.2}
Hereafter, we evaluate the spectral index gradient across the galaxy and, in doing so, we separately analyze the properties of the central region. For this analysis we used the Effelsberg map at 2.7\,GHz and all the other radio maps projected and convolved to $5 ^{\prime}$ as explained in Sect. \ref{sec:ancillaryData}. We fitted a simple power law with two free parameters using the integrated flux densities in each given region instead of single pixel values, following the same procedure for the extraction of the spectral index map.

For the galaxy central region, we evaluated the spectral index within four concentric rings centered at $(Ra, Dec) = (10.70, 41.28)$ deg (see Table \ref{Tab. Core Rings} for more details). For each ring, the total flux density was evaluated as:
\begin{equation}
    F =  \left(\bar{S}_{in} - \bar{S}_{ext} \right) \Omega_{in}
,\end{equation}
where $ \bar{S}_{in} $ is the brightness inside the integration region, $ \bar{S}_{ext} $ is the background brightness and $\Omega_{in}$ is the solid angle subtended by the region. The thermal brightness uncertainty was computed as explained in \cite{flux_unc} and then combined with the calibration error\footnote{The calibration errors are considered to be $5 \%$ for all the maps.}.

\begin{table}
\scalebox{1}{
\begin{tabular} { c c c c c }
\hline
Ring n. & Galactoc. & Spectral & $\sigma$ \\
        & Distance(kpc) &  Index & S.Index \\
\hline\hline
1 & 0-0.34 & 0.62 & 0.14\\
\hline
2 & 0.34-1.02 & 0.63 & 0.10\\
\hline
3 & 1.02-1.70 & 0.68 & 0.09\\
\hline
4 & 1.70-2.38 & 0.77 & 0.09\\
\hline
\end{tabular}
}
\caption{Spectral index values in the central region as a function of galactocentric distance.}
\label{Tab. Core Rings}
\end{table}

\begin{figure}[h!]
    \centering
    {\includegraphics[scale=0.4]{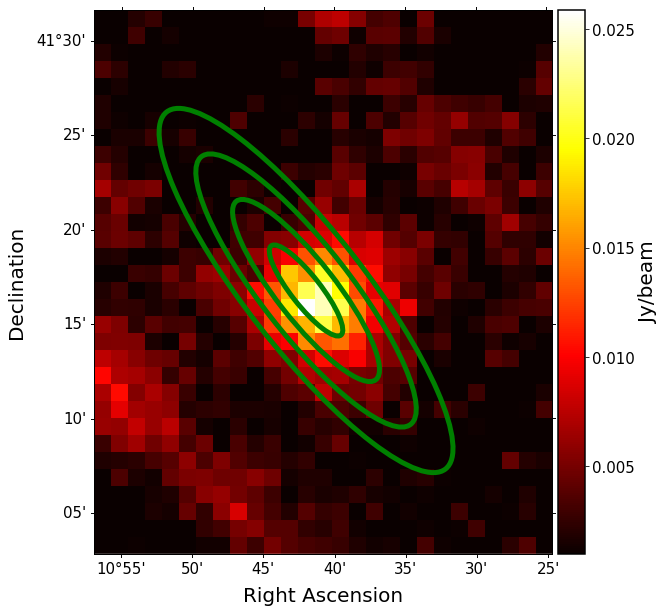}}
    \quad
    {\includegraphics[scale=0.6]{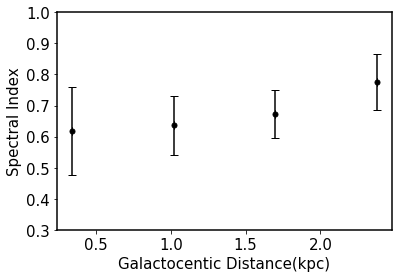} }
    \caption{Upper panel: Four elliptical rings used to compute the spectral index trend with the distance from the center of M31. Lower panel: Spectral index average radial profile calculated within the rings shown above.}
    \label{fig:S_index_radius}
\end{figure}

The spectral index as a function of distance from the galaxy center and the four rings where it was calculated are shown in Fig. \ref{fig:S_index_radius}. For each ring we used a position angle of $-52$ deg in the $(Ra, Dec)$ system and an eccentricity equal to 0.26.
The spectral index within the central region is found to be approximately constant within the uncertainties. We further investigate the central region emission in Sect. \ref{sub_s:M31 core} by decomposing thermal and nonthermal emission.

We also repeated the same analysis for the whole galaxy: in this case we used seven elliptical rings, centered at $(Ra, Dec) = (10.70, 41.28)$ deg, with the same position angle and eccentricity as those used for the central region. The details of the seven regions and the spectral index values are reported in Table \ref{Tab:GalaxyEllipticalRings}.

\begin{table}
\scalebox{1}{
\begin{tabular} {c c c c c }
\hline
Ring n. & Galactoc. & Spectral & $\sigma$ \\
        & Distance(kpc) &  Index & S.Index \\
\hline\hline
1 & 0-1.13 & 0.66 & 0.05\\
\hline
2 & 1.13-3.40 & 0.90 & 0.05\\
\hline
3 & 3.40-5.66 & 1.05 & 0.04\\
\hline
4 & 5.66-7.92 & 0.94 & 0.04\\
\hline
5 & 7.92-10.18 & 0.83 & 0.04\\
\hline
6 & 10.18-12.45 & 0.87 & 0.05\\
\hline
7 & 12.45-14.14 & 1.01 & 0.05\\
\hline
\end{tabular}}
\caption{Spectral index values as a function of the galactocentric distance for the whole galaxy.}
\label{Tab:GalaxyEllipticalRings}
\end{table}
 
\begin{figure}
    \centering
    {\includegraphics[scale=0.4]{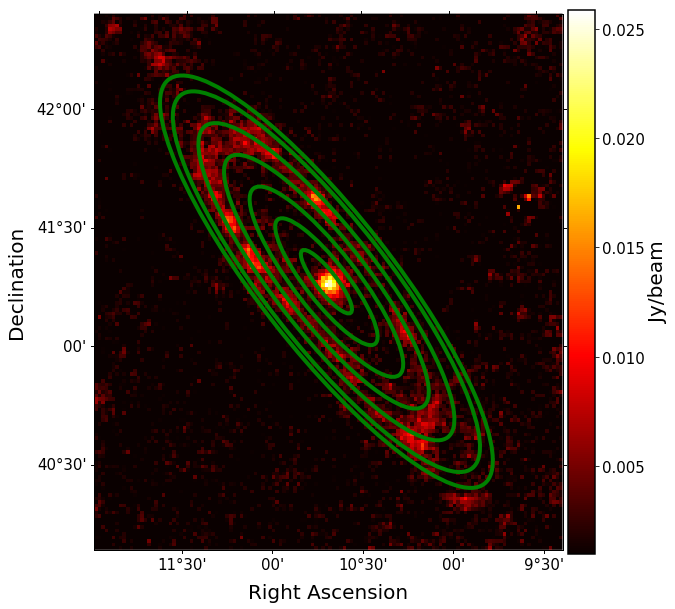}}
    \quad
    {\includegraphics[scale=0.6]{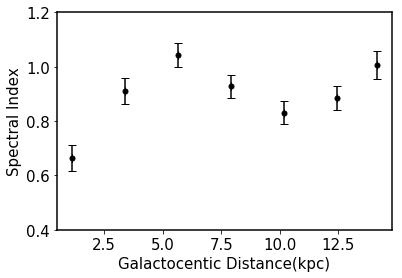} }
    \caption{Upper panel: Seven elliptical rings used to compute the spectral index trend with the distance from the galaxy center. Lower panel: Spectral index average radial profile calculated within the rings shown above.}
    \label{fig:S_index_radius_allGal}
\end{figure}

The seven elliptical rings and the spectral index as a function of the distance from the galaxy center are  shown in Fig. \ref{fig:S_index_radius_allGal}. 
The spectral index shows a clear positive gradient in the innermost 6 kpc (first three rings). In the outermost rings, the gradient changes sign twice, at $\sim 6$ (decreasing) and at $\sim 10$ kpc (increasing again), even though the values are compatible with a flat trend between $\sim 8$ kpc and $\sim 12$ kpc. Overall the 10 kpc ring presents smaller values with respect to the peak reached within the third ring. The average value of $\sim 0.9$ found at distances greater than $\sim 4$ kpc strongly hints at synchrotron emission being the predominant mechanism at these frequencies.
The point at highest galactocentric distance corresponds to the edge of the ring: as expected, because of the aging of electrons, we found a steeper spectral index value. The same spectral index trend and similar values have been found by \cite{ber03}.

\section{Thermal versus nonthermal emission}
\label{sec:6}

\begin{figure}
    \centering
    {\includegraphics[scale=0.4]{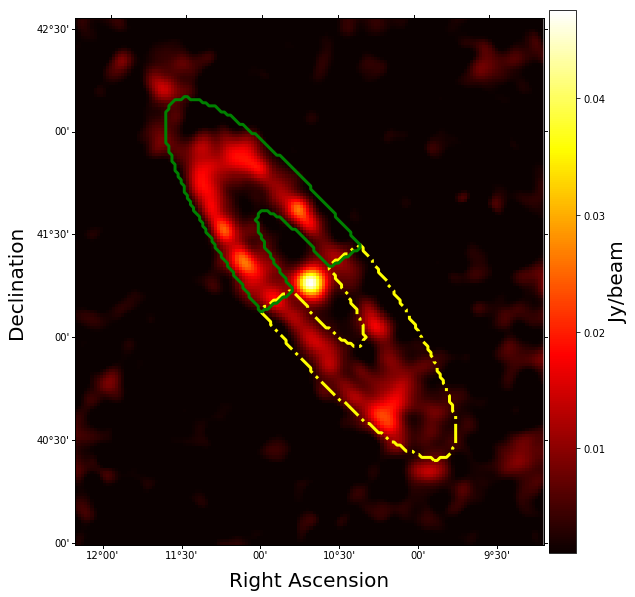}}
    \caption{Two regions we selected as {the north} and {south ring} for disentangling the total thermal and nonthermal emission. }
    \label{fig:2 ring regions}
\end{figure}

In this section we report a full decomposition of the M31 emission into thermal and nonthermal components. We start by analyzing the central region and the full 10 kpc ring individually, and then the northern and the southern halves of the 10 kpc ring individually as well (Fig. {\ref{fig:2 ring regions}}). It is interesting to conduct a separate analysis on the latter two regions because they appear to have different properties in  the spectral index map.

For this decomposition, we used the maps reported in Table \ref{Tab: Fluxes}, all projected to the SRT C-band map coordinate system and convolved to a resolution of $5^\prime$.
For all the regions mentioned above, the flux densities were fitted with a synchrotron+free-free model with three free parameters: the free-free and synchrotron amplitudes at 1\,GHz, and the synchrotron spectral index.

\begin{equation}
    F_{sync} (\nu) = A_{sync} \nu ^{-\alpha _{sync}}
    \label{Eq:Sync}
,\end{equation}

\begin{equation}
    F_{f-f} (\nu) = A_{f-f} T_{e} ^{-0.5} g_{ff}
    \label{Eq:Free-F}
.\end{equation}

We used Eqs. \ref{Eq:Sync} and \ref{Eq:Free-F} to describe synchrotron and free-free emissions,  respectively. The electron temperature $T_e$ in Eq. \ref{Eq:Free-F} has been fixed at $8000 \, K$ throughout the paper. Fixing the electron temperature is not fundamental in the decomposition between thermal and nonthermal emission because the free-free amplitude is left as a fit parameter, but it is important in the SFR extraction. In the latter case, we fix $T_e = 8000 \, K$ as we expect the 10 kpc ring to host most of the M31 star forming regions \citep{for13} and we know that in our Galaxy the electron temperature at around a galactocentric distance of 10 kpc  is $\sim 8000 \, K$ \citep{pal04}.

A summary of the flux densities we used as data points for the fit, divided by region, is reported in Table \ref{Tab: Fluxes}, where the reported flux density uncertainties are a combination of statistical and calibration errors. A summary of the fit results is reported in Table \ref{Tab. Global}.

\begin{table*}[tb]
    \centering
    \begin{center}
        \begin{tabular} {c c c c c c}
        \hline
        Map & Frequency(GHz) & Central Region [Jy] & 10 kpc Ring [Jy] & North Ring [Jy] & South Ring [Jy]\\
        \hline\hline
        VLA+EFF21 & 1.46 & $0.252 \pm 0.020$ & $3.47 \pm 0.17$ & $1.858 \pm 0.093$ & $1.526 \pm 0.077$\\
        \hline
        EFF11 & 2.7 & $0.184 \pm 0.010$ & $2.04 \pm 0.10$ & $1.130 \pm 0.057$ & $0.861 \pm 0.043$\\
        \hline
        EFF6 & 4.85 & $0.113 \pm 0.006$ & $1.20 \pm 0.06$ & $0.671 \pm 0.034$ & $0.506 \pm 0.025$\\
        \hline
        SRT & 6.6 & $0.087 \pm 0.005$ & $0.90 \pm 0.05$ & $0.493 \pm 0.025$ & $0.388 \pm 0.020$\\ 
        \hline
        \hline
        \end{tabular}
    \end{center}
\caption{Flux densities used as data points for the fit. The flux density uncertainties are a combination of statistical and calibration errors.}
\label{Tab: Fluxes}
\end{table*}

Finally, in Sect. \ref{sec:6_th&nth} we carry out a detailed separation into components, pixel by pixel, for the whole galaxy and produce thermal and nonthermal emission maps.

\subsection{Central region}
\label{sub_s:M31 core}

In order to evaluate the total flux density in the central area, we used a circular region centered at $(Ra, Dec) = (10.70, 41.28)$ deg with a radius of $6^\prime$. 

\begin{figure}[h!]
    \centering
    {\includegraphics[scale=0.55]{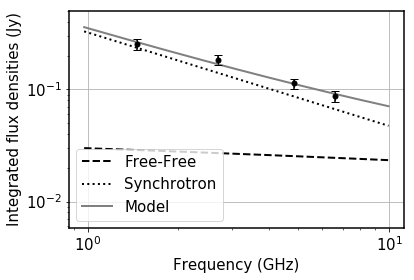} }
    \caption{Fit to the flux densities of the central region (see Table \ref{Tab. Global}). The fit is characterized by $\chi^2=4.0$. The error bars in the plot correspond to 2$\sigma$.}
    \label{fig:core.fit}
\end{figure}

In Fig. \ref{fig:core.fit} we show the data points and the thermal, nonthermal, and total emission fit models. The final $\chi^2$ is 4.0.
According to the fit, at 1\,GHz, the thermal emission within the central region is about $ \sim 8 \%$ of the total emission, while extrapolating the flux density to the SRT C-band frequency we conclude that the total thermal emission in the central region, at 6.6\,GHz, is $\sim 26 \%$ of the total emission.

\subsection{10 kpc ring}
\label{sec:6.2}

For the 10 kpc ring, the flux density was extracted inside a region bounded by two ellipses, one with a major semi-axis of $25'$ and another with major semi-axis of $65'$, both with the same eccentricity  of 0.26 and  the  same position angle of $-52$ deg in the $(Ra, Dec)$ coordinate system.

\begin{figure}[h!]
    \centering
    {\includegraphics[scale=0.55]{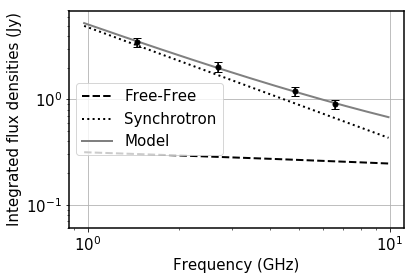} }
    \caption{Fit to the flux densities of the ring (see Table \ref{Tab. Global}). The fit is characterized by $\chi^2=0.6$. The error bars in the plot correspond to 2$\sigma$. } 
    \label{fig:ring fit}
\end{figure}

In Fig. \ref{fig:ring fit} we plot the data points and the fit model that yields a total $\chi^2$ of 0.6. The synchrotron spectral index is found to be compatible with the value found by \cite{ber03},  $\alpha_{sync}= 1.0 \pm 0.1,$ and with previous studies of the M31 SED.

We find that, at 1\,GHz, the thermal emission within the ring is about $7 \%$ of the total emission.
At 6.6\,GHz, the whole ring emission is dominated by nonthermal emission, contributing to $\sim 70\%$ of the total emission budget. The $\sim 30\%$ thermal fraction is similar to the value of 26\% that we extracted in Sect. \ref{sub_s:M31 core} for the central region, indicating that the total contribution of the thermal emission at 6.6\,GHz in not negligible. 

\subsection{Northern ring region}
\label{sec:6.3}

\begin{figure}[h!]
    \centering
    {\includegraphics[scale=0.55]{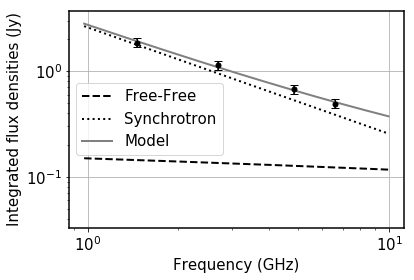} }
    \caption{Fit to the flux densities of the northern half of the 10kpc ring (see Table {\ref{Tab. Global}}). The fit is characterized by $\chi ^2 = 1.5$. The error bars in the plot correspond to 2$\sigma$. }
    \label{fig:up ring}
\end{figure}

In the following, we present a repetition of the previous analysis, this time focusing on the northern half of the 10 kpc ring shown by the green region in Fig. \ref{fig:2 ring regions}. For this purpose, we used the same ellipses as in Sect. \ref{sec:6.2}.
Data points and the fit model are reported in Fig. \ref{fig:up ring}. The fit is characterized by a $\chi ^2 = 1.5 $.

We note that in the upper ring the free-free emission at 1\,GHz is only a fractional part of the total emission. Moreover, we have only a marginal detection for the free-free emission: the 1\,GHz free-free amplitude is $\sim 1.5 \sigma$, which is compatible with zero emission.\newline

The thermal flux density, extrapolated to {6.6\,GHz} is about one-third of the synchrotron emission, while at 1\,GHz it is 16 times weaker.

\subsection{Southern ring region}
\label{sec:6.4}

We repeated here the analysis we carried out in Sect. \ref{sec:6.3},
this time focusing on the southern half of the 10 kpc ring, the region delimitated by yellow dash-dotted lines in Fig. {\ref{fig:2 ring regions}}. Again, for this purpose, we used the same ellipses as in Sect. {\ref{sec:6.2}}.

\begin{figure}
    \centering
    {\includegraphics[scale=0.55]{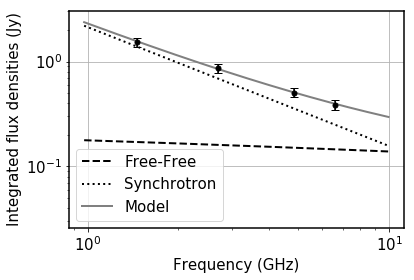} }
    \caption{Fit to the flux densities of the southern half of the 10kpc ring (see Table {\ref{Tab. Global}}). The fit is characterized by $\chi ^2 = 0.3$. The error bars in the plot correspond to 2$\sigma$.}
    \label{fig:down ring}
\end{figure}

\begin{figure*}[!h]
    \centering
    \begin{tabular}{cc}
        \includegraphics[scale=0.4]{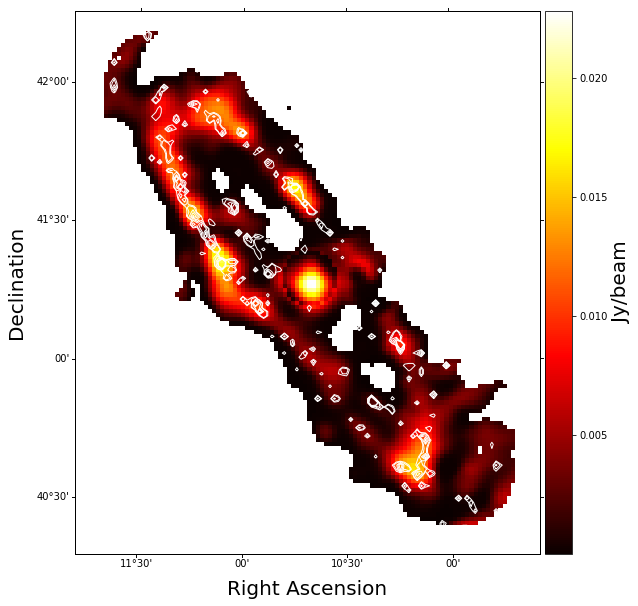} &
        \includegraphics[scale=0.4]{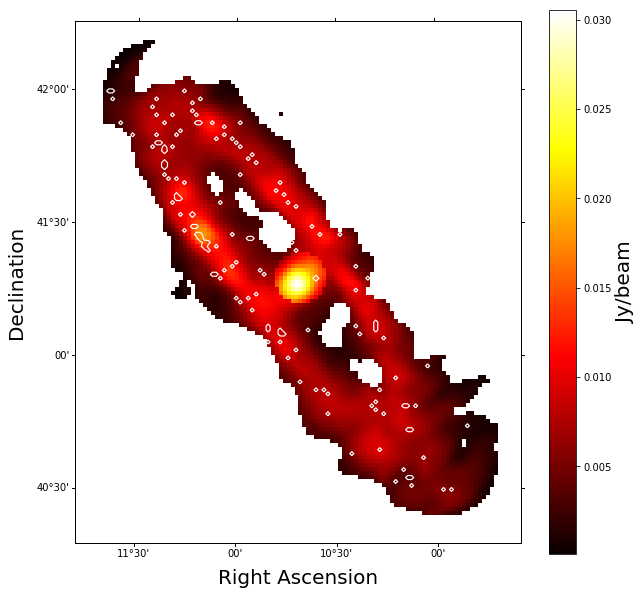}\\
    \end{tabular}
    \caption{Left panel: the thermal (free-free) emission map at 6.6\,GHz with {\sc Hii} contours. Contour levels indicate 2, 3, 6, 9, and 12 {\sc Hii} regions inside the C-band pixel. Right panel: Same as left panel, but for the synchrotron component (nonthermal) with SNR contours. Contour levels show 1, 2, and 3 SNRs inside the C-band pixel. The resolution of both thermal and nonthermal maps is $5^{\prime}$.}
    \label{fig:emissionmap}
\end{figure*}

From an inspection of the spectral index map, we expect to find a synchrotron spectral index which is higher than what we found for the northern region. Also, looking at the distribution of  {\sc Hii} regions within the ring (Fig. \ref{fig:emissionmap}), we expect a total thermal emission fraction at 6.6\,GHz which is compatible with what we extracted for the northern ring. Figure \ref{fig:down ring} shows the best-fit curve: the fit model is characterized by $\chi ^2 = 0.3$.

As expected, the synchrotron spectral index value found for the southern region is steeper than the one we found for the northern region. However, due to the high uncertainties, the two synchrotron spectral indices turn out to be in agreement.

We note from the extrapolated flux densities at $6.6 \,$ GHz that within the southern ring the nonthermal emission is essentially twice the thermal emission. The integrated thermal emission in the two half ring regions is of the same order of magnitude, as predicted, while nonthermal emission was found to be higher in the northern half of the ring. The thermal emission fraction at $6.6 \, $GHz in the southern half was found to be $38 \%$, while in the northern half it makes up $26 \% $ of the total budget.

\begin{table*}[tb]
    \centering
    \begin{center}
        \begin{tabular} {c c c c c c}
        \multicolumn{6}{c}{Fit results} \\
        \hline  \\
        \multicolumn{2}{c}{Fit parameters} & Central region & 10 kpc ring & North reg. & South reg.\\ 
        \hline \\
        
        \multicolumn{2}{c}{$\alpha_{sync}$} & $0.83^{+0.19}_{+0.10}$ & $1.06^{+0.18}_{-0.12}$ & $1.01^{+0.17}_{-0.10}$ & $1.13^{+0.23}_{-0.16}$ \\ \\
        
        \multicolumn{2}{c}{$A_{sync}^{1GHz}$ [Jy]} & $0.322^{+0.026}_{-0.027}$ & $4.84^{+0.32}_{-0.31}$ & $2.58^{+0.16}_{-0.16}$ & $2.13^{+0.15}_{-0.14}$ \\ \\
        
        \multicolumn{2}{c}{$A_{ff}^{1GHz}$ [Jy]} & $0.029^{+0.029}_{-0.021}$ & $0.31^{+0.24}_{-0.21}$ & $0.16^{+0.14}_{-0.11}$ & $0.18^{+0.11}_{-0.11}$ \\ \\
        
        \multicolumn{2}{c}{$F_{tot}^{6.6GHz}$ [Jy]} & $0.091 \pm 0.028$ & $0.92 \pm 0.27$ & $0.51 \pm 0.14$ & $0.40 \pm 0.13$ \\ \\
        
        \multicolumn{2}{c}{$F_{ff}^{6.6GHz}$ [Jy]} & $0.024 \pm 0.020$ & $0.26 \pm 0.18$ & $0.13 \pm 0.10$ & $0.15 \pm 0.09$ \\ \\
        
        \multicolumn{2}{c}{$F_{sync}^{6.6GHz}$ [Jy]} & $0.067 \pm 0.019$ & $0.66 \pm 0.19$ & $0.39 \pm 0.10$ & $0.25 \pm 0.9$
        \\ \\
        
        \multicolumn{2}{c}{$ \chi ^2$} & $ 4.0 $ & $ 0.7 $ & $ 1.5 $ & $ 0.3 $
        \\ \\
        \hline
        \end{tabular}
    \end{center}
\caption{Summary table of thermal and nonthermal emission decomposition done in the central region, the 10 kpc ring and in the NE and SW halves of the ring. In the first three rows, fit results are reported  while in rows 4, 5, and 6   the total, thermal, and nonthermal emission at $6.6 \,$ GHz are shown, respectively. In the last row we report the goodness of the fit for 1 degree of freedom.}
\label{Tab. Global}
\end{table*}

\subsection{Thermal and nonthermal maps}
\label{sec:6_th&nth}

As anticipated by the spectral index map, we found a typical synchrotron spectral index ($0.5 \leq \alpha \leq 0.8$) in the central region, and a steeper spectrum with $\alpha \geq 0.8$ for the 10 kpc ring. We can deduce from this that the galaxy presents sources of active electron injection in the central region, while the ring mainly hosts old electrons.

By using the free-free and synchrotron emission models in Eqs. \ref{Eq:Free-F} and \ref{Eq:Sync} to reproduce the emission at the four different considered frequencies, we derived the thermal and the nonthermal contributions to the total flux density at a resolution of $5^\prime$. We decided to leave the synchrotron and free-free emission amplitude at 1\,GHz as
free parameters, while the synchrotron spectral index has been fixed to 1.06 in the ring and to 0.83 in the central region. Fixing the latter improves sensitivity on synchrotron and free-free amplitude values, removing the intrinsic degeneration between free-free and synchrotron emission models.
Following this procedure, we were able to build spatially resolved emission maps for the thermal and nonthermal emission, which we present in Fig. \ref{fig:emissionmap}.\newline

The thermal map shown in the left panel of Fig. \ref{fig:emissionmap} shows a concentrated thermal emission in the southern and central part of the ring and diffuse emission in the northern ring as expected given the distribution of  {\sc Hii} regions \citep[e.g.,][]{azi11}. The  regions of high thermal emission in the ring match the ones identified in previously published M31 thermal maps \citep{ber03}, and we generally find good agreement in the whole ring and in the central region. \cite{tab13} separated free-free and synchrotron emission in M31 using a template for the free-free emission from H$\alpha$ emission line. Despite their different frequency, angular resolution, and decomposition technique, we noted good agreement in the location of the brightest thermal spots.

\begin{figure}[h!]
    \centering
    {\includegraphics[scale=0.4]{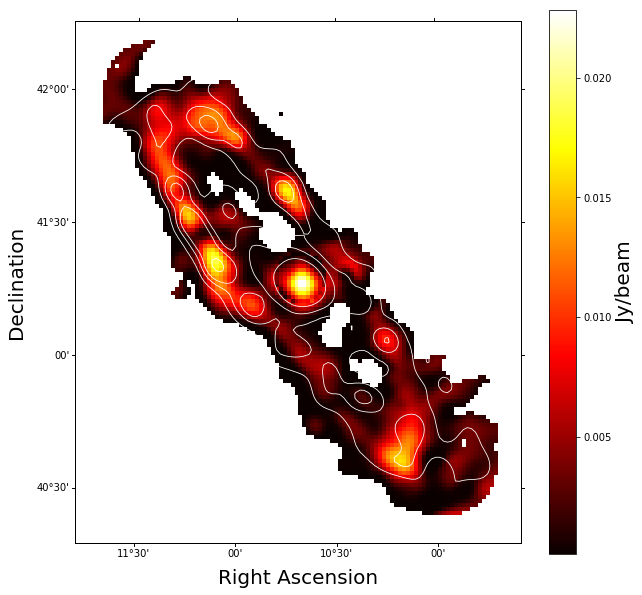} }
    \caption{Thermal map ($FWHM=5'$) with H$\alpha$ contours by \cite{devereux}.}
    \label{fig:Halpha+Thermal}
\end{figure}

The value of the thermal peak emission in the ring is $\sim 19$ mJy/beam, whereas the nonthermal intensity only reaches $\sim 10$ mJy/beam. The thermal peak is located in the central part of the ring, $(Ra; Dec) = (11.17; 41.35)$ deg. In the southern ring, the thermal peak is $\sim 16$ mJy/beam, while nonthermal emission is about one-third of that.
High thermal emission is also found within the central region, in agreement with the results presented in \cite{ber03}. On the other hand, the H$\alpha$ map extracted by \cite{devereux} shows a bright bulge. These results seem to suggest that the M31 bulge is  also a star-forming region, despite no evidence of star formation in the central region being found  by
\cite{for13}. In the following section, we investigate the SFR in more detail, starting from the given thermal map. Globally, the average free-free noise values resulting from the pixel-by-pixel MCMC fit were found to be in the range $0.5-2 \,$ mJy/beam.
We stress the perfect agreement between our thermal map and the H$\alpha$ map by \cite{devereux} convolved to $5^{\prime}$; see Fig. \ref{fig:Halpha+Thermal}.\newline

The nonthermal emission map is shown in the right panel of Fig. \ref{fig:emissionmap}. This map is much smoother compared to the thermal one, especially within the ring, confirming the results at several wavelengths and angular resolutions obtained by \cite{ber03, tab13, beck20}.
The nonthermal emission amplitude varies in the ring between $\sim 5$ mJy/beam and $\sim 19$ mJy/beam, while the thermal map shows values even two orders of magnitude lower than the peak in the central ring. This nonthermal uniformity can be explained with the spread of the relativistic electrons from their birth regions into the whole ring along the magnetic field lines, as shown in \cite{ber03}. Further details about the magnetic fields and the polarized emission in M31 based in the SRT C-band data will be presented in Murgia et al. (in prep.). The average nonthermal noise was found to be in the range $0.5-1 \,$ mJy/beam.

As sources of free-free and synchrotron emission, it is important to correlate the thermal and nonthermal maps with the position of the {\sc Hii} regions and SNR or SNRs candidates, respectively. In the left panel of Fig. \ref{fig:emissionmap},  the {\sc Hii} region contours are overlayed on the thermal map at 6.6\,GHz. We used contours relative to 2, 3, 6, 9, and 12 {\sc Hii} regions inside the C-band pixel. We note the good agreement between the brightest thermal regions and the areas where more than two {\sc Hii} regions are located. The only exception is the central region of M31. Here we find a non-negligible thermal emission but no {\sc Hii} region counterpart.

In the right panel of Fig. \ref{fig:emissionmap}, the SNR region contours are overlayed on the nonthermal map at 6.6\,GHz. We find excellent agreement between the location of the SNRs and nonthermal emission at 6.6\,GHz. Contour levels are 1, 2, and 3 SNRs inside the C-band pixel.

\section{Star formation rate in M31}
\label{sec:7}

Here, we use the thermal map derived in Sect. \ref{sec:6} to compute the total SFR of M31. In addition, we also generate a SFR map for the galaxy.\newline
The SFR and thermal emission are related by a linear model. The thermal flux density at a given frequency $\nu$ is proportional to the number, $N_c$, of Lyman continuum photons (Lyc) absorbed by the gas in the {\sc Hii} regions. The relation between $N_c$ and the thermal flux density is reported below \citep{mezger74}:
\begin{equation}
    N_c[s^{-1}]=4.761 \cdot 10^{48} \alpha (\nu,T_e)^{-1} \nu [GHz]^{0.1} T_e^{-0.45} S_{th} [Jy] D[kpc]^2
,\end{equation}

where $S_{th} $ is the thermal flux density at frequency $\nu$, and we assumed the M31 distance to be $D=780$ kpc. For the electron temperature we used $T_e=8000$ K while $\alpha(\nu,T_e)$ is a slow varying function, tabulated in \cite{mez67}. We considered $ \alpha (\nu,T_e)=1$ at radio wavelengths. This gives N$_{c}$ = 3.11 $\times$ 10$^{52}$ s$^{-1}$. Given $N_c$, the SFR can be found using a simple relation. In fact, the ratio of SFR to Lyman continuum photon rate is given by \cite{kennicutt94} and  \cite{kennicutt98}:
\begin{equation}
    SFR\left[M_{sun} \cdot yr^{-1} \right] = 7.5 \cdot 10^{-54} \cdot Nc\;\;\; \left[\mbox{photons} \cdot \mbox{s}^{-1} \right]
    \label{eq:1}
.\end{equation}
The left hand side of Eq. \ref{eq:1} is the continuous SFR required to maintain a steady-state population of ionizing stars that produces an observed ionizing photon rate $N_c$. \newline

\begin{figure*}[!h]
    \centering
    \includegraphics[scale=0.4]{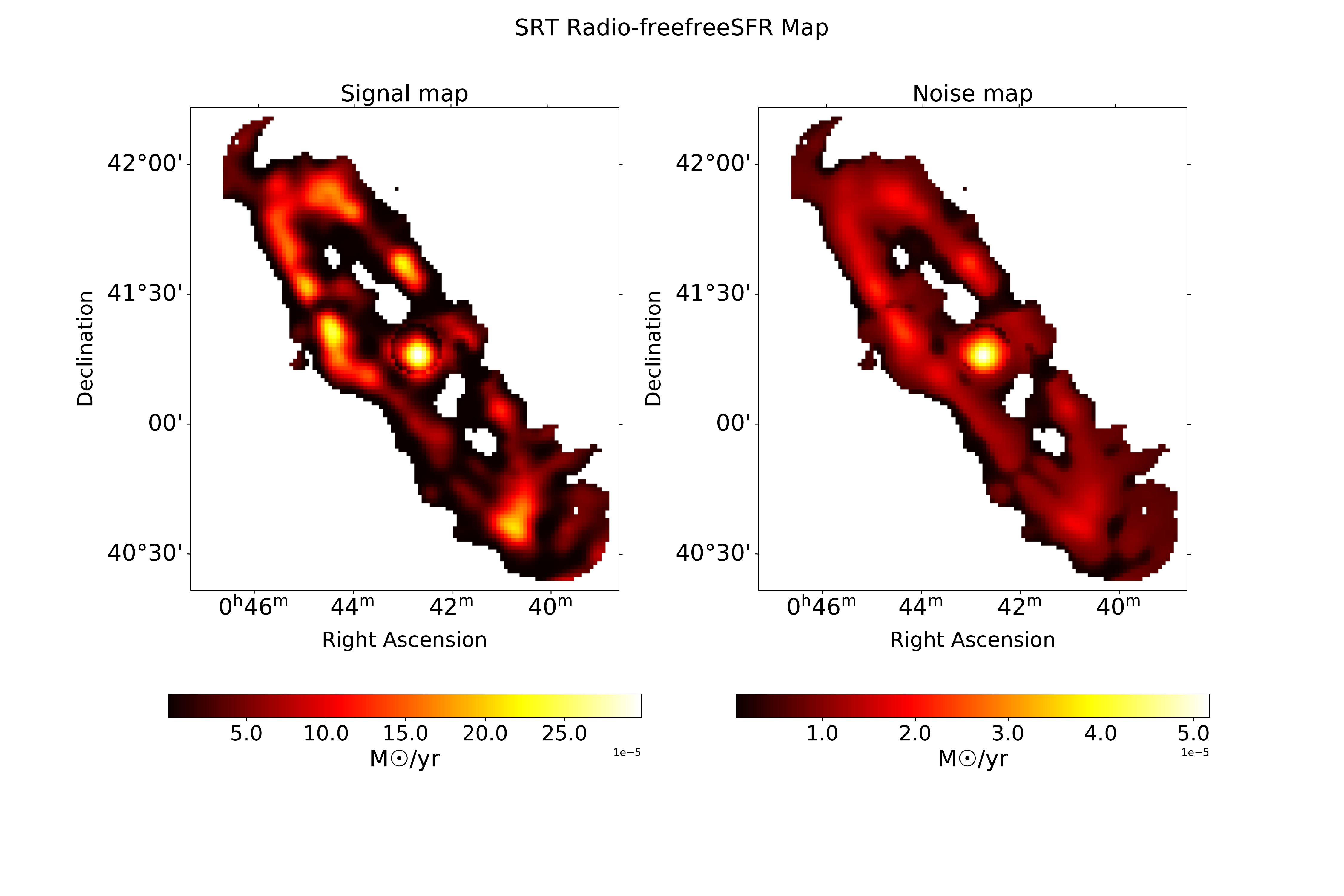} 
    \caption{SFR (left) and SFR noise (right) maps extracted from radio and free-free emission. The radio--free-free map used as a starting point is the one described in Sect. \ref{sec:6}.}
    \label{fig:SFR_map_SRT}
\end{figure*}

Using the conversion factor from $N_{c}$ to SFR from Eq. \ref{eq:1} and given the thermal map extracted in Sect. \ref{sec:6}, we created a full SFR map. The output map is reported in Fig. \ref{fig:SFR_map_SRT}. 
From this map we extracted a total star formation rate of $SFR_{radio - freefree} = 0.19 \pm  0.01$  M$_{\odot}$ yr$^{-1}$, which is consistent with the findings of \cite{for13} and Xu $\&$ Helou (1996). This total SFR value was found by integrating the SFR map in Fig. \ref{fig:SFR_map_SRT} in all the available pixels (i.e., in all the pixels with $S/N>3$), which corresponds to an integration radius of $R_{max}\sim15 \,$ kpc. Here, and for the following integrated SFR values in this section, the reported error is a sum of the SFR map statistical uncertainty and the calibration error.
 \newline

The obvious advantage of using a radio tracer of the recent star formation activity is that it is almost unaffected by dust extinction, unlike other classical tracers at optical and UV wavelengths (e.g., the H$\alpha$ luminosity and the near- or far-UV luminosity). The main issue with methods that use ionization of hydrogen, and the related free-free emission, to trace star formation from massive stars is the direct absorption of Lyman continuum photons by dust. In this case, the ionizing photons are removed altogether and are no longer available to ionize hydrogen. Thus, no emission from free-free continuum emission will result \citep{ken12}.

The actual impact of Lyc photon absorption by dust has been notoriously difficult to establish because of the absence of a ``ground truth'' (or reference) with which to compare measurements. Models are therefore needed, and these show that the level of Lyc absorption depends on the assumptions of the geometry of the nebulae (Dopita et al. 2003, and references therein).
In {\cite{tabataberk2010},} a lower limit was put on the M31 ring SFR in the radial range $6 < R <17$ kpc, after dust attenuation correction. This resulted in $SFR_{H\alpha}[6 < R <17$ kpc$]=0.27$ M$_{\odot}$ yr$^{-1}$. We extracted the total SFR in a similar region, finding $SFR_{radio - freefree}[6 < R < 15$ kpc$] = 0.16 \pm  0.01$ M$_{\odot}$ yr$^{-1}$. In \cite{Kang_2009}, the authors extracted the SFR of young star forming regions (<10 Myr) from UV observations as a function of the galactocentric distance within a $R_{max}=27$ kpc; their Fig. 13 suggests that the SFR in the region $6 <R<15$ kpc corresponds to $\sim47\%$ of the total, while $31\%$ can be found in the region $15 < R <17$ kpc. Using these estimates, we extrapolated the total SFR from free-free emission within the annular region $6 < R < 17$ kpc  from our data, finding $SFR_{radio - freefree}[6 < R < 17$ kpc$] = 0.27 \pm  0.01$ M$_{\odot}$yr$^{-1}$, which is in very good agreement with the value found in {\cite{tabataberk2010}}.

\begin{figure}[h!]
    \centering
    {\includegraphics[scale=0.45]{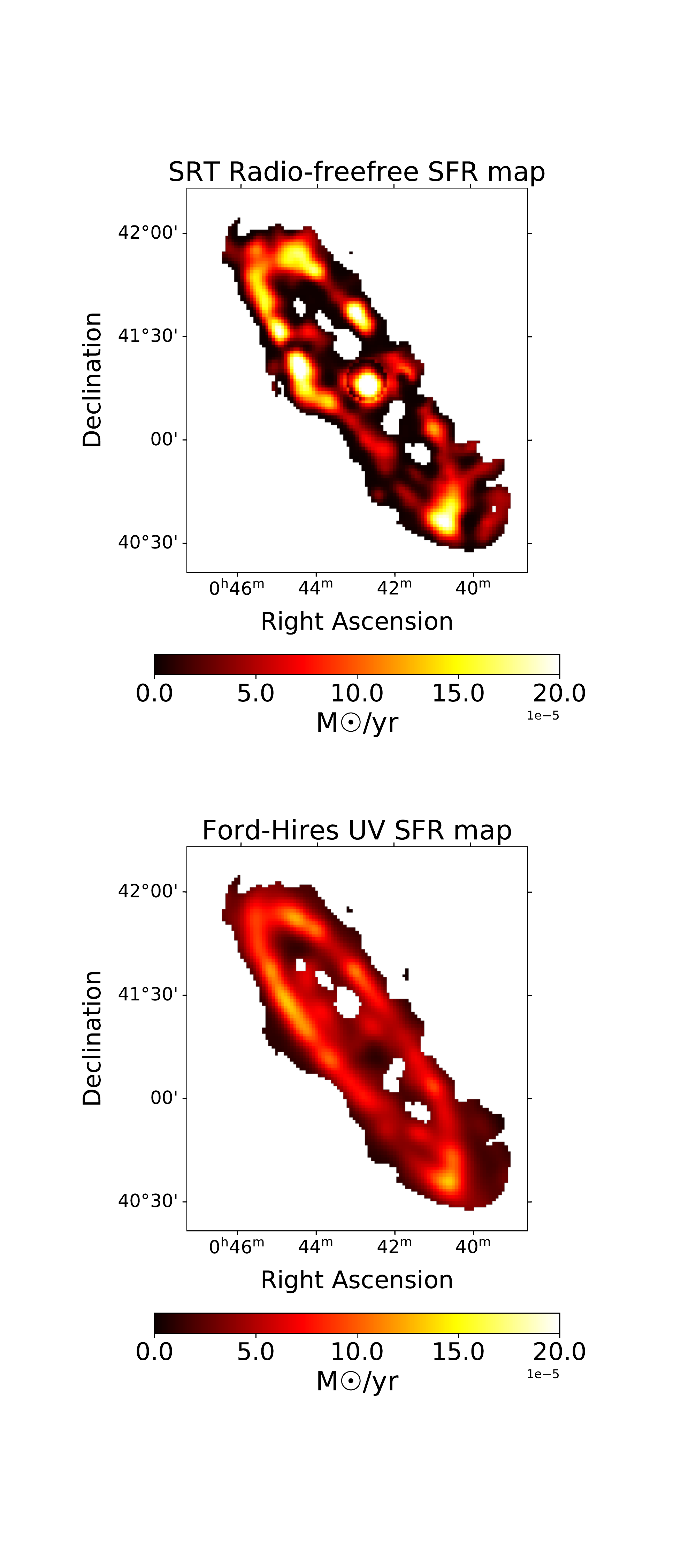} }
    \caption{Two SFR maps extracted from two different tracers: the free-free-based SFR map (top) with a resolution of $5'$, and the Ford-Hires UV+MIR SFR map (bottom) convolved to $5'$ angular resolution.}
    \label{fig:SFR_SRTvsFH}
\end{figure}

In the following we compare the SFR map derived from SRT data with the one from \cite{for13} (F13 hereafter). The latter was obtained by combining far-UV (FUV) and 24 $\mu$m (mid-infrared (MIR)) data, which accounts for both optically visible and dust obscured star formation. Furthermore, the contribution to the emission in the FUV and to the heating of dust emitting in the MIR coming from the old stellar populations was carefully taken into account. This is an issue in M31 in particular, given the very luminous and dense bulge. As we discuss later, this is an issue for the SFR calculated from the free-free emission as well. The two maps are shown in Fig.~\ref{fig:SFR_SRTvsFH}.

In order to be able to compare the two maps, we converted the UV SFR map, originally in units of SFR surface density, into units of M$_\odot$ yr$^{-1}$ , multiplying the value of each pixel by the projected pixel area in kpc$^2$.
With respect to the UV+MIR SFR map, the SRT map shows a less smooth star formation distribution, reaching higher peaks. This is due to the different star formation tracers used to extract the two maps: the UV and 24 $\mu$m are more uniformly distributed than the HII regions responsible for free-free emission. Moreover, in the case of free-free emission, the gas is ionized by stars that are younger than the stars responsible for the heating in the UV and MIR emission. Therefore, while the SRT map represents the present SFR, the F13 data show the SFR for older stars.
Another morphological difference that immediately emerges from the comparison of the two maps is the star formation in the central region. The map extracted from radio--free-free emission shows a peak of SFR in the central area of the galaxy, while the F13 map shows almost no star formation in the central region. The total SFR integrated limited to the central area is $Centre\_SFR_{radio - freefree} =  (12.8\pm0.7)\cdot 10^{-3}  $  M$_{\odot}$ yr$^{-1}$ for the SRT map. We also extracted the same quantity from the F13 map, and found  $Centre\_SFR_{UV} =  (3.5\pm0.2)\cdot 10^{-3}  $  M$_{\odot}$ yr$^{-1}$.  This value has been extracted from both maps integrating on an elliptical region of ellipticity $e=0.26$ at the center of the SRT map, with the same position angle as the galaxy, namely $\theta= -52 \deg, $ and a semi-major axis of $10'$ (2.3 kpc). No noise cut was applied to the two maps for this purpose.

The excess of thermal emission we find in the central region is not due to star formation activity.  The center of M31 has been known to host ionized gas for at least four decades now, the presence of which has been proven by several works \citep[e.g.,][]{deharveng82,jacoby85,devereux}. \cite{deharveng82} suggested that horizontal-branch stars could be a likely source of ionizing photons as the detected UV emission quite closely follows the optical one, the latter surely being dominated by old stellar populations. Line flux density ratio diagnostics, such as {\sc [Nii]}/H$\alpha$ \citep[found to be $\sim 3$ on average, see][]{ rubin71}, also support an origin different from star formation. Using HST images, \cite{king92} noted that, within the bulge, there are no bright point-like sources that could be identified as O stars, and \cite{devereux} suggest that the peculiar filamentary morphology displayed by the H$\alpha$ emission within the bulge is unlikely attributable to star forming regions. The presence of these filaments indicates the existence of shocks, which represent a possible origin of ionized gas. Results supporting this view were obtained by analyzing IR emission, using both SED fitting \citep[see][]{smith12,viaene14} and radiative transfer models \citep{viaene17}. All such studies were in agreement in attributing both the dust heating and the UV luminosity to the old stellar populations hosted within the bulge. The presence of this kind of stars in the central region is confirmed by the high correlation we find in the center between our thermal map at 6.6\,GHz and the {\it Spitzer} map at 3.6 $\mu$m, as discussed later in Sect. \ref{subs:tt_plot_Total}. Taking into account this set of considerations, in the following we mask the central region and do not include it in the SFR investigation.

\begin{figure*}[!h]
    \centering
    \includegraphics[scale=0.4]{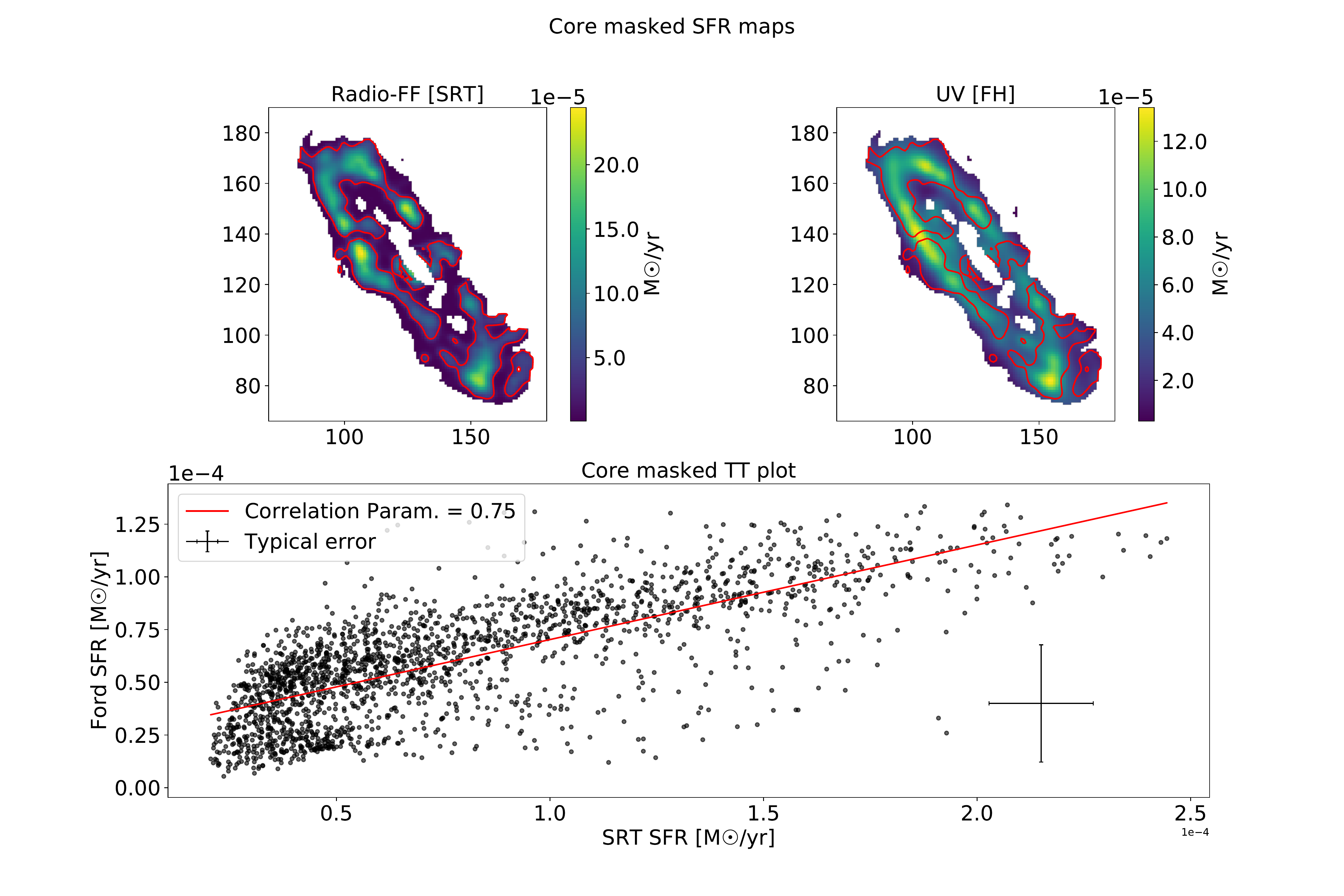} 
    \caption{ T-T plot between the radio--free-free SFR map and the F13 map with the central area masked. The red contours on the top two maps highlight the regions with $S/N>3$ in our SFR map (top-left panel), and therefore the pixels that have been considered in this analysis. The plot shows a good correlation between the two maps in the ring region with a correlation parameter of $r_{ring}=0.75$.}
    \label{fig:TT_plots}
\end{figure*}

In order to quantify the level of correlation between the SFR map generated from radio and free-free emission and the same map extracted from UV emission, we created a T-T plot. For this purpose, all the pixels with a S/N of lower than 3 in the radio--free-free thermal map have been excluded from the analysis. The plot is reported in Fig. \ref{fig:TT_plots}. The two maps show good correlation in the ring region, with a correlation coefficient of $r_{ring} = 0.75$. As expected  \citep[see][]{smith12,for13,viaene14}, the ring is the region with the strongest star formation activity. 

\begin{figure*}[h!]
    \centering
    \begin{tabular}{cc}
    \includegraphics[scale=0.35]{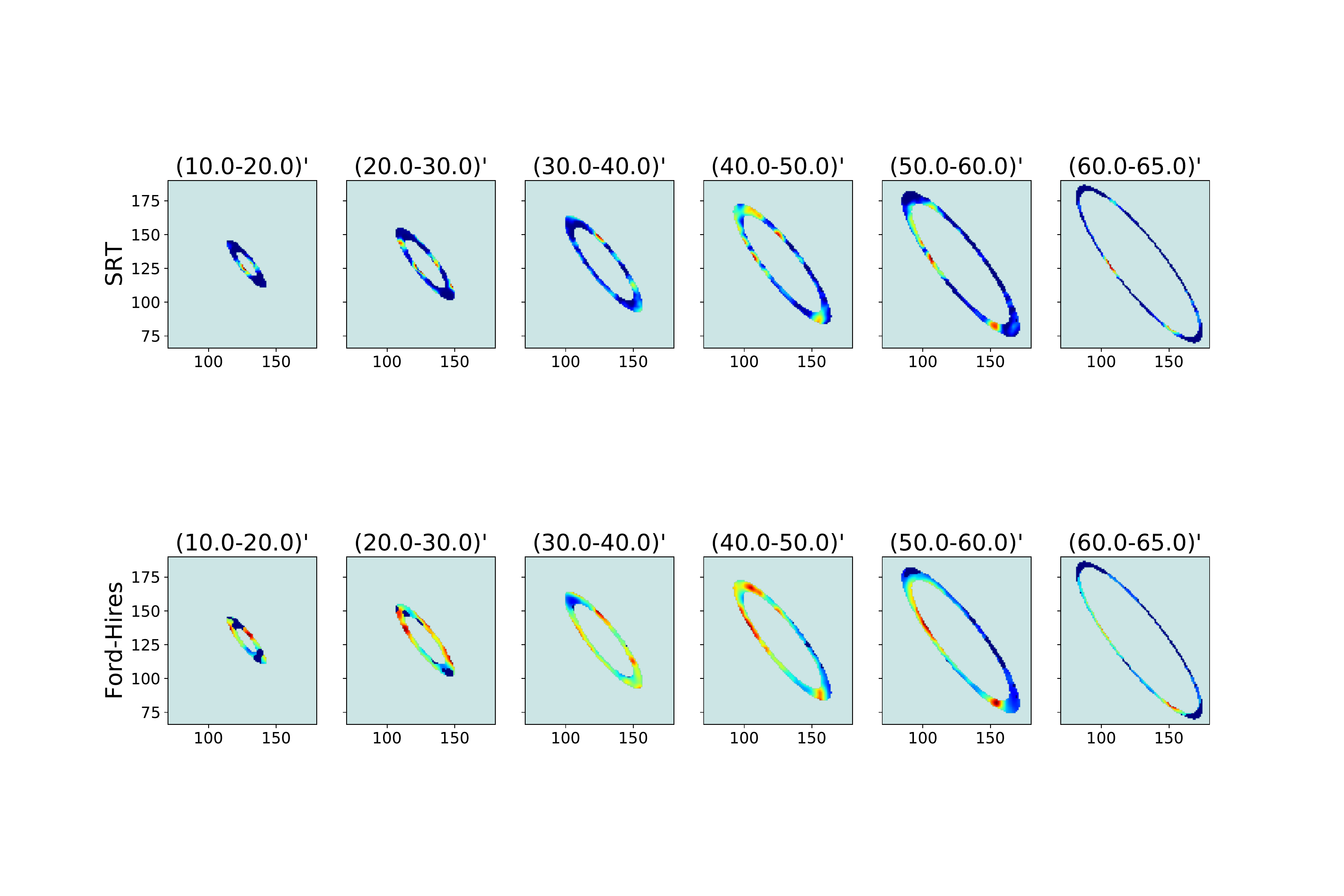}  \\
    \includegraphics[scale=0.30]{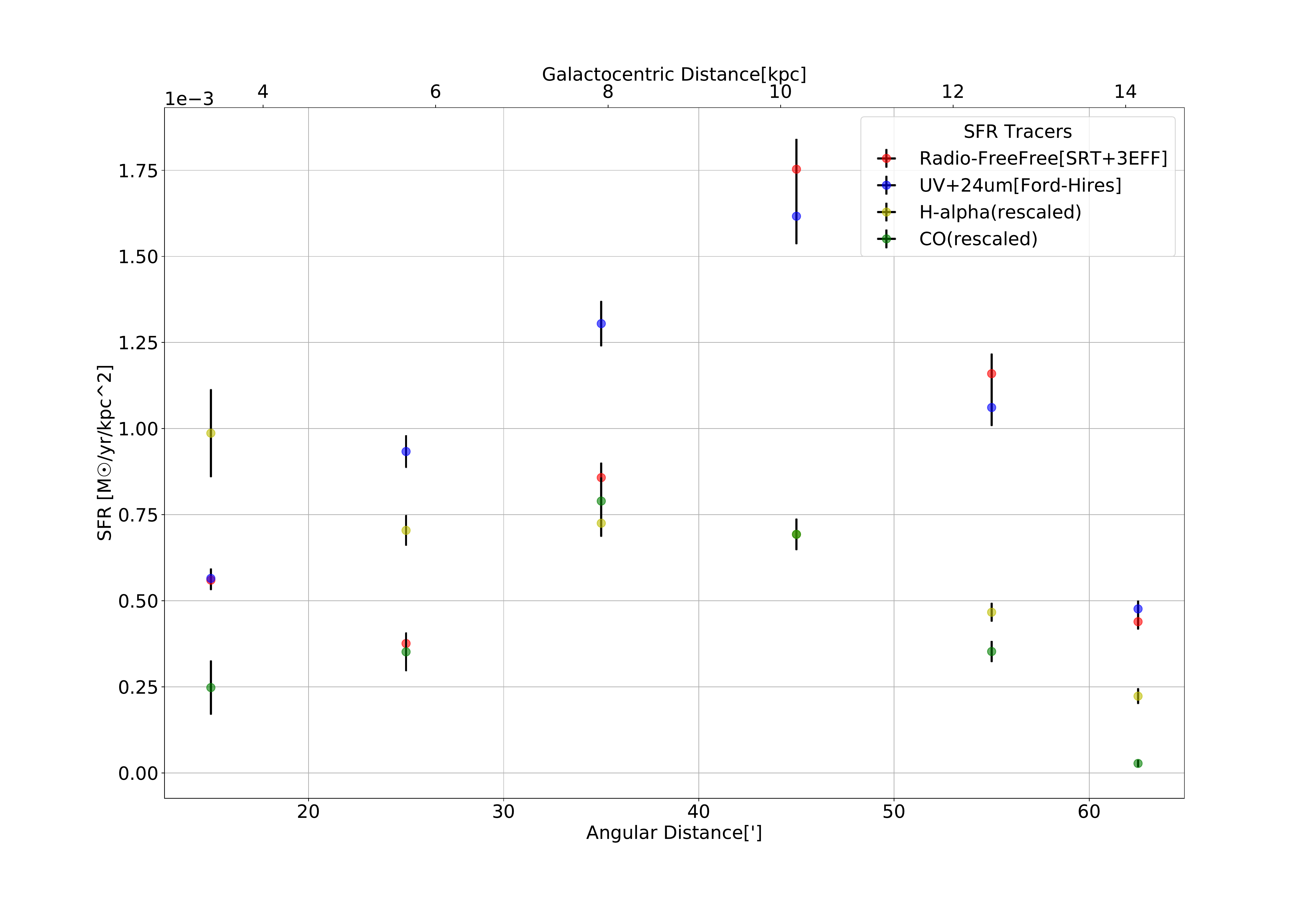}
    \end{tabular}
    \caption{Top panel: Regions of integration used to extract the integrated SFR values reported in the bottom panel and in Table \ref{Tab:SFR_vs_ring_table}. Bottom panel: Integrated SFR in annular regions of increasing galactocentric distance for four different tracers. For the radio--free-free and UV SFR maps, the plot illustrates the SFR value integrated in annular regions of increasing galactocentric distance (see top panel) and renormalized by the area of the region. For H$\alpha$ and CO maps, we report the flux density renormalized to the mean SRT-integrated SFR value in order to be able to visually compare the four different trends. The SFR values from the radio--free-free map are also reported in Table \ref{Tab:SFR_vs_ring_table}.}
    \label{fig:SFR_vs_ring}
\end{figure*}

We also analyzed the SFR radial trend (see Fig. \ref{fig:SFR_vs_ring} and Table \ref{Tab:SFR_vs_ring_table}) by computing the integrated SFR in annular regions of increasing distance from the center of the galaxy. An image of the regions of integration is reported in the top panel of Fig. \ref{fig:SFR_vs_ring}, while Table \ref{Tab:SFR_vs_ring_table}  reports the radial SFR trend extracted from our map. In this step of the analysis, we compared our radio--free-free SFR map with the F13 SFR map as well as with the flux density extracted from the H$\alpha$ \citep{devereux} and CO \citep{nieten06} maps, these latter two being proportional to the recent SFR. For these two maps, the flux density values have been renormalized to the radio--free-free mean SFR value in order to be able to visually compare them to the other two radial trends in the plot.

\begin{table}[htp!]
\scalebox{0.75}{
\begin{tabular}{| *{6}{c|} }
\hline
Ring n.
    & \multicolumn{2}{c|}{Inner major semi axis}
        & \multicolumn{2}{c|}{Outer major semi axis} 
            & SFR$_{radio-ff}$ \\
\hline
    &  [arcmin]  & [kpc] & [arcmin] & [kpc] & $10^{-3}$[M$_\odot$ yr$^{-1}$ kpc$^{-2}$] \\
\hline
\hline

2 & 10 & 2.3 & 20 & 4.5 & $0.56\pm0.03$ \\
\hline
3 & 20 & 4.5 & 30 & 6.8 & $0.38\pm 0.02$\\
\hline
4 & 30 & 6.8 & 40 & 9.1 & $0.86\pm0.04$  \\
\hline
5 & 40 & 9.1 & 50 & 11.3 & $1.75\pm0.09$\\
\hline
6 & 50 & 11.3 & 60 & 13.6 & $1.16\pm 0.06$ \\
\hline
7 & 60 & 13.6 & 65 & 14.7 &$0.44\pm 0.02$ \\
\hline
\end{tabular}}
\caption{Total SFR values found from the radio--free-free map by integrating in the annular regions shown in Fig. \ref{fig:SFR_vs_ring}. The ring number follows the same numbering as that used in Table \ref{Tab:GalaxyEllipticalRings}.}
\label{Tab:SFR_vs_ring_table}
\end{table}

The radial trend is similar for all the star formation tracers, with a SFR peak at $~45'$ (or equivalently $10$ kpc) from the center of the galaxy. The SFR derived from the F13 map is compatible with the one calculated from radio--free-free everywhere in the galaxy apart for the second and third regions of integration. Finding a lower SFR from the radio--free-free map in these two regions is expected. Indeed, in the F13 map the arms of the galaxy are visible at this distance from the center of the galaxy, while we do not detect the same structure in the SRT map.

\section{Radio versus IR cross-correlation}
\label{sec:8}

A linear correlation is known to exist between the radio continuum and MIR galactic emission \citep{jon85, helou1985, yun01}. 
We investigate this correlation with {\it Spitzer}-MIPS $24 \mu$m on SRT total, thermal, and nonthermal maps within M31. Moreover, we used the {\it Spitzer}-IRAC map at 3.6 $\mu$m to investigate how the presence of old stars affects both thermal and nonthermal emission within the M31 central region.\newline 

\subsection{The relation between radio continuum emission and warm dust}
\label{subs:tt_plot_Total}

In this section, we present a  study of the relation between the spatially resolved 6.6\,GHz and the warm dust emission as sampled by the map at 24 $\mu$m, observed with the MIPS instrument on board the SST. The latter map has been transformed to the SRT map coordinate system and then convolved from its angular resolution of $0.1^\prime$ to $2.9^\prime$. 

\begin{figure}[htp!]
    \centering
    {\includegraphics[scale=0.6]{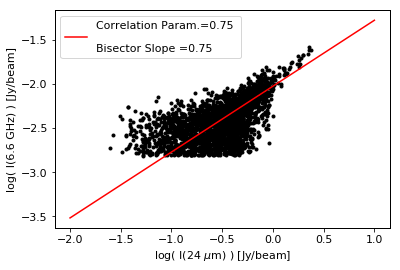} }
    \caption{Correlation between the {\it Spitzer} 24 $\mu m$ maps and the SRT 6.6 GHz. The best linear fit is indicated with the red line. The log-log slope for the whole galaxy is $0.75 \pm 0.01$. Only pixels above $3\sigma$ have been used.}
    \label{fig:TT_cband_s24}
\end{figure}

Taking into account only pixels with $ S/N > 3 $ we obtained the correlation shown in Fig.~\ref{fig:TT_cband_s24} over the whole galaxy. The correlation parameter over the whole galaxy is found to be $\sim 0.75$. Here and in the following, for each correlation between radio and MIR data, we extracted the power-law exponent (b) of the MIR (X)-radio (Y) correlation fitting $ log(Y) = a + b * log (X)$. In each case, we fitted a power law determining the ordinary least-squares bisector in the log-log plane \citep{isobe90}.
For the whole galaxy, we found $b = 0.75 \pm 0.01 $ indicating that the correlation is nonlinear.

\begin{figure}[htp!]
    \centering
    {\includegraphics[scale=0.55]{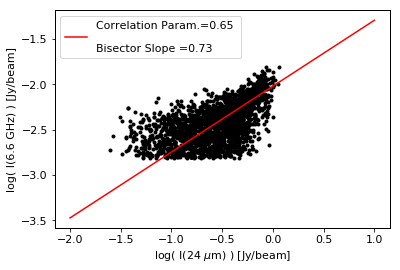} }
    \quad
    {\includegraphics[scale=0.55]{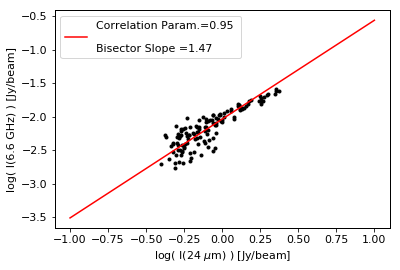} }
    \caption{Upper panel:  24 $\mu$m--radio correlation for pixels within the 10 kpc star forming ring. The log-log slope for the ring is $0.73 \pm 0.01$.
    Lower panel: Same correlation as in the left panel but for pixels within the central region. The log-log slope for the central region is $1.47 \pm 0.04$. Only pixels above $3\sigma$ have been used.}
    \label{fig:TT_cband_s24_Ring&Core}
\end{figure}

We note that the points at higher flux densities in Fig. \ref{fig:TT_cband_s24} belong to the central region of the  galaxy. In order to perform a more detailed analysis, we repeated the study of the radio--MIR correlation, separating pixels belonging to the ring and to the central region. This was done by exploiting the same regions as we did in Sect. \ref{sec:6} to disentangle thermal and nonthermal emission within the central region and the ring. The results are shown in Fig. \ref{fig:TT_cband_s24_Ring&Core}.
The correlation is significantly better in the central region (correlation parameter of $\sim 0.95$) and consequently gets worse for pixels within the ring (with a correlation parameter of $\sim 0.65$), even though a clear trend is still present. We find that the power-law exponent for the ring only is compatible with that found for the whole galaxy, while for the central region we find $b = 1.47 \pm 0.04$ indicating that the relation between MIR and radio is more than linear.

\subsection{The infrared to thermal and nonthermal correlation}

In Sect. \ref{subs:tt_plot_Total} we showed that a good correlation exists between the 24 $\mu$m and the 6.6\,GHz emission. Now we analyze this correlation further by investigating the correlation between MIR and both the thermal and nonthermal emission at 6.6\,GHz, with the aim of gaining more insight into the origin and characteristics of the two kinds of emission. We used the thermal and nonthermal maps extracted with the MCMC fit in Sect.~\ref{sec:6}. Considering that the thermal and nonthermal maps have a resolution of $5^\prime$, the {\it Spitzer} map has been convolved to the same resolution starting from the original $0.1^\prime$ map. With the same fit we also extracted thermal and nonthermal noise maps. These noise maps have been used to select only the pixels where thermal or nonthermal detection is $>3\, \sigma$. The correlation factors over the whole galaxy were found to be $\sim 0.72$ and $\sim 0.84$ , respectively,  for thermal and nonthermal emission.

\begin{figure}[htp!]
    \centering
    {\includegraphics[scale=0.55]{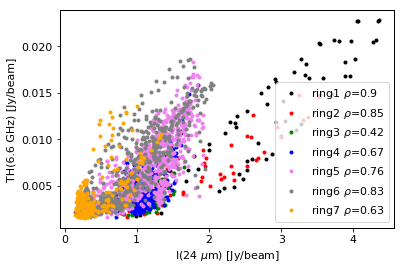}}
    \quad
    {\includegraphics[scale=0.55]{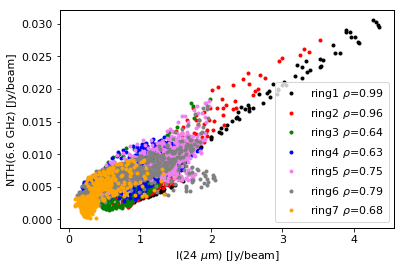} }
    \caption{Top panel: M31 thermal--MIR correlation. Bottom panel: M31 nonthermal--MIR correlation. With $\rho$ we indicate the Pearson correlation coefficient within each elliptical ring. Only pixels above $3\sigma$ have been used.}
    \label{fig:Correlation_Th_NTh_s24}

\end{figure}

Figure \ref{fig:Correlation_Th_NTh_s24} presents a pixel-by-pixel correlation between the thermal and nonthermal emission and the 24 $\mu$m emission, taking into account the seven elliptical rings used in Sects. \ref{sec:5} and \ref{sec:7}. In general, we note a strong correlation in the central region (rings 1 and 2), then the correlation decreases around the 5 kpc structure (ring 3), increases around the 10 kpc ring (rings 4, 5, and 6) and then decreases again out to the 15 kpc structure (ring 7).

We find that, in general, the correlation between nonthermal and MIR is very close to linear for all the considered regions while it is clear that the thermal T-T plot shows two different trends for the central region and the ring. This could suggest different emission mechanisms within the two regions.
As we did in Sect. \ref{subs:tt_plot_Total}, it can be interesting to disentangle the emission from the ring and the central region and analyze possible changes in the correlation parameters found above. 

\begin{figure}[htp!]
    \centering
    {\includegraphics[scale=0.5]{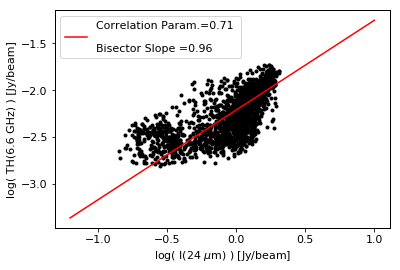}}
    \quad
    {\includegraphics[scale=0.5]{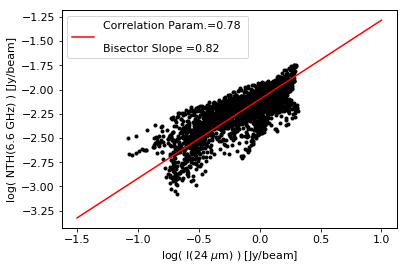} }
    \caption{Top panel:  M31 ring thermal--MIR correlation.\newline Bottom panel:  M31 ring nonthermal--MIR correlation. Log-log slopes for thermal and nonthermal emission in the ring are respectively $0.96 \pm 0.02$ and $0.82 \pm 0.02$. Only pixels above $3\sigma$ have been used.}
    \label{fig:Correlation_Th_NTh_s24_ring}
\end{figure}

The correlation plots for both thermal and nonthermal emission, including only the pixels in the ring region, are reported in Fig. \ref{fig:Correlation_Th_NTh_s24_ring}. In this case, the correlation parameters, of   $\sim 0.78$ for nonthermal emission and $0.71$ for thermal emission, are very close to the ones we found for the whole galaxy. Correlations between thermal and nonthermal emission with MIR data in the M31 ring have been found before \citep{berk13}. The latter authors showed that the log-log slope  of the correlation between nonthermal and MIR depends on frequency, finding $0.61 \pm 0.02$ and $0.97 \pm 0.09$  using 20.5 and 6.3 cm data, respectively. At 4.5 cm, we found a log-log slope of $0.82 \pm 0.02$. This value is intermediate with respect to the previous ones. We noticed that the log-log slope between nonthermal and MIR in the ring is strongly affected by data cut-off, and for this reason we cannot make a direct comparison with the frequency dependence reported in \cite{berk13}. For instance, taking into account only pixels above five times the linear offset we find a log-log slope of $1.07 \pm 0.10$.
The power-law exponent extracted for the thermal--MIR correlation, of namely $0.96 \pm 0.02$, is not significantly lower than the one extracted by previous authors at 20.5 cm: $1.08 \pm 0.02$. The power-law exponent we found for the M31 ring between thermal and MIR emission confirms the linear relation derived by \cite{xu92} in the Large Magellanic Cloud.

The correlation between the IR thermal dust emission and nonthermal synchrotron emission within galaxies is most likely linked to the formation of massive stars, as explained in \cite{har75} and \cite{helou1985}. In Sect. \ref{sec:7}, we showed that the M31 ring presents active star forming activity. The dust grains absorb UV and optical photons from young massive stars, and re-radiate them at MIR wavelengths. The same young massive stars are the progenitors of supernovae and supernovae remnants that can accelerate the electrons responsible for synchrotron radiation. This mechanism therefore provides a natural explanation to the strong correlation between nonthermal and MIR emissions. The linear correlation between thermal radio and MIR emissions in the M31 ring is not surprising considering that both of them are associated with young ionizing stars.

\begin{figure}[htp!]
    \centering
    {\includegraphics[scale=0.5]{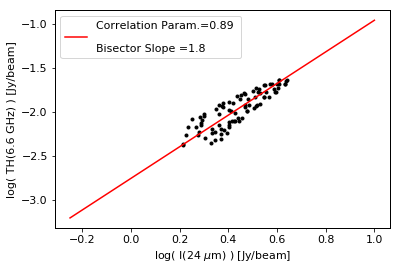}}
    \quad
    {\includegraphics[scale=0.5]{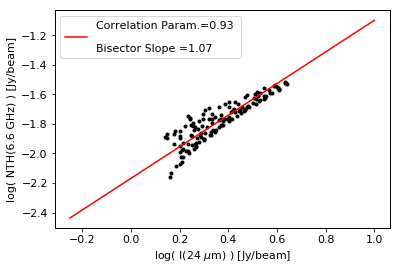} }
    \caption{Top: M31 central region thermal--MIR correlation. Bottom: M31 central region nonthermal--MIR correlation. It is evident, especially from the bottom panel, that the scatter plots show sharp edges. This comes naturally from the fact that the central region points are distributed over a circular area with sharp edges.
    Log-log slopes for thermal and nonthermal emission in the central region are respectively $1.80 \pm 0.06$ and $1.07 \pm 0.02$. Only pixels above $3\sigma$ have been used.}
    \label{fig:Correlation_Th_NTh_s24_core}
\end{figure}

We find a stronger correlation between the central region and the ring. The radio--MIR correlation parameter for thermal emission within the central region is $\sim 0.89$ and this is $\sim 0.93$ for nonthermal emission. Figure \ref{fig:Correlation_Th_NTh_s24_core} shows the two radio--MIR correlation plots for the  central region of M31. The power-law exponents between thermal or nonthermal emission and MIR emission were found to be  $1.80 \pm 0.06$ and $1.07 \pm 0.02$,  respectively.

Unlike the ring, the central
region of M31 does not show star forming activity  \citep{for13}. This means that the correlation between nonthermal and MIR emission cannot be explained by the presence of young massive stars. To understand the correlation with the M31 central region, we used the IRAC-{\it Spitzer} map at 3.6 $\mu$m, which is a very good tracer of low-mass, old stars. We know that the central region contains a dense and compact star cluster at its very center \cite{hst93}. \cite{knu06} found that the M31 bulge and inner disk contain a great number of old($>5 Gyr$), nearly solar-metallicity stars. 

The correlation parameters between radio-thermal or nonthermal and 3.6 $\mu m$ emission were found to be of 0.93 and 0.89,  respectively, confirming the expected strong correlations. For the central region, the correlation between the free-free and the 24 $\mu$m emission gives a slope that is significantly higher than 1. The deviation from linear in this region is in line with what we find in Sect. \ref{sec:7}: an excess of thermal emission in the center that is likely not due to star formation activity. The absence of young ionizing stars does not guarantee that the correlation between thermal and MIR emissions is linear. The high density of stars in the bulge, although of small mass,  can provide an explanation to the correlation observed between the free-free and 24 $\mu$m emissions even if the explanation of the power-law exponent of close to 2 remains an open question.\newline
It is known that young stars and binary systems can generate synchrotron emission \citep{blo11}. \cite{knu06} found that the M31 bulge only hosts old stars, and so only binary systems can provide an explanation for synchrotron emission. However, there is no evidence for the presence of a large number of binary systems in the M31 bulge. Most probably, as mentioned in Sect. \ref{sec:7}, the presence of nonthermal emission within the central region of M31  is due to the existence of shocks which accelerate cosmic ray electrons. 
Another possibility is that the given nonthermal emission is generated by energetic events at the center of the galaxy, similar to the ones that are responsible for the Haze emission within our Galaxy \citep{Fink04, pla13_haze}. The physical mechanism at the origin of the Haze nonthermal emission is not unanimously agreed upon in the literature, and a variety of scenarios have been suggested. One of the most intriguing proposals is that this signal is generated by cosmic rays injected in the galactic halo by annihilation or decay of dark matter \citep{hoop07, egor16}.
A radio emission from M31 Haze has never been detected, but its $\gamma$-ray counterpart, the Fermi-Bubbles, have been observed \citep{FBubM31-16}. In addition to this, what was found for the Haze synchrotron spectral index at the center of the Milky Way is $\alpha_{sync} \simeq 0.8$ (Guidi et al. in prep.), which is very similar to what we find for the central region of M31 $\alpha_{sync}=0.83$ (see Sect. \ref{sub_s:M31 core}). This is also compatible with previous single-dish telescope observations that measure, for the Galactic center, a synchrotron emission characterized by a broken power law, transitioning from $\alpha = 0.25 $ to $\alpha = 1.1$ at a break frequency of approximately $ \sim$ 3 GHz \citep{hey19}.  Furthermore, this kind of radio nonthermal emission due to energetic events at the center of the galaxy has
also been observed outside the Milky Way \citep{Li19}, so it is not only a Galactic mechanism. In this latter study, the best-fit spectral slope in C-band was found to be $1.11 \pm 0.34$, which is compatible with our value. Nonthermal emission in the M31 central region seems to be supported by some polarization in its emission measured with Effelsberg \citep{ber03}. This emission will be  investigated further in a follow-up paper, which will exploit polarized data (Murgia et al. in prep.).

\section{Logarithmic infrared-to-radio ratio (q)}
\label{sec:9}

Another possible way to study the radio--IR correlation uses the logarithmic IR--radio ratio, defined as:
\begin{equation}
    q_{\lambda} \equiv log\left( \frac{F_{\lambda}(Jy/beam)}{F_{21 cm}(Jy/beam)}  \right)
.\end{equation}
Below we present our study of this parameter for M31, correlating our and EFF20 data with MIR data within different annular regions and the whole galaxy. Our results are compared with those of \cite{mur06}, where four galaxies were observed both at radio and MIR wavelengths. Then, we compute the IR--radio correlation parameter $q_{FIR}$, following the convention of {\cite{helou1985}}. Finally, the result is compared with the value found in the literature over several galaxies.

\subsection{$q_{24 \mu m}$}
\begin{figure}
    \centering
    {\includegraphics[scale=0.3]{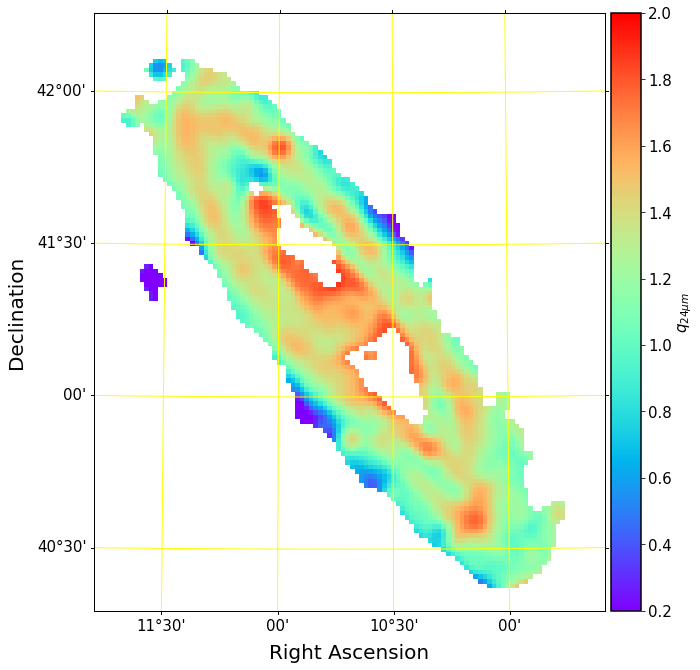} }
    \caption{Logarithmic IR--radio ratio map obtained with EFF20 and {\it Spitzer} $24 \mu m $ maps, both convolved to $3^{\prime}$.}
    \label{fig:q_ratio}
\end{figure}

Firstly, we focused on MIR, correlating radio and 24 $\mu$m data. To extract $q_{24\mu m}$, we used the EFF20 map and the {\it Spitzer} 24 $\mu$m map both convolved to $3.0$\arcmin. 
This ratio map is shown in Fig. \ref{fig:q_ratio} and was extracted using a 3$\sigma$ cut-off in the EFF20 map convolved to $3^{\prime}$. We note that within the ring, the logarithmic IR-to-radio ratio index assumes greater values in the star forming regions. If MIR and radio emission are not linearly correlated, $q_{24\mu m}$ depends on the radio intensity. Indeed, if $F_{\lambda} \propto a F_{21 cm}^{1+b} $ and b is nonzero, then q assumes the form $q_{24 \mu m} = log(a) + b \cdot log(F_{21 cm})$. In Fig. \ref{fig:q_ratio} it is evident, especially in the 10 kpc ring, that $q_{24 \mu m}$ assumes higher values where radio emission is greater, confirming that MIR and radio emissions are not exactly linearly correlated.

\begin{figure}
    \centering
    {\includegraphics[scale=0.6]{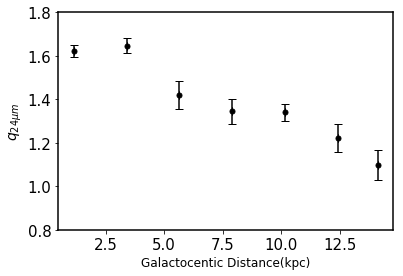} }
    \caption{The $q_{24 \mu m}$ parameter vs. galactocentric distance.}
    \label{fig:q24_trend}
\end{figure}

Using the aperture photometry masks reported in Table \ref{Tab:GalaxyEllipticalRings}, we extracted the mean ($q_{24 \mu m}$) and dispersion ($\sigma _{24 \mu m}$) of the logarithmic IR--radio ratio for M31 at several galactocentic distances. The dispersion is the standard deviation within the region normalized with the square root of the pixel number in the single beam. The results are presented in Table \ref{Tab:q_ratio_af} and shown in Fig. \ref{fig:q24_trend}. We also report in the last column of Table \ref{Tab:q_ratio_af} the $q_{24 \mu m}$ radial values extracted using the SRT $6.6 \, GHz$ map convolved to 3$^{\prime}$ and extrapolated to 21 cm.

\begin{table}
\centering
    \begin{tabular} {c c | c }
    \hline
    Ring no. & $q_{24 \mu m \, EFF20 \, \, 3.0^{\prime}}$ & $q_{24 \mu m \, SRT_{extrap.}}$ \\
    \hline\hline
    1 & $ 1.62 \pm 0.03 $ & $ 1.62 \pm 0.03 $ \\
    \hline
    2  & $ 1.65 \pm 0.04 $ & $ 1.63 \pm 0.04 $ \\
    \hline
    3  & $ 1.42 \pm 0.07 $ & $ 1.45 \pm 0.07 $ \\
    \hline
    4 & $ 1.34 \pm 0.06 $ & $ 1.41 \pm 0.07 $\\
    \hline
    5  & $ 1.34 \pm 0.06 $ & $ 1.38 \pm 0.06  $\\
    \hline
    6  & $ 1.22 \pm 0.07 $  & $ 1.26 \pm 0.07 $\\
    \hline
    7 & $ 1.10 \pm 0.07 $ & $ 1.10 \pm 0.07 $\\
    \hline
    \end{tabular}
    \caption{Log IR--RC estimate for different galaxy regions using EFF20 map convolved to $3^{\prime}$ and SRT $6.6 \, GHz$ map extrapolated to 1.4 GHz. A decreasing $q_{24 \mu m}$ trend can be seen as well as an increasing $\sigma _{24 \mu m}$ trend both for real and extrapolated data.}
    \label{Tab:q_ratio_af}
\end{table}

In both cases we find a clear radial trend, the logarithmic IR-to--radio ratio decreases with increasing galactocentric distance. A similar trend was also found by \cite{mur06} who, using {\it Spitzer} MIPS and WSRT data, observed a decrease in 24 $\mu m$ to 21 cm ratios with increasing radius. As explained by these latter authors, this decreasing trend can be justified by smaller scale lengths for the IR disks than the radio disks. The radio image of a galaxy can be thought of as the smoothed version of the MIR one.
 \citet{mur06}  also observed an increase in the logarithmic IR-to-radio ratio dispersion with increasing  galactocentric radius. This result is 
also observed in M31. This seems to confirm that it is not an artefact due to lower S/N  at larger radii. This increasing trend is most likely due to the decrease in surface brightness and {\sc Hii} region density moving out radially into the disks. Remarkably, in the outer regions, the presence of a small brighter region or an {\sc Hii} can cause a larger dispersion.

We estimated the global $q_{24 \mu m}$  by repeating the aperture photometry analysis on the whole galaxy. Using the SRT extrapolated map we find $ q_{24 \mu m} = 1.30 \pm 0.09$, while using the EFF20 map at its proper angular resolution,   we find $ q_{24 \mu m} = 1.26 \pm 0.09$.  
These values are in agreement with the ones derived by \cite{mur06} within four different galaxies.

\subsection{$q_{FIR}$}
In this section we extract another correlation parameter between FIR and radio continuum data. Following the convention of \cite{helou1985}:
\begin{equation}
    q_{FIR} \equiv log_{10} \left( \frac{FIR}{3.75 \cdot 10^{12}\, Wm^{-2}} \right)-log_{10} \left( \frac{S_{1.4\,GHz}}{Wm^{-2}Hz} \right)
,\end{equation}
where FIR, in units of W m$^{-2}$, is the total FIR flux density calculated as:
\begin{equation}
    FIR \equiv 1.26 \cdot 10^{-14} \left( \frac{2.58\, F_{60 \mu m}+F_{100 \mu m}}{Jy} \right)
,\end{equation}
and where $F_{60 \mu m}$ and $F_{100 \mu m}$ are flux densities. \cite{helou1985} found that the mean $q_{FIR}$ over a large sample of different galaxies is $\sim 2.3$ with a scatter of $\sigma_{q_{FIR}} \le 0.2$.

We used the $IRAS$ data at 60 $\mu m$ with an angular resolution of $4.0$\arcmin, the {\it Herschel} data at 100 $\mu$m with an angular resolution of $0.21$\arcmin , and the SRT map at 1.385 GHz made by \cite{mel18} with an angular resolution of $13.9$\arcmin. In \cite{bat19}, SRT contours at 1.385 GHz were overlayed on the 6.6 GHz map (their Fig. 1) and this map convolved to $51.3 \, \arcmin$ is shown in Fig. 2. 
After the resolution matching between the previous two IR maps and the SRT L-band map, we extracted the galaxy flux densities  by integrating over an ellipse of major semi-axis $65^{\prime}$, eccentricity 0.26, and  position angle $-52$ deg in the $RA$-$DEC$ coordinate system. We find $q_{FIR} = 2.41 \pm 0.04$, a value that is compatible with the one reported in \cite{helou1985} ($q_{FIR,1985} = 2.37$).

\section{Conclusions}

We used the 64 m SRT to observe M31 at 6.6\,GHz down to a sensitivity of 0.43 mJy/beam averaging over the whole bandwidth of 1.2\,GHz. We mapped the galaxy with an angular resolution of $2.9^\prime$ over a field of $RA \times DEC = \ang{2.4} \times \ang{3.1}$ along the minor and major axis, respectively, corresponding to a physical size of $ 32$ kpc $\times \, \,42$ kpc (Fig. \ref{Fig:M31_Letter}). We subtracted a total of 93 compact sources detected within the given field with a S/N greater than 3 (Fig. \ref{fig:SRT+SOURCES and subtracted}). The large number of detected point sources gave us the opportunity to create a catalog and to study the differential source counts, comparing the results with a theoretical model (Fig. \ref{fig:Source+DeZotti}). With more than 100 SNR or SNR candidates known in the M31 field, we compared our source sample with the SNR positions. We find 5 SNR candidates in our catalog within a $1.45 ^{\prime}$ search radius. \newline
Our map at 6.6\,GHz and the ancillary Effelsberg maps at 4.85 and 1.46\,GHz were used to extract the spectral index map at 3' angular resolution (Fig. \ref{fig:S_indexMap_Eff+SRT}). Including also the 2.7\,GHz Effelsberg map we studied the spectral index trend within the central region and over the whole galaxy (Figs. \ref{fig:S_index_radius} and \ref{fig:S_index_radius_allGal}) and we disentangled the thermal and nonthermal integrated emission in the northern and southern ring regions, separately (Figs. \ref{fig:up ring} and \ref{fig:down ring}). We find that the nonthermal emission is relatively homogeneous and smooth over the whole ring, while the thermal emission at 6.6\,GHz is characterized by well-defined bright regions following the  distribution of {\sc Hii} regions. Finally, we extracted a thermal and nonthermal map with an angular resolution of $5^{\prime}$ (Fig. \ref{fig:emissionmap}). Using the thermal map at 6.6\,GHz, we extracted the SFR from the whole galaxy (Fig. \ref{fig:SFR_map_SRT}). We studied the SFR trend versus galactocentric radius and the integrated SFR over the whole galaxy (Fig. \ref{fig:SFR_vs_ring}). We compared our SFR map with the one by \cite{for13}, finding a linear correlation within the ring between the two maps (Fig. \ref{fig:TT_plots}). We estimated the integrated M31 SFR to be $0.19 \pm 0.01 M\ensuremath{_\odot}/yr$, a compatible value with \cite{for13}. \newline
Finally, we correlated microwave data with IR observations of M31. Correlation in the ring of M31 between thermal and MIR emission is characterized by a power law of $0.96 \pm 0.02$, indicating that the relationship is linear. Interestingly, in the central region we find a deviation from linear in the thermal--MIR correlation. This can be explained by the thermal emission mechanism in the central region not being totally related to star formation activity.
We extracted the logarithmic IR-to-radio ratio $q_{24 \mu m}$. We find a decreasing trend with increasing galactocentric radius and an increasing trend for the dispersion ratio (Fig. \ref{fig:q24_trend}).
Finally, we confirmed the \cite{helou1985} correlation with FIR data, using the maps at 60 $\mu m$ and 100 $\mu m$ and radio-continuum data at 1.4\,GHz, finding $q_{FIR} = 2.41\pm 0.04$ while the expected value is $\sim 2.3$.

We summarize the main results obtained in this work as follows.\begin{enumerate}
    \item The SRT map at 6.6\,GHz over $\sim 7.4$ deg$^2$ contains 93 compact sources with $S/N > 3$. We compared the normalized differential counts  with the theoretical model by \cite{bonato}, and found excellent agreement between experimental data and theoretical predictions.
    \item In the central region, we find $\alpha_{sync} \sim 0.8$, and therefore a younger accelerated electrons population with respect to the ring. The population of cosmic ray electrons was confirmed to be old in the M31 ring, $\alpha_{sync} \sim 1.1$. 
    \item We find that the galaxy presents an active injection of relativistic electrons in the northeast region of the ring and also in the central region, while the southwest region of the ring mainly hosts old electrons.
    \item We find smooth nonthermal emission in the ring which confirms that nonthermal emission profiles are uniform along the ring and that magnetic fields in the ring must be almost constant in strength. Thermal emission was found to be much more irregular,  with variations even of two orders of magnitude in the ring. 
    \item The total SFR integrated within a radius of 15 kpc is $0.19$ M$_{\odot}$ yr$^{-1}$, which is  in agreement with the findings of \cite{for13}. The different star formation morphology in the inner regions between 2 kpc and 5 kpc is due to the nondetection of the M31 spiral arms with radio observations of the M31 spiral arms. On the other hand, the disagreement in the central region at galactocentric distances smaller than ~2 kpc is due to an excess of radio thermal emission in this region of the galaxy. This probably originates from old low-mass stars rather than star forming regions.
    \item Thermal emission in the M31 central region was found to be strongly correlated with IR 24 $\mu m$ data and 3.6 $\mu m$. These correlations seem to suggest that thermal emission within the central region is due to a high number of old stars. In any case, old stars cannot be responsible for nonthermal emission, so this is most likely due to shocks.
    \item The correlation between MIR and free-free emission within the ring and the central region shows two different slopes, which could mean that these emissions have different origins. Radio-thermal emission in the ring is confirmed as being due to star forming processes, while in the central region, free-free emission can be correlated to a high density of Sun-like old stars. 
     \item We found a high correlation between nonthermal and MIR emission in the ring region, with a correlation parameter of $r=0.78$. This correlation is most likely linked  to the formation of massive stars. 
     \item We found a nonthermal--MIR correlation in the central region that is even stronger than the same correlation found in the ring, with $r=0.93$. This was unexpected, as the bulge is  a low-SFR region. Polarized data and further radio observations will be necessary to investigate the origin of this nonthermal emission in the central region of M31.
    \item The correlation between FIR and radio-continuum was computed following the convention by \cite{helou1985}. We find $q_{FIR} = 2.41 \pm 0.04$, in agreement with 2.3 (i.e., the average value over a sample of galaxies).
    \end{enumerate}

This work provides a brand new insight into M31, unveiling the galaxy morphology and physical properties at a frequency that is unprecedented. The study will be continued using new data, already in our hands, for the same region of sky.

\section*{Acknowledgements}
We acknowledge the support by 2016 Sapienza Ateneo project.\newline
The Sardinia Radio Telescope (SRT) is funded by the Ministry of University and Research (MIUR), Italian Space Agency (ASI), and the Autonomous Region of Sardinia (RAS), the European Union (EU) and is operated for the National Institute for Astrophysics (INAF) by Cagliari Observatory (OAC).\newline
This research made use of Astropy,\footnote{http://www.astropy.org} a community-developed core Python package for Astronomy \citep{ast13, ast18}.\newline
This research made use of Montage. It is funded by the National Science Foundation under Grant Number ACI-1440620, and was previously funded by the National Aeronautics and Space Administration's Earth Science Technology Office, Computation Technologies Project, under Cooperative Agreement Number NCC5-626 between NASA and the California Institute of Technology.\newline
J.F. acknowledges financial support from the UNAM- DGAPA-PAPIIT IN111620 grant, México. \newline
We thank the referee for the useful comments that helped us improving the paper. \newline
We thank Mark Halpern for useful suggestions which improved the paper.

\bibliography{AA}   
\bibliographystyle{aa}

\begin{appendix}
\section{Compact sources}
\label{ps}

We present in this section the fit of the spectra of 93 point sources detected in the SRT C-band map; see Fig. \ref{fig:all_fit}. Each panel is composed of ancillary radio flux densities (red points) and SRT C-band flux densities extracted in this work (green points), and the best fit is represented by the blue line. The experimental data have been fitted with the modified power law reported in Eq. \ref{eq:modif-power-law}. Moreover, at the top of each panel we report the best-fit parameters for each spectra, at the bottom the point source coordinates, and on the right the ``local'' spectral indexes at 0.15, 1.4, and 6.6 GHz. Above each panel is reported the point source identification in the NVSS catalog that we constructed during the point sources extraction; see Sect. \ref{sec:4} for further details.\newline

Table \ref{Tab:SRT_Sources_catalogue} shows all the details of our point-source sample. The SRT catalog includes the source identification in our NVSS catalog in the M31 field, the name of the nearest object (i.e., the most probable match), the source position on the sky, the SRT C-band flux density, the modified power-law fit parameters, and the spectral index measured between 1.4 and 6.6\,GHz $\alpha _{1.4GHz} ^{6.6\,GHz}$. 
The data at 1.4\,GHz refer to the NVSS catalog. The given position is the position of the peak of the 2D Gaussian that best fits the SRT source emission. The value of the flux density corresponds to the value of the Gaussian peak.

Finally, the $Nearest-object$ is the name of the closest object that appears in other catalogs, together with the object type ( [*] for star or point source, [G] for galaxy, [UvS] for ultraviolet source, [IrS] for infrared source, [SNR] for supernova remnant, and [*Cl] for star cluster). The search was made using the NED/IPAC\footnote{https://ned.ipac.caltech.edu/} extragalactic database and using a search radius of 1 beam ($2.9^{\arcmin} $). We decided to report the B3, 4C, or NVSS source name if it is available, otherwise we used the WISEA, SSTLS2, or 2MASS catalog.


\begin{figure*}[b]
\centering
\subfigure{\includegraphics[scale=0.23]{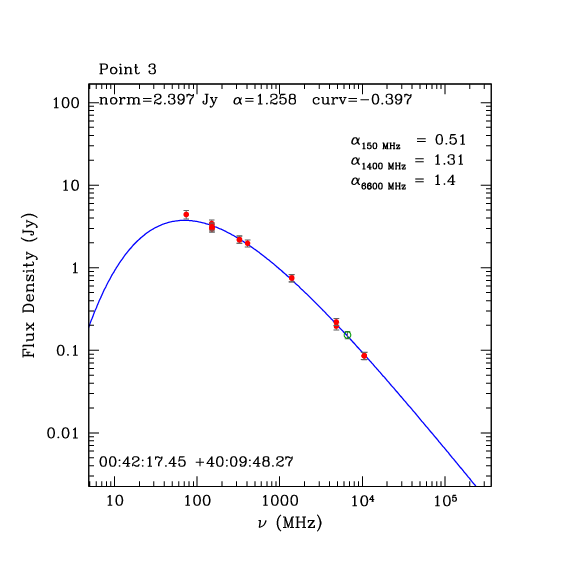} }
\subfigure{\includegraphics[scale=0.23]{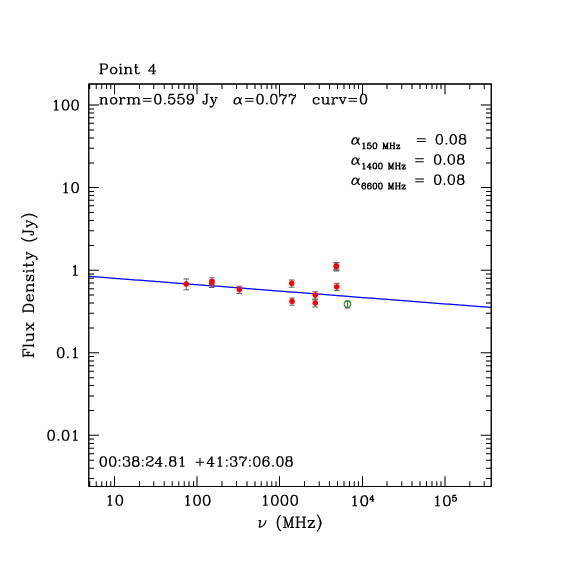} }
\subfigure{\includegraphics[scale=0.23]{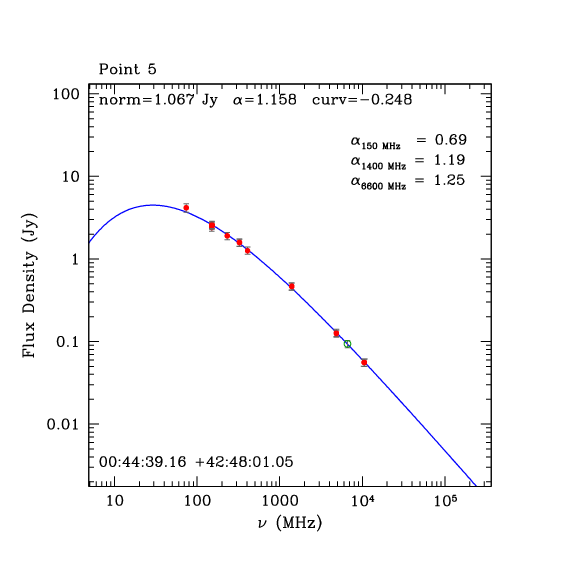} }
\subfigure{\includegraphics[scale=0.23]{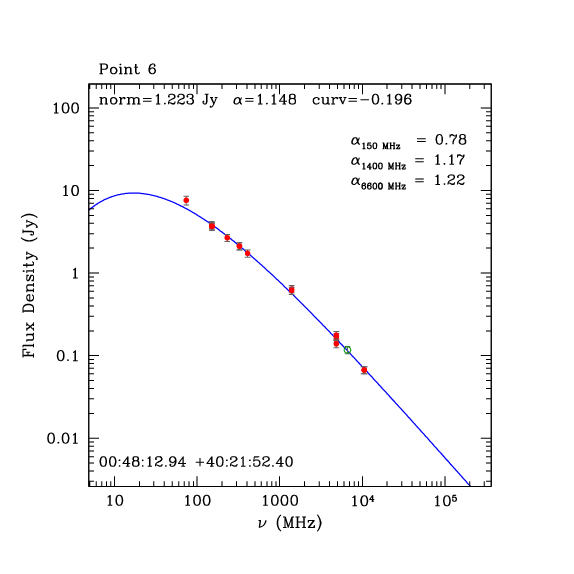} }
\subfigure{\includegraphics[scale=0.23]{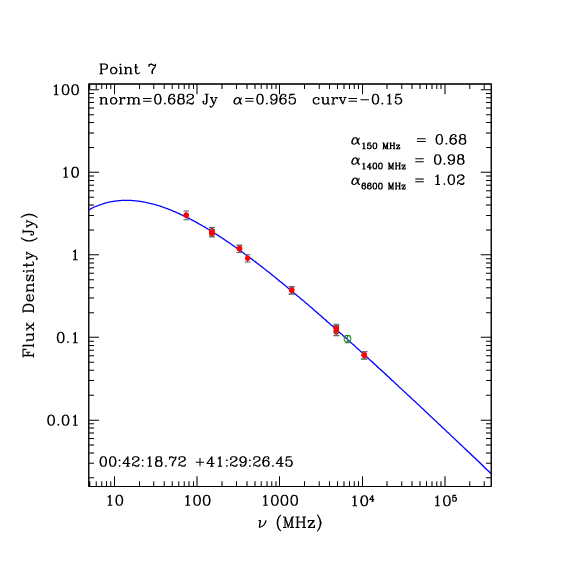} }
\subfigure{\includegraphics[scale=0.23]{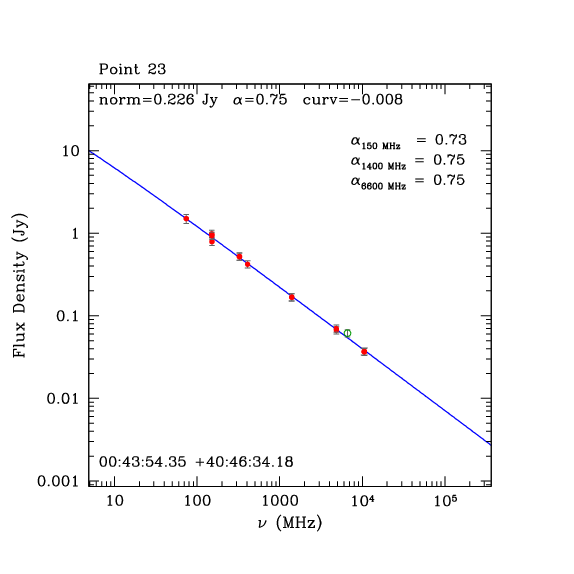} }
\subfigure{\includegraphics[scale=0.23]{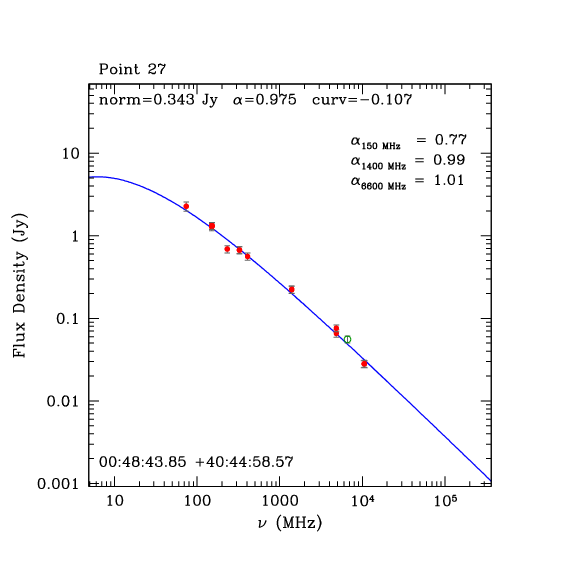} }
\subfigure{\includegraphics[scale=0.23]{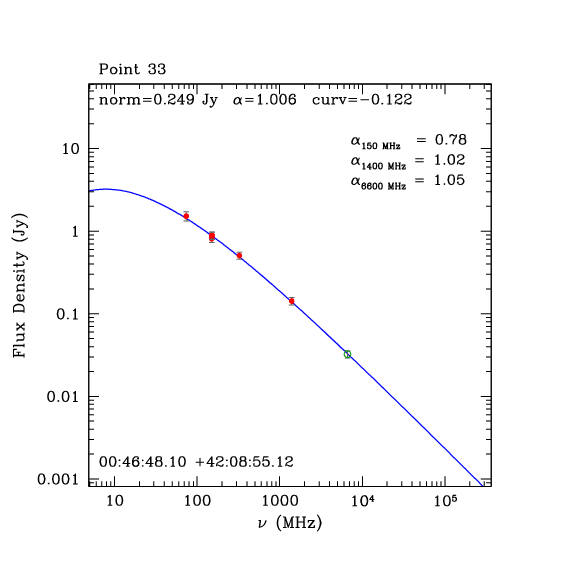} }
\subfigure{\includegraphics[scale=0.23]{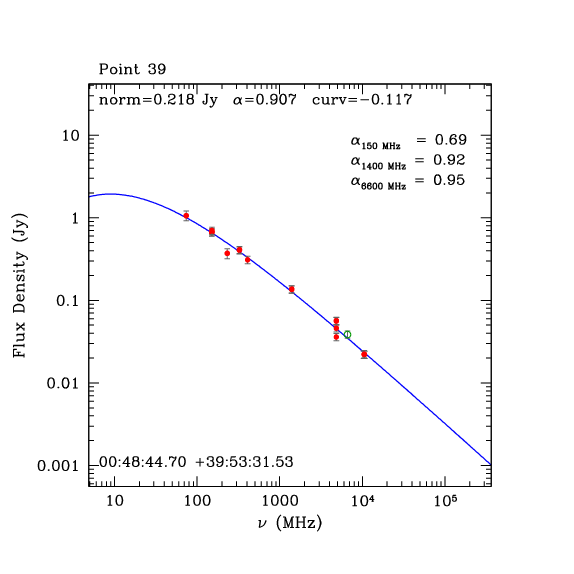} }
\subfigure{\includegraphics[scale=0.23]{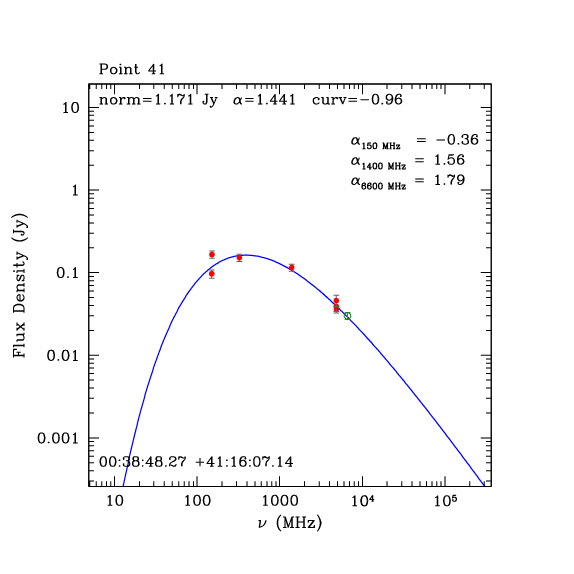} }
\subfigure{\includegraphics[scale=0.23]{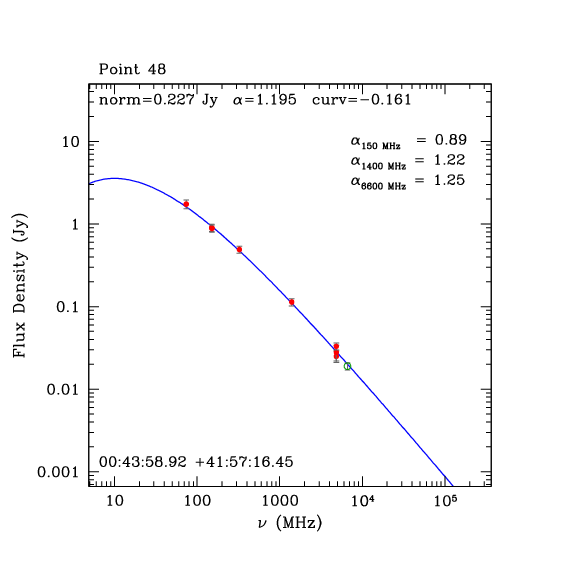} }
\subfigure{\includegraphics[scale=0.23]{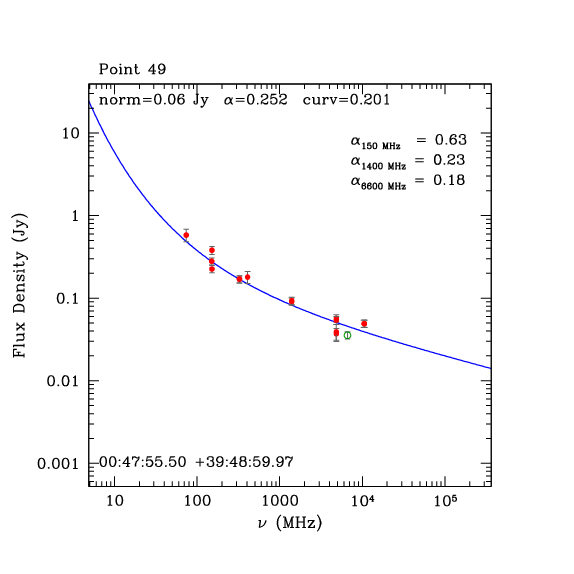} }
\subfigure{\includegraphics[scale=0.23]{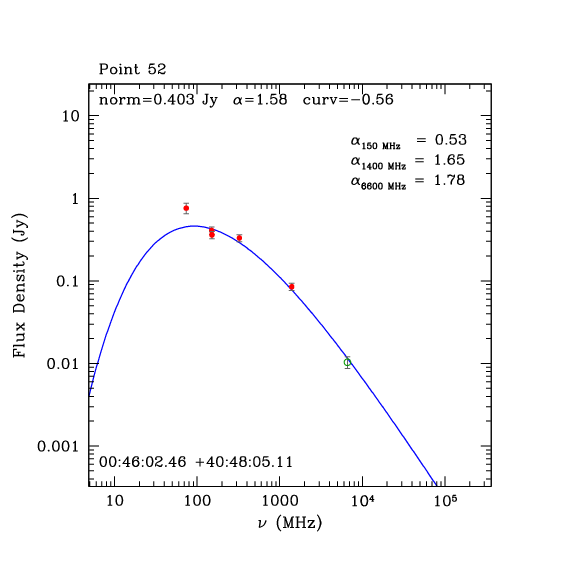} }
\subfigure{\includegraphics[scale=0.23]{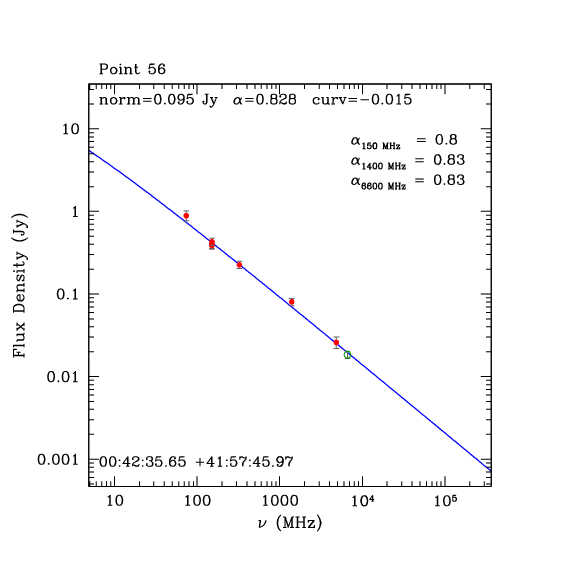} }
\subfigure{\includegraphics[scale=0.23]{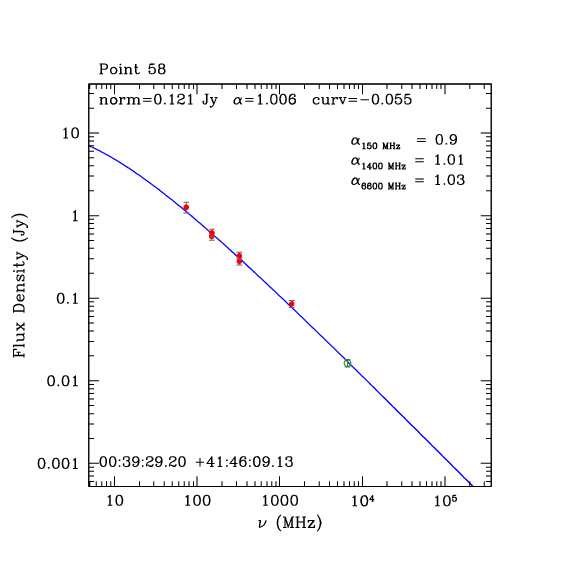} }

\end{figure*}

\begin{figure*}[h!]
\centering
\subfigure{\includegraphics[scale=0.23]{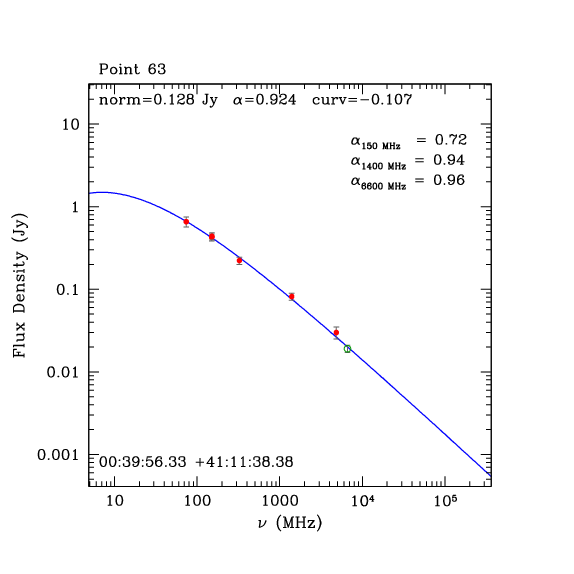} }
\subfigure{\includegraphics[scale=0.23]{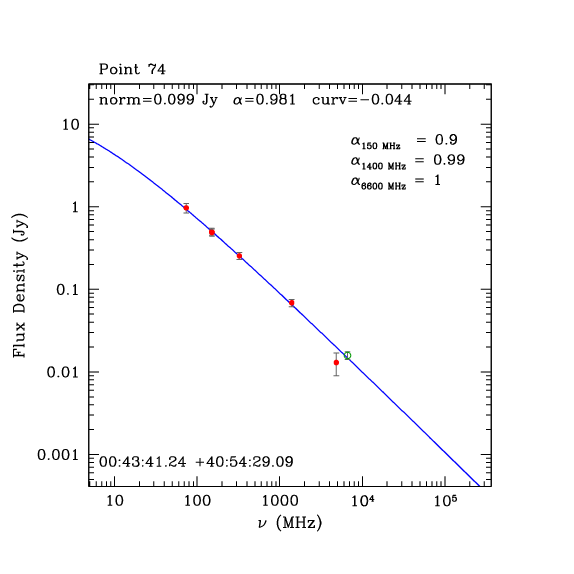} }
\subfigure{\includegraphics[scale=0.23]{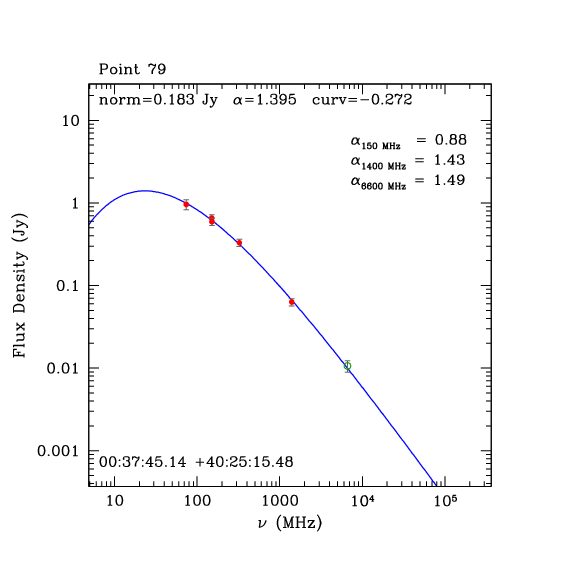} }
\subfigure{\includegraphics[scale=0.23]{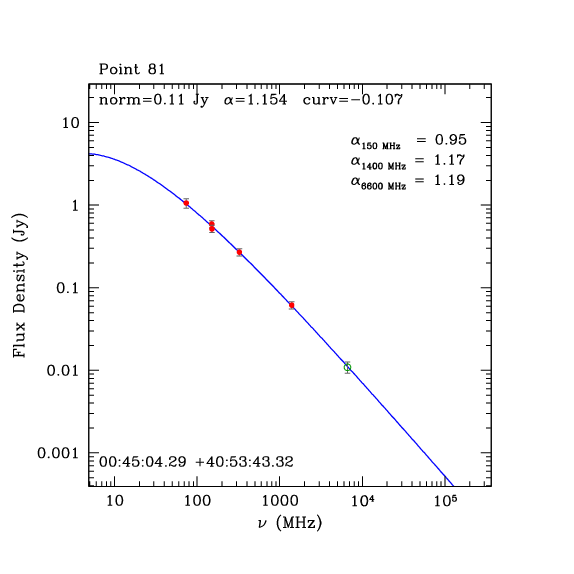} }
\subfigure{\includegraphics[scale=0.23]{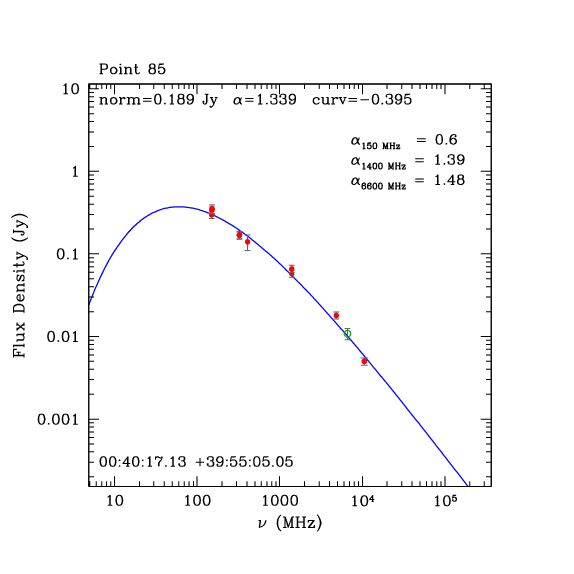} }
\subfigure{\includegraphics[scale=0.23]{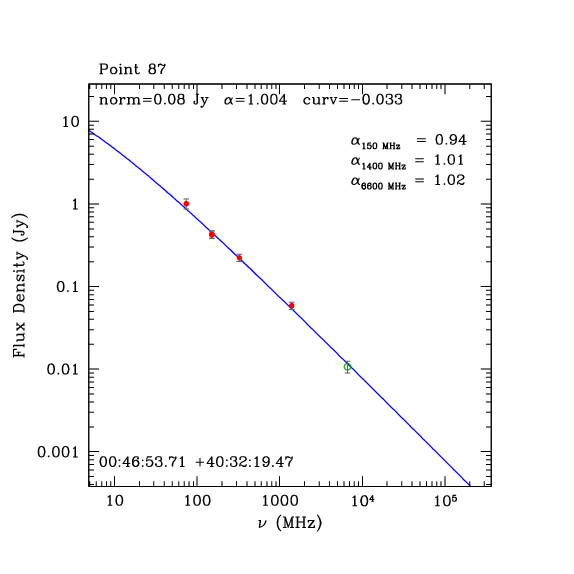} }
\subfigure{\includegraphics[scale=0.23]{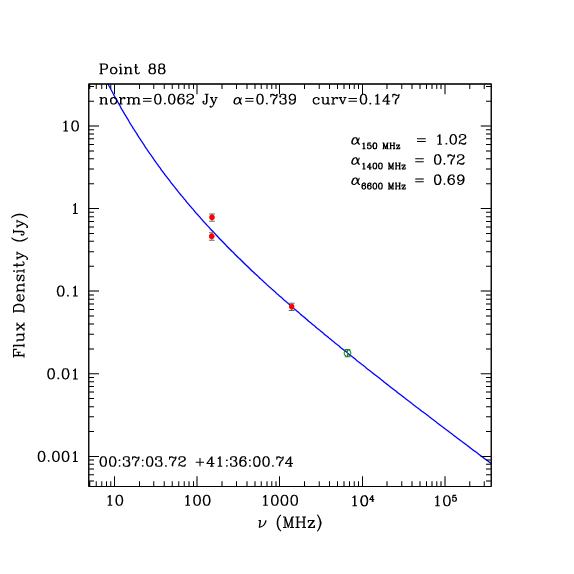} }
\subfigure{\includegraphics[scale=0.23]{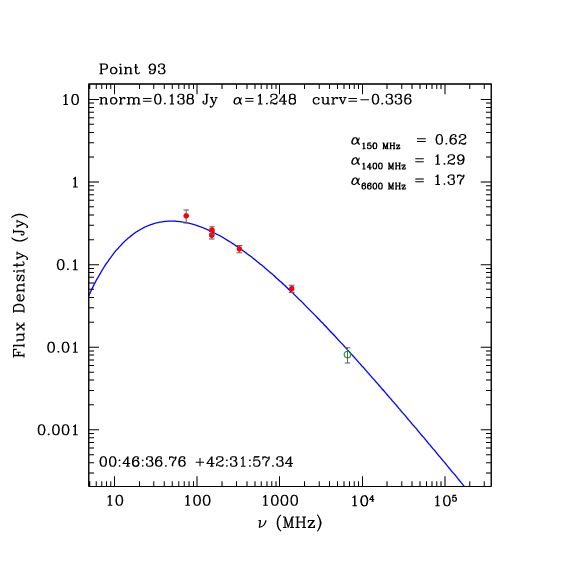} }
\subfigure{\includegraphics[scale=0.23]{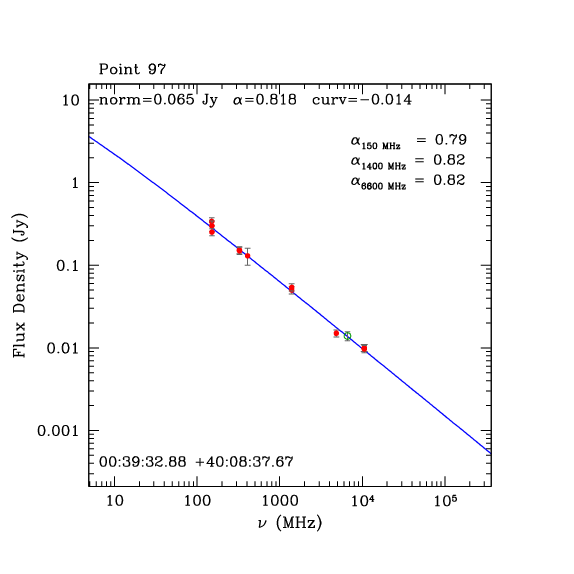} }
\subfigure{\includegraphics[scale=0.23]{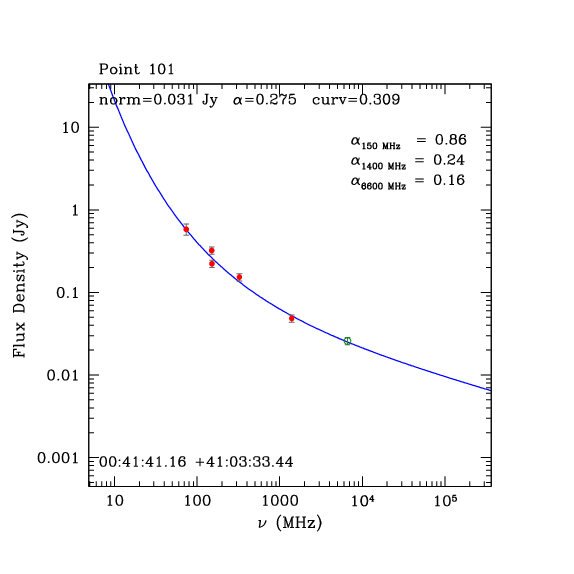} }
\subfigure{\includegraphics[scale=0.23]{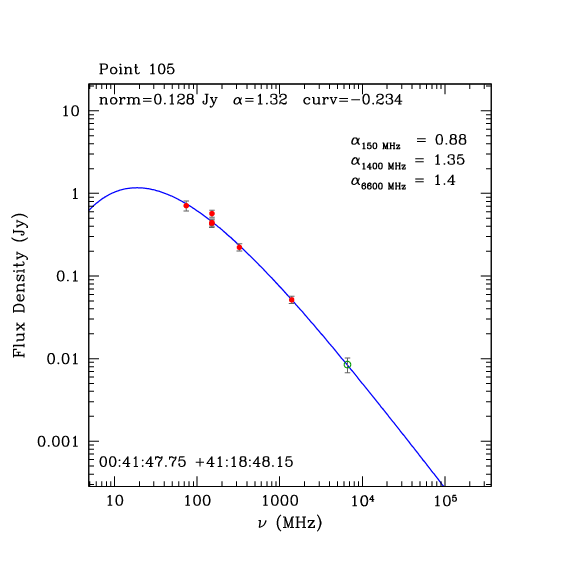} }
\subfigure{\includegraphics[scale=0.23]{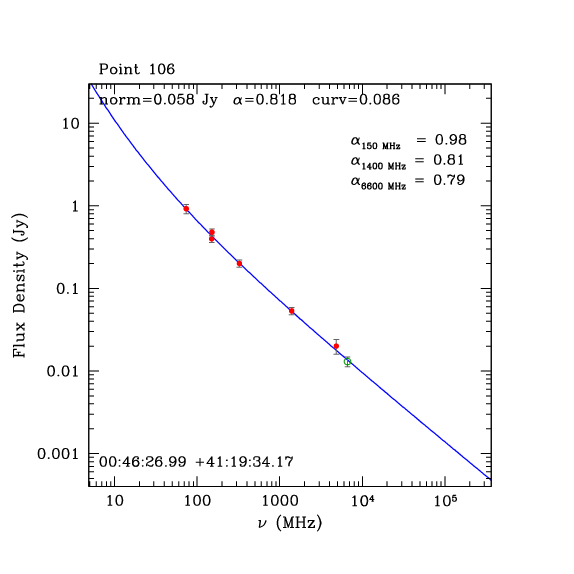} }
\subfigure{\includegraphics[scale=0.23]{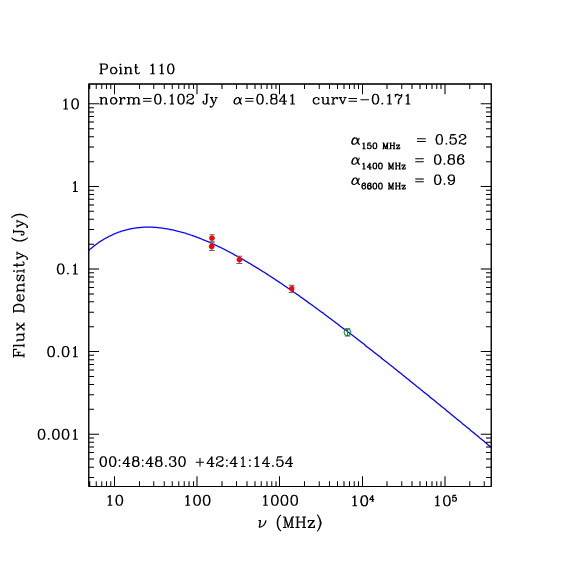} }
\subfigure{\includegraphics[scale=0.23]{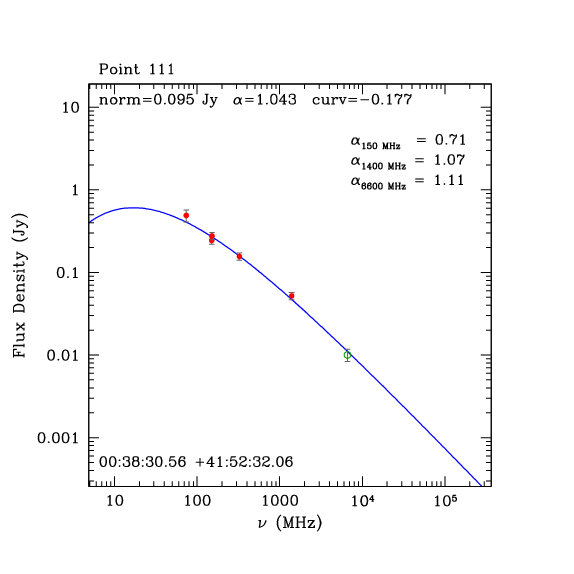} }
\subfigure{\includegraphics[scale=0.23]{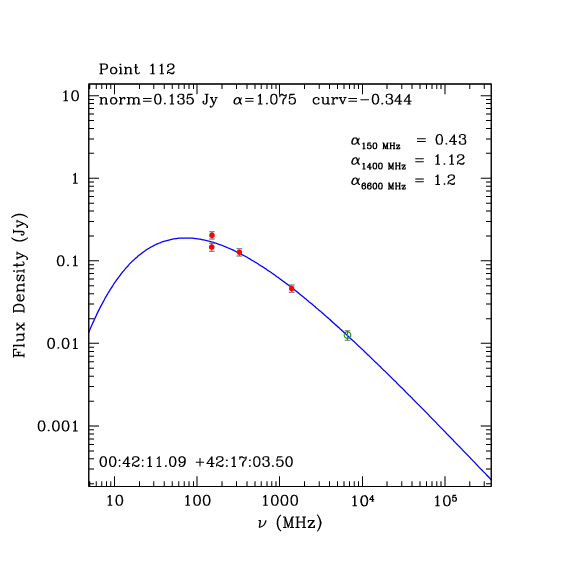} }

\end{figure*}

\begin{figure*}[h!]
\centering
\subfigure{\includegraphics[scale=0.23]{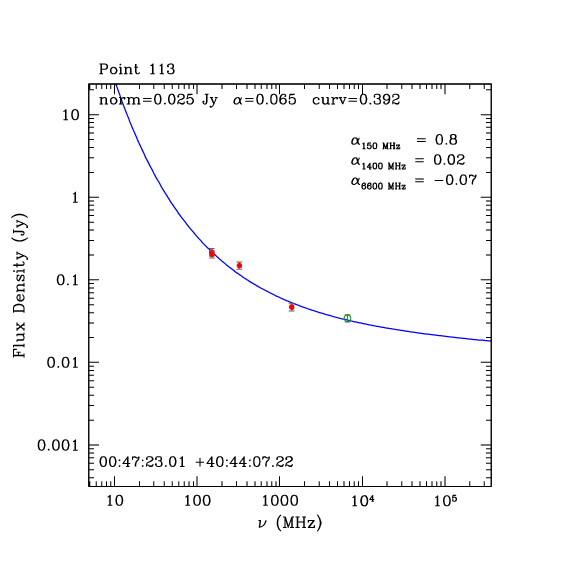} }
\subfigure{\includegraphics[scale=0.23]{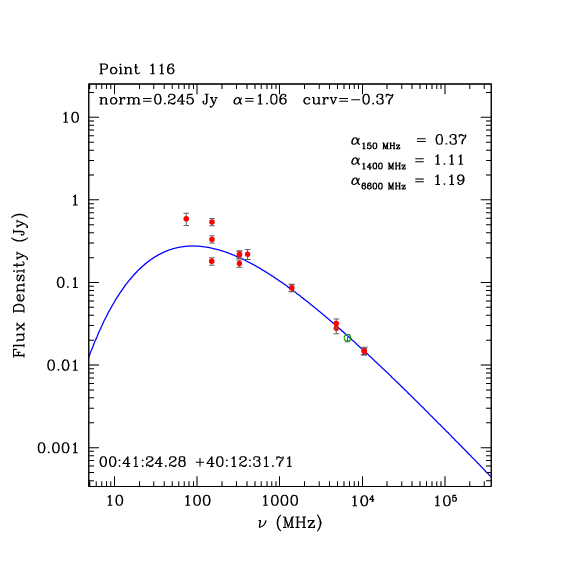} }
\subfigure{\includegraphics[scale=0.23]{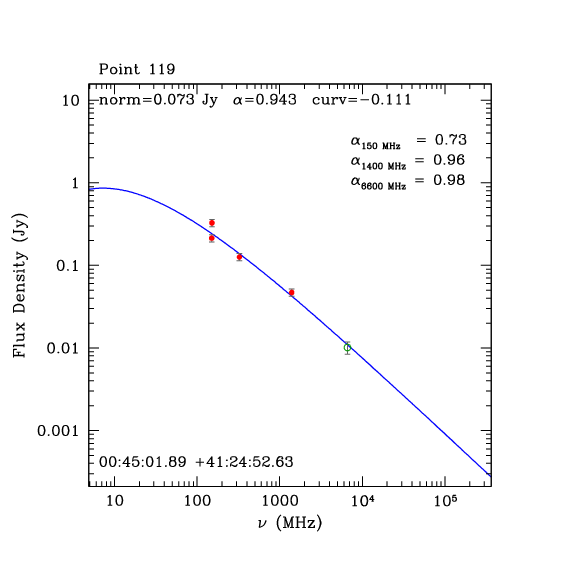} }
\subfigure{\includegraphics[scale=0.23]{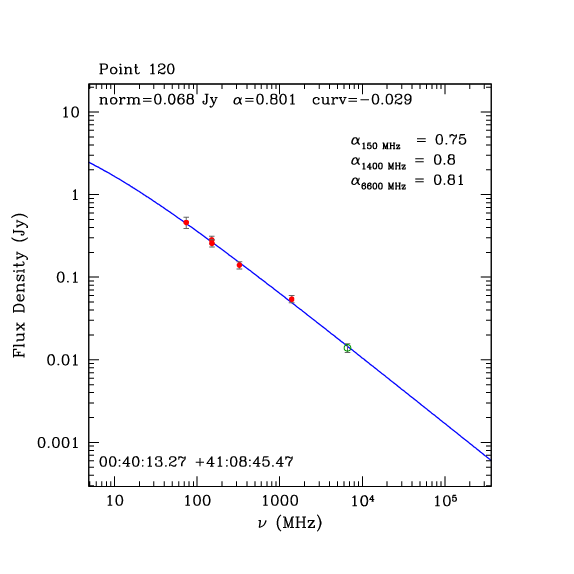} }
\subfigure{\includegraphics[scale=0.23]{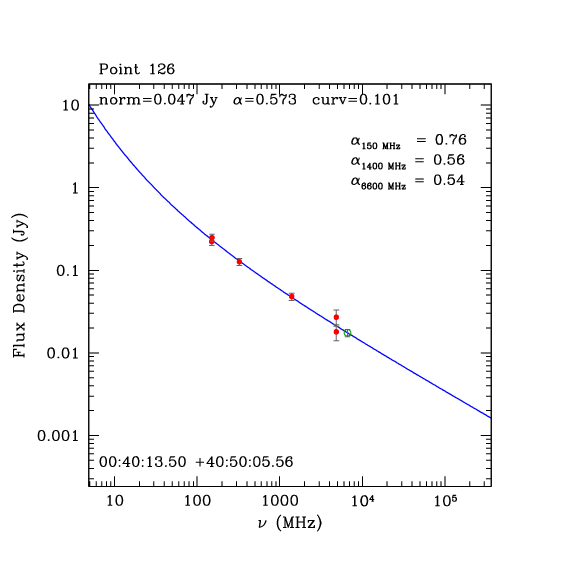} }
\subfigure{\includegraphics[scale=0.23]{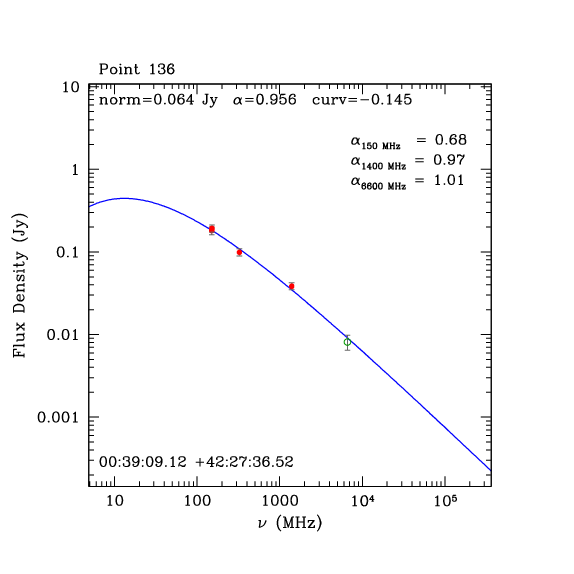} }
\subfigure{\includegraphics[scale=0.23]{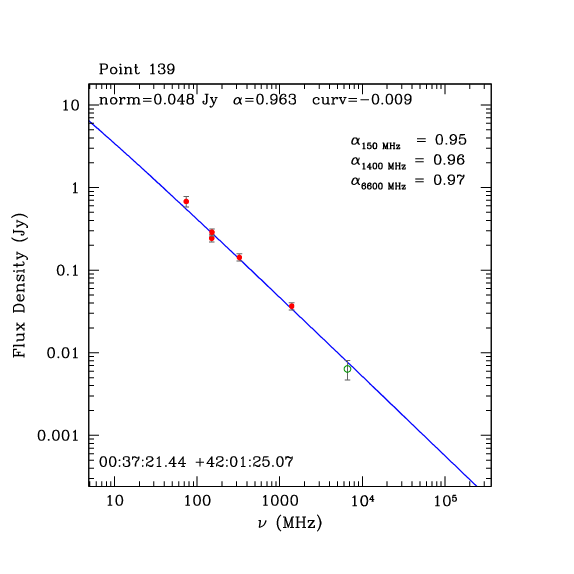} }
\subfigure{\includegraphics[scale=0.23]{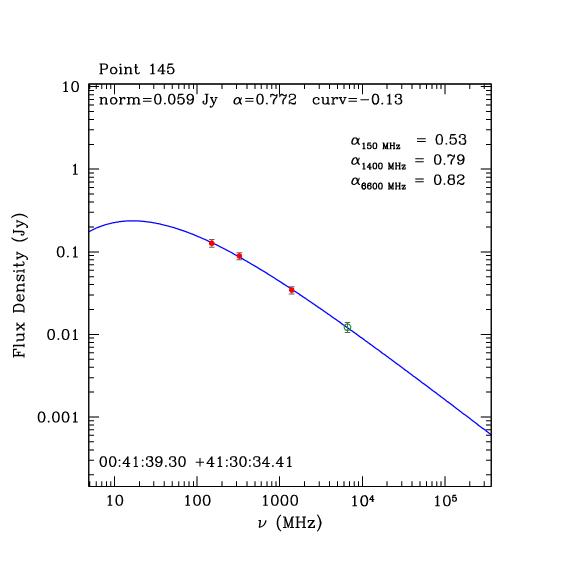} }
\subfigure{\includegraphics[scale=0.23]{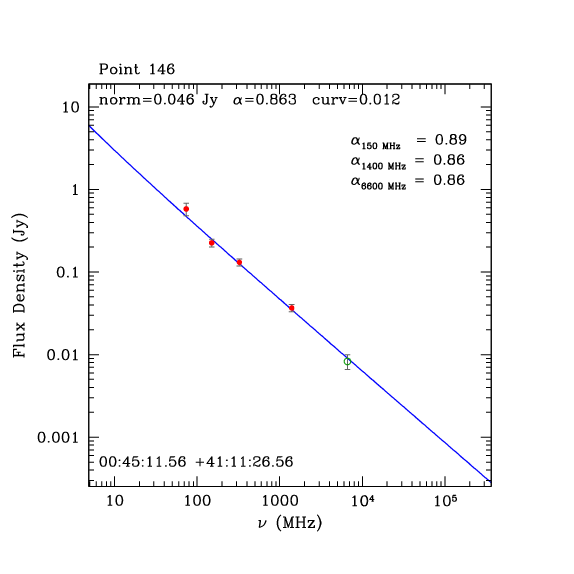} }
\subfigure{\includegraphics[scale=0.23]{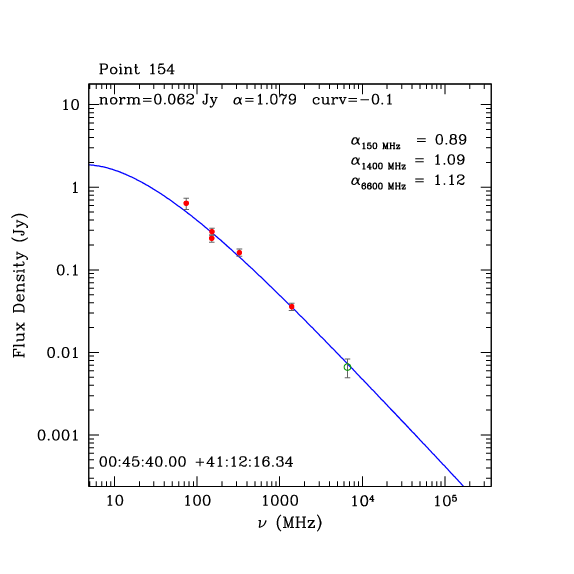} }
\subfigure{\includegraphics[scale=0.23]{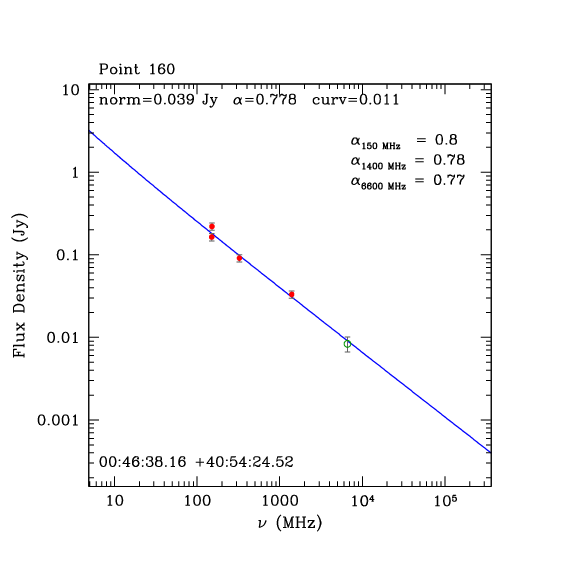} }
\subfigure{\includegraphics[scale=0.23]{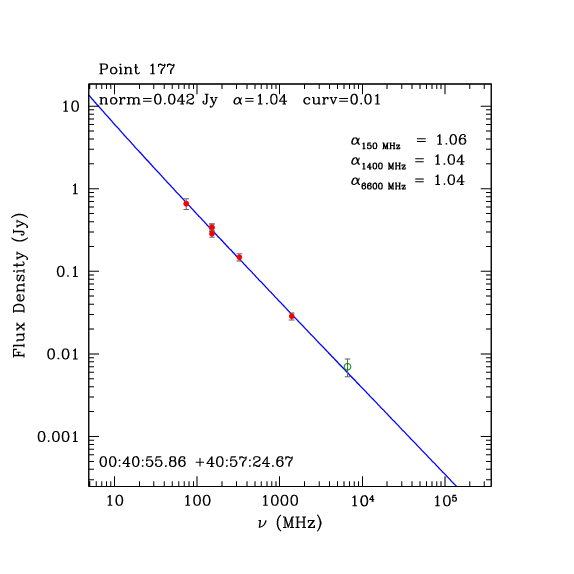} }
\subfigure{\includegraphics[scale=0.23]{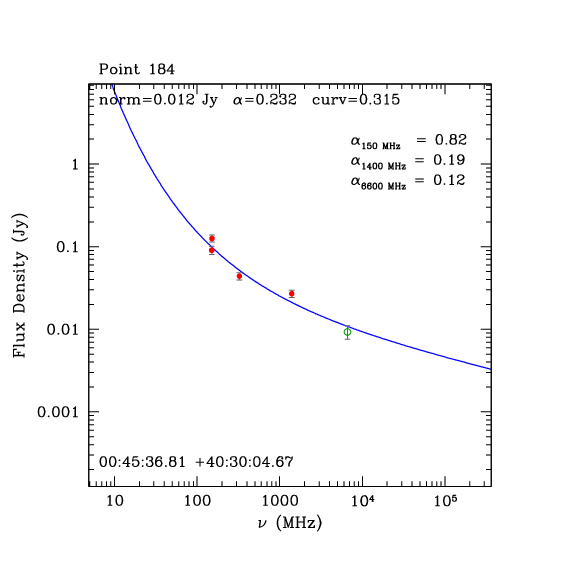} }
\subfigure{\includegraphics[scale=0.23]{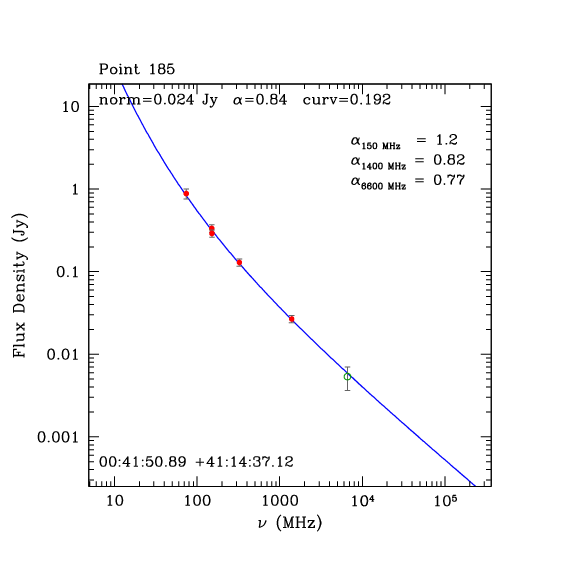} }
\subfigure{\includegraphics[scale=0.23]{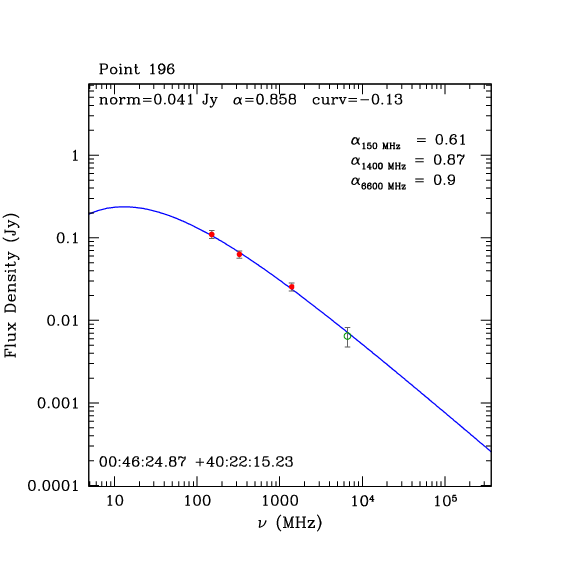} }

\end{figure*}

\begin{figure*}[h!]
\centering
\subfigure{\includegraphics[scale=0.23]{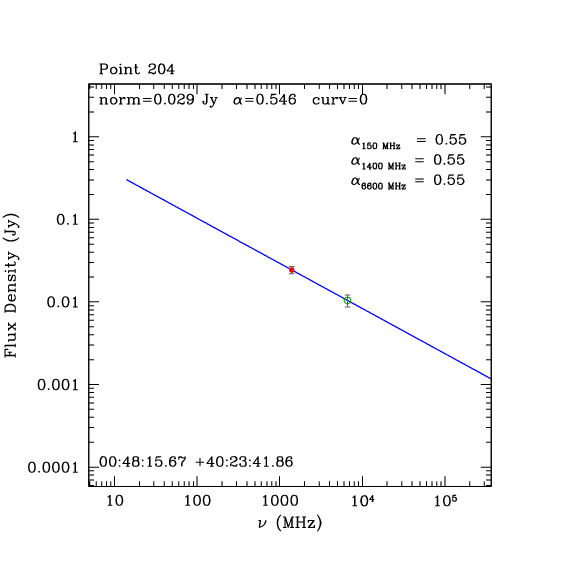} }
\subfigure{\includegraphics[scale=0.23]{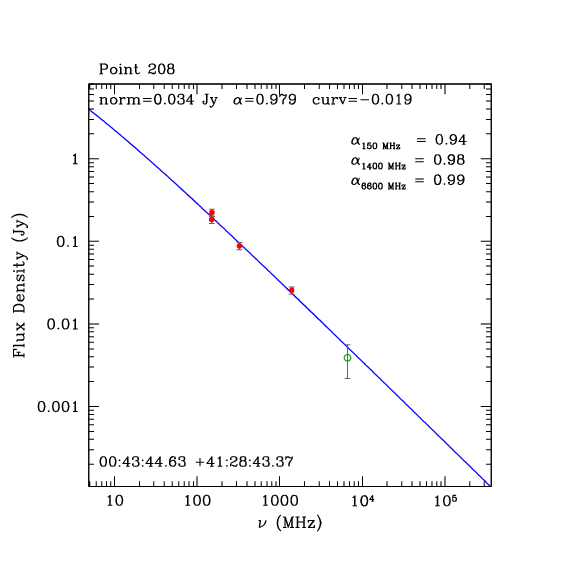} }
\subfigure{\includegraphics[scale=0.23]{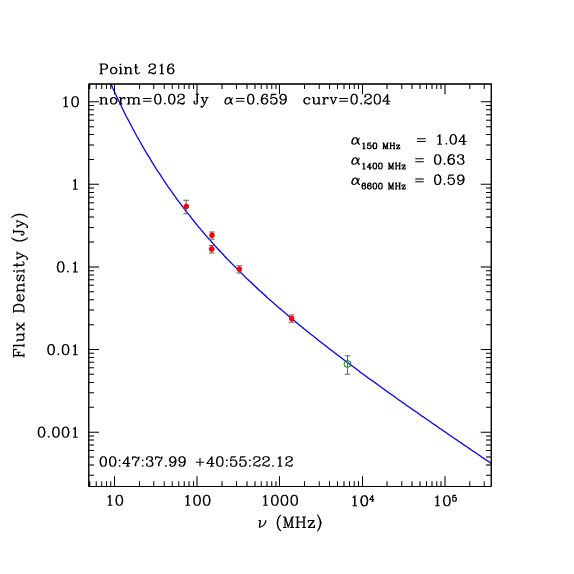} }
\subfigure{\includegraphics[scale=0.23]{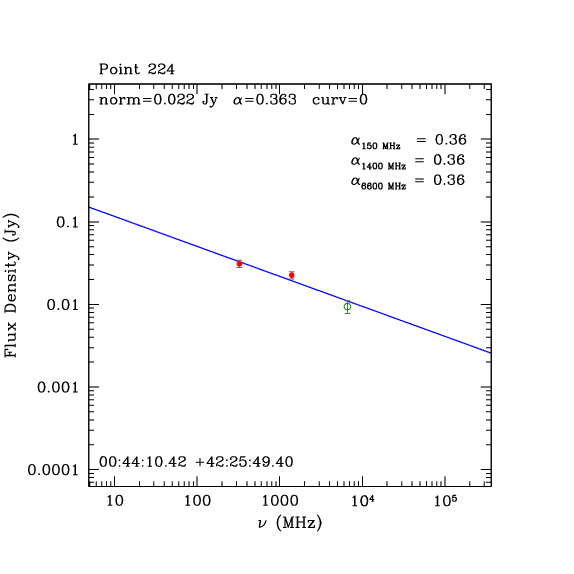} }
\subfigure{\includegraphics[scale=0.23]{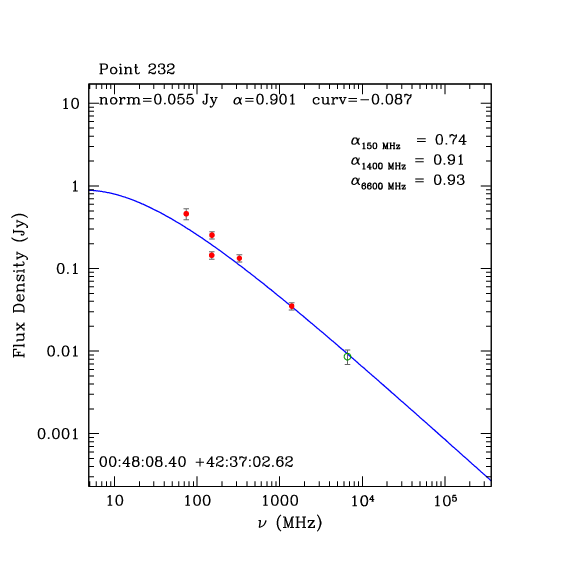} }
\subfigure{\includegraphics[scale=0.23]{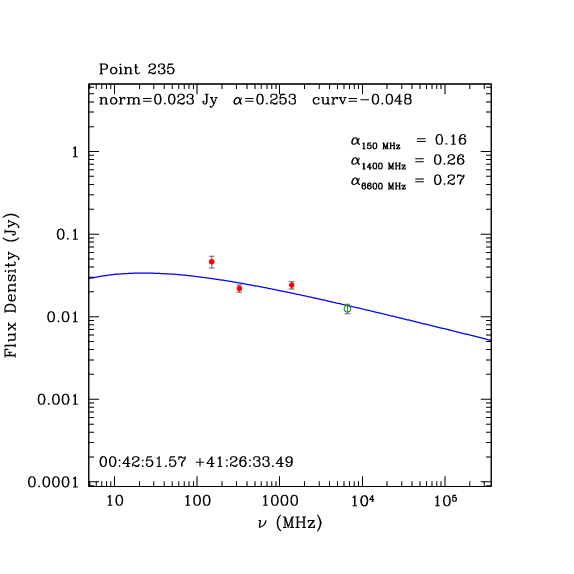} }
\subfigure{\includegraphics[scale=0.23]{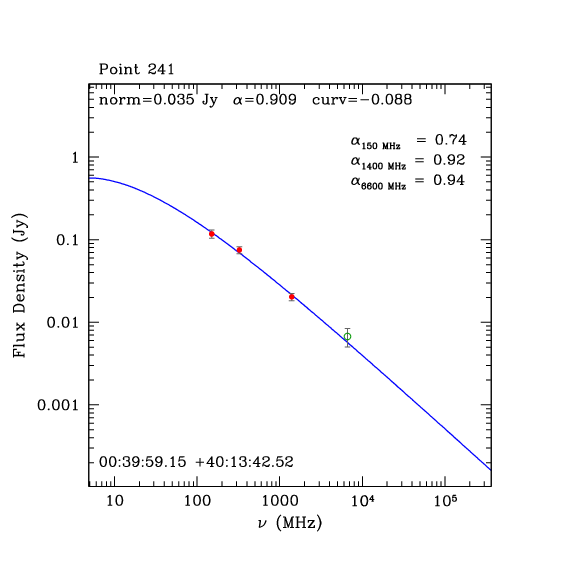} }
\subfigure{\includegraphics[scale=0.23]{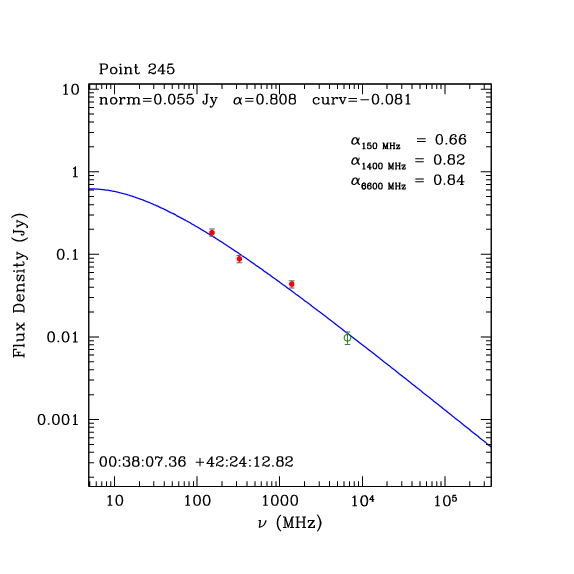} }
\subfigure{\includegraphics[scale=0.23]{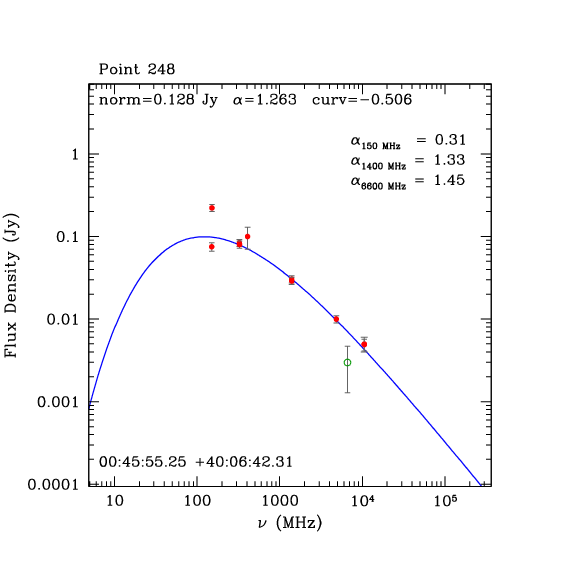} }
\subfigure{\includegraphics[scale=0.23]{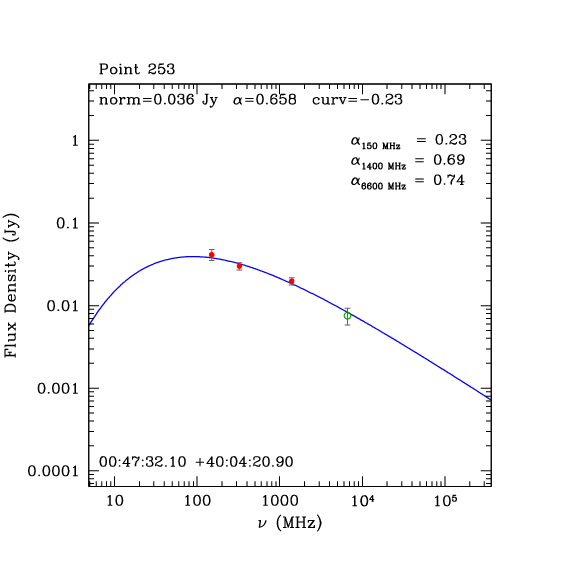} }
\subfigure{\includegraphics[scale=0.23]{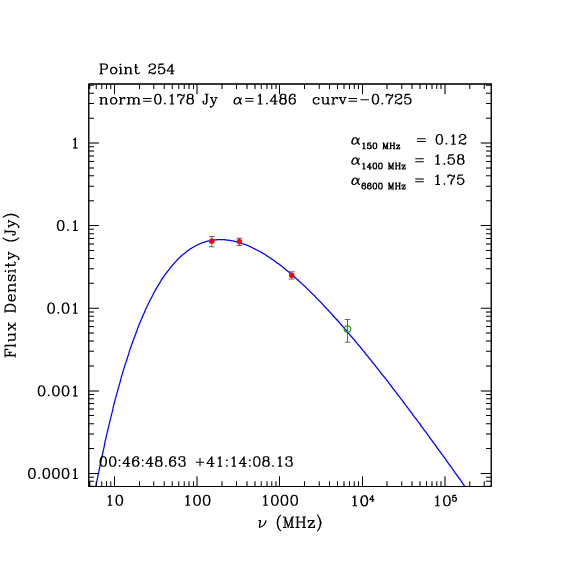} }
\subfigure{\includegraphics[scale=0.23]{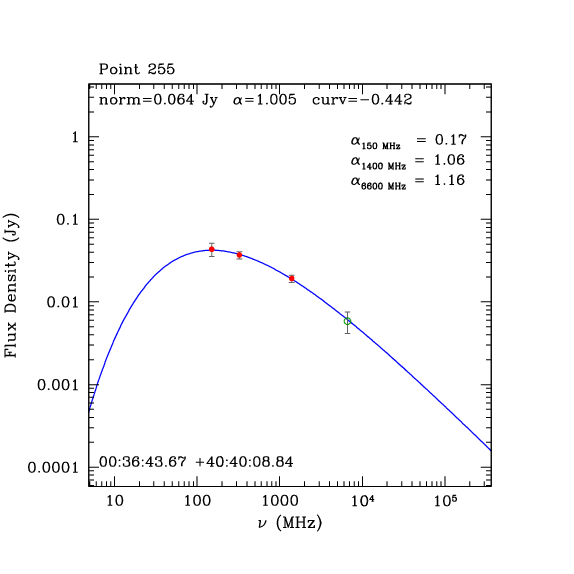} }
\subfigure{\includegraphics[scale=0.23]{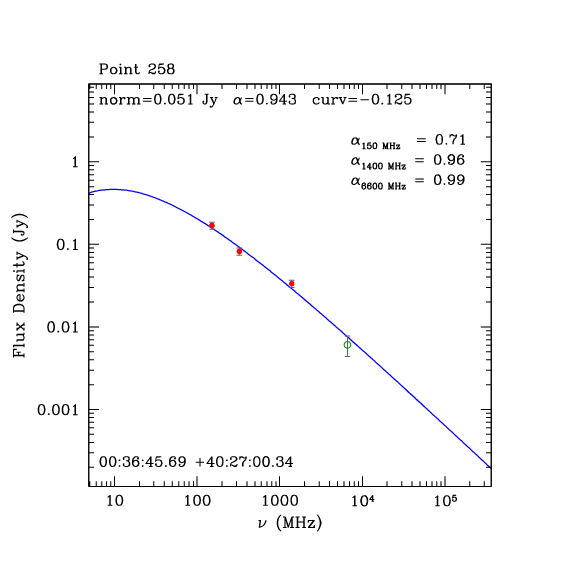} }
\subfigure{\includegraphics[scale=0.23]{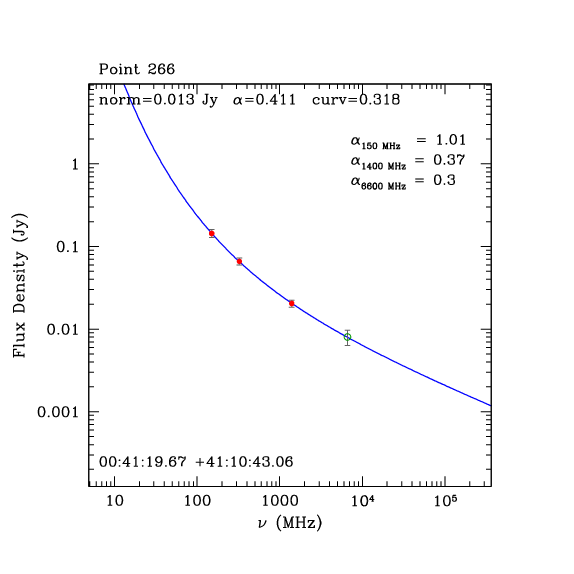} }
\subfigure{\includegraphics[scale=0.23]{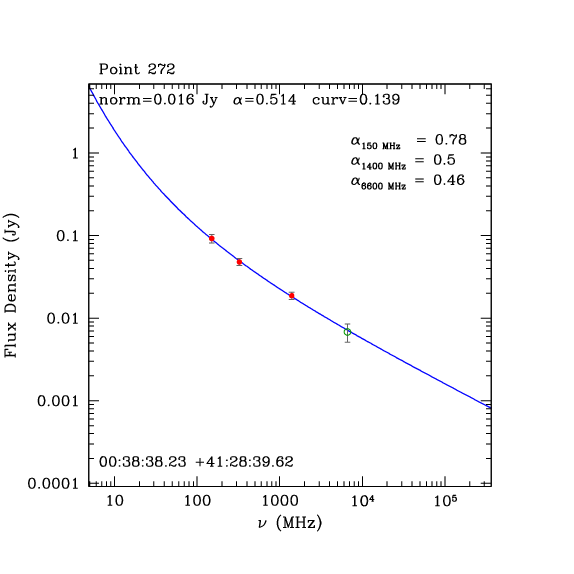} }

\end{figure*}

\begin{figure*}[h!]
\centering
\subfigure{\includegraphics[scale=0.23]{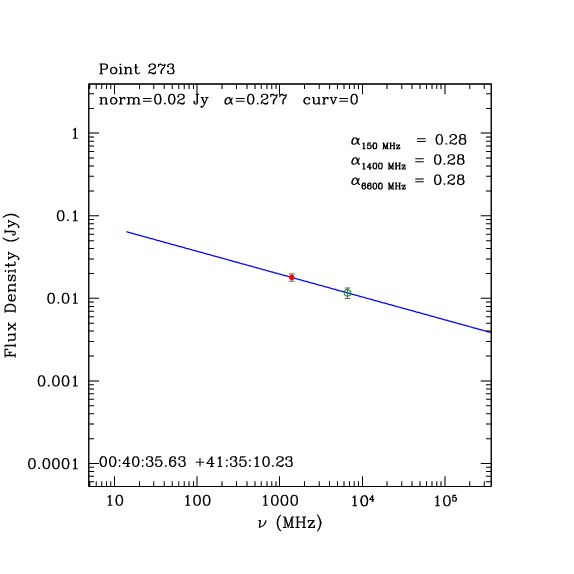} }
\subfigure{\includegraphics[scale=0.23]{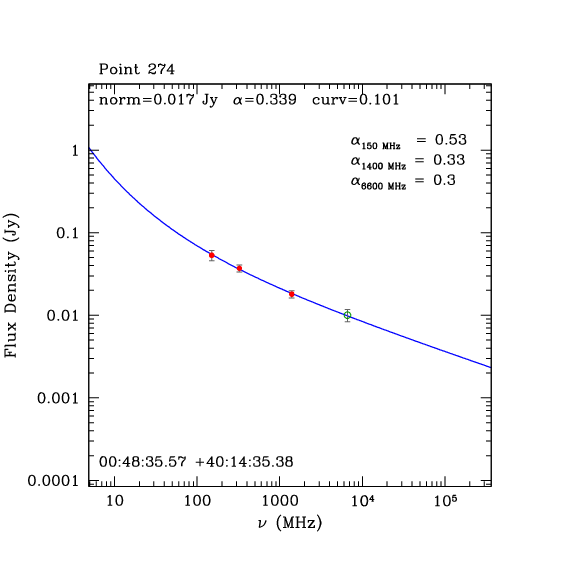} }
\subfigure{\includegraphics[scale=0.23]{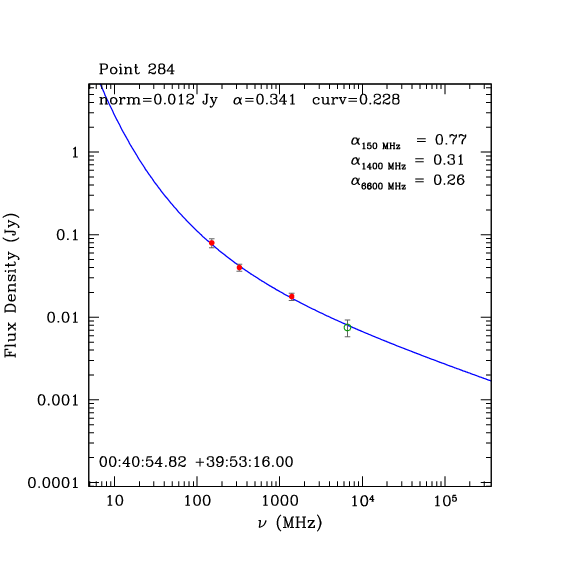} }
\subfigure{\includegraphics[scale=0.23]{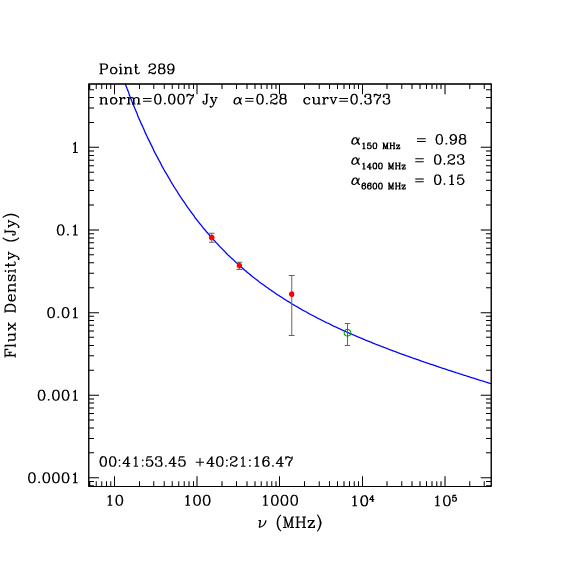} }
\subfigure{\includegraphics[scale=0.23]{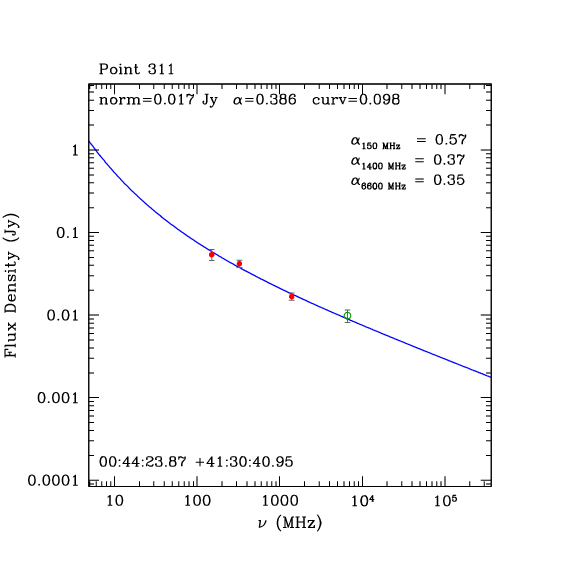} }
\subfigure{\includegraphics[scale=0.23]{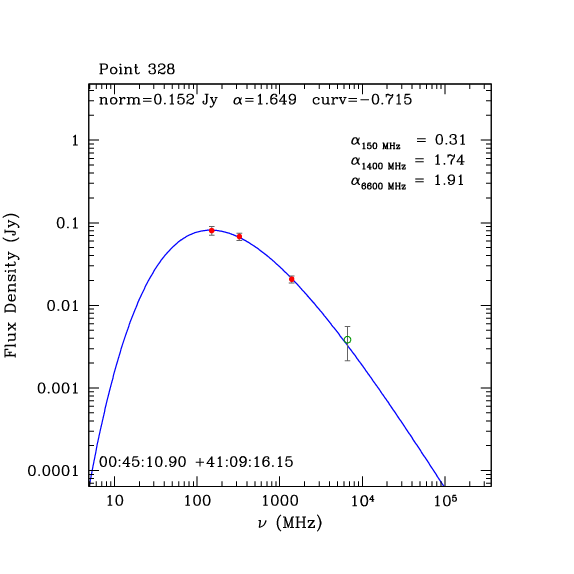} }
\subfigure{\includegraphics[scale=0.23]{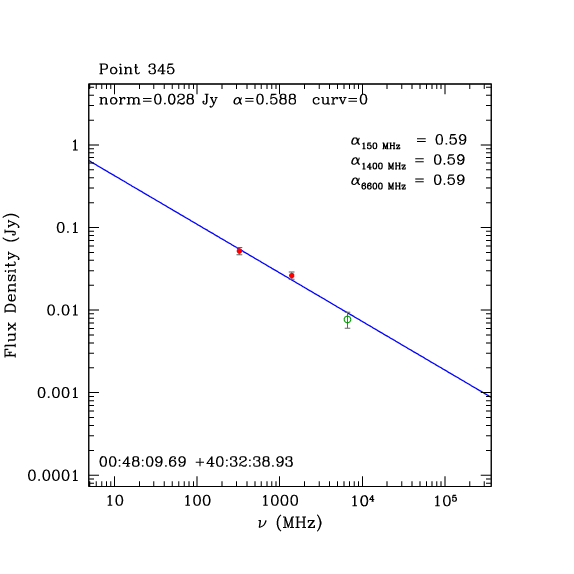} }
\subfigure{\includegraphics[scale=0.23]{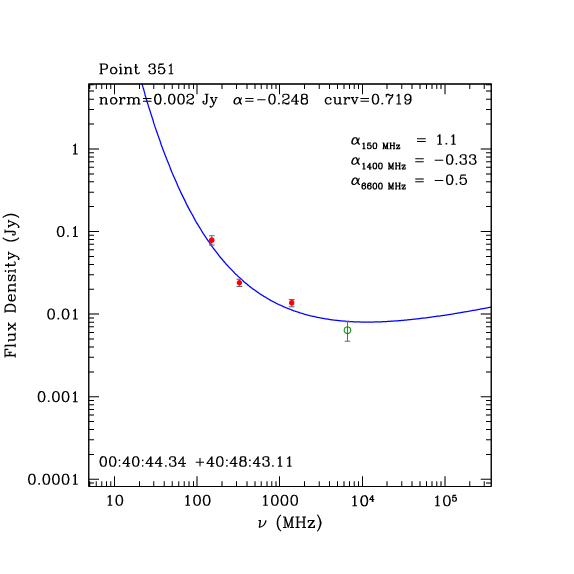} }
\subfigure{\includegraphics[scale=0.23]{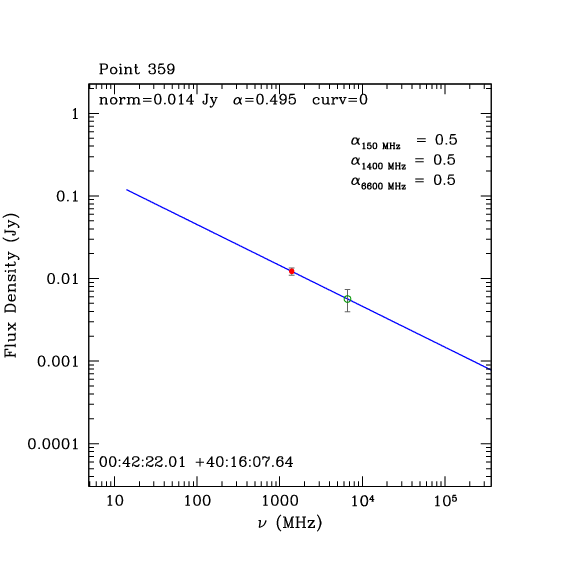} }
\subfigure{\includegraphics[scale=0.23]{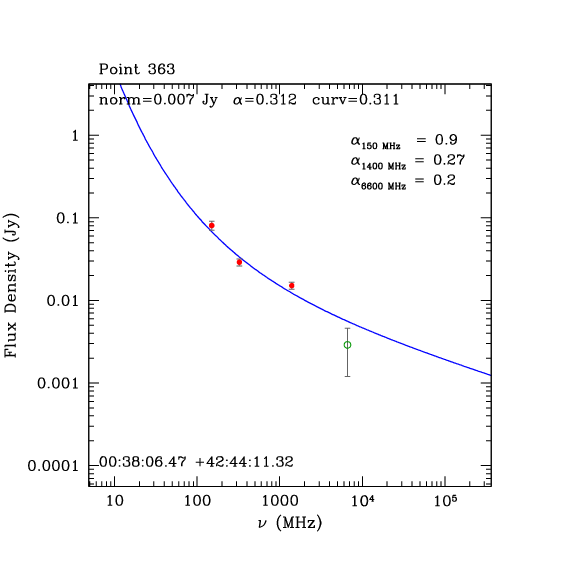} }
\subfigure{\includegraphics[scale=0.23]{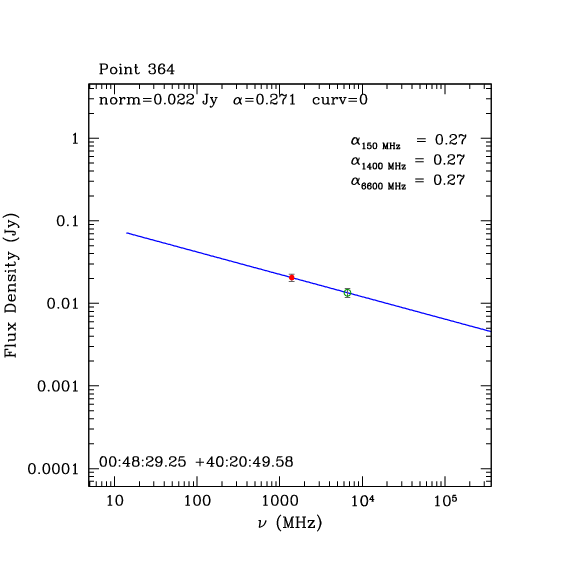} }
\subfigure{\includegraphics[scale=0.23]{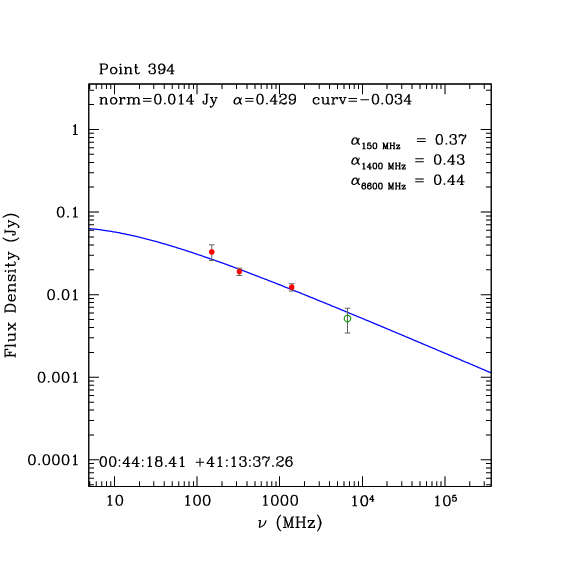} }
\subfigure{\includegraphics[scale=0.23]{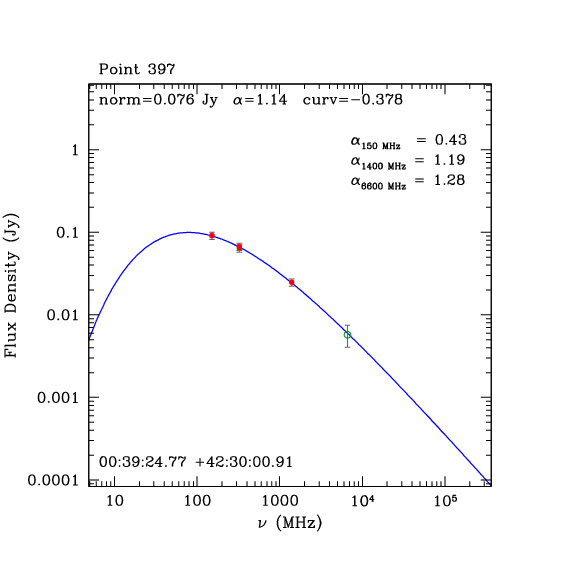} }
\subfigure{\includegraphics[scale=0.23]{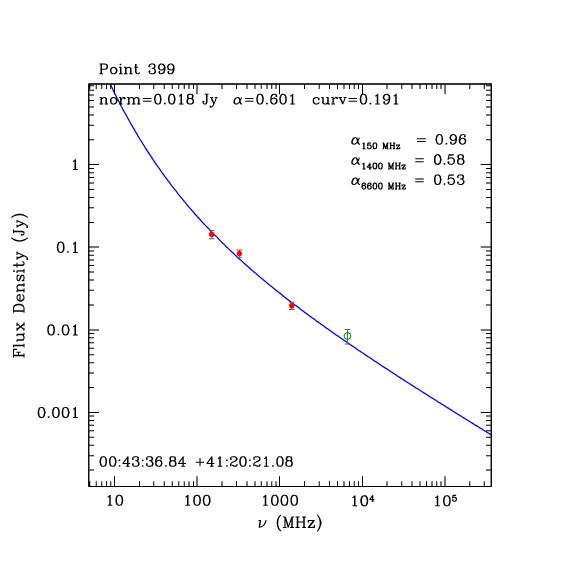} }
\subfigure{\includegraphics[scale=0.23]{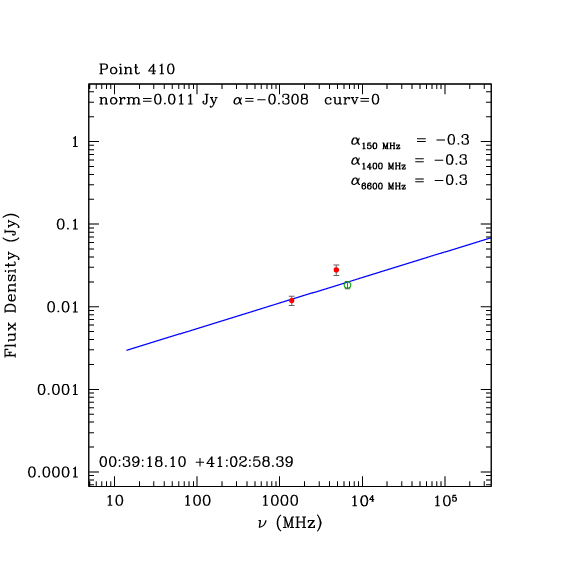} }

\end{figure*}

\begin{figure*}[h!]
\centering
\subfigure{\includegraphics[scale=0.23]{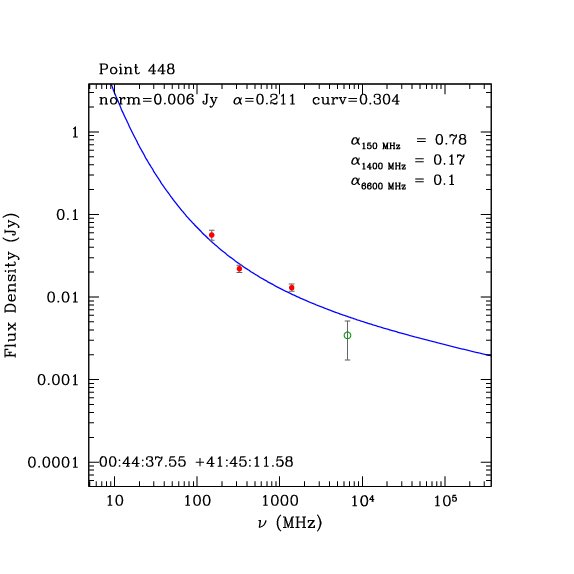} }
\subfigure{\includegraphics[scale=0.23]{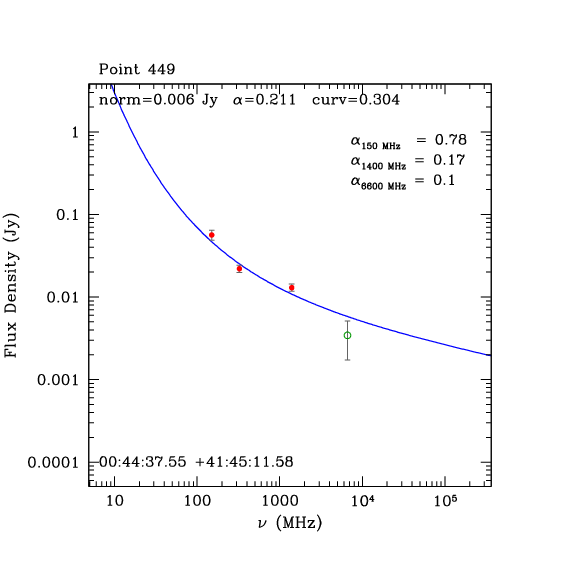} }
\subfigure{\includegraphics[scale=0.23]{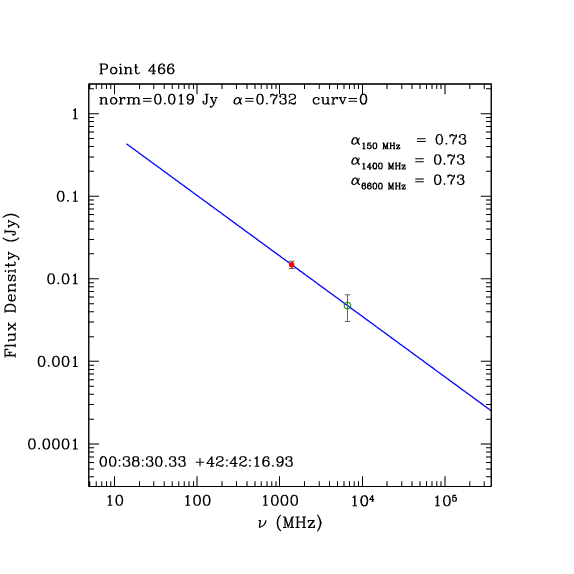} }
\subfigure{\includegraphics[scale=0.23]{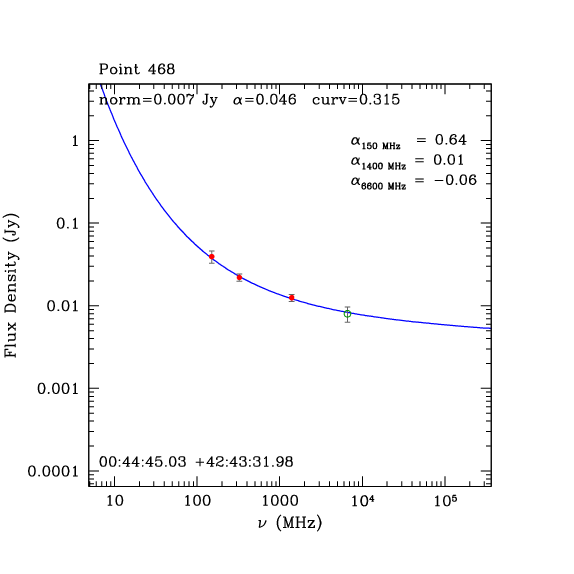} }
\subfigure{\includegraphics[scale=0.23]{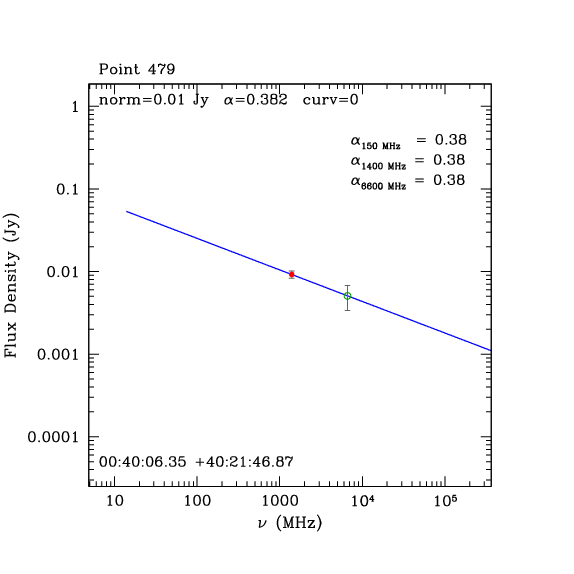} }
\subfigure{\includegraphics[scale=0.23]{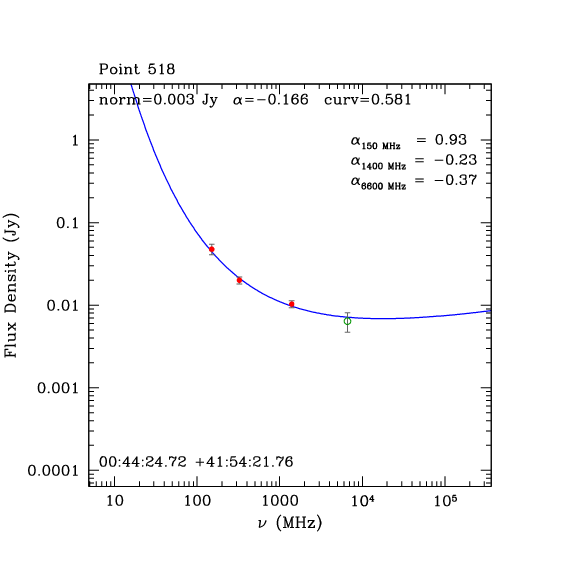} }
\subfigure{\includegraphics[scale=0.23]{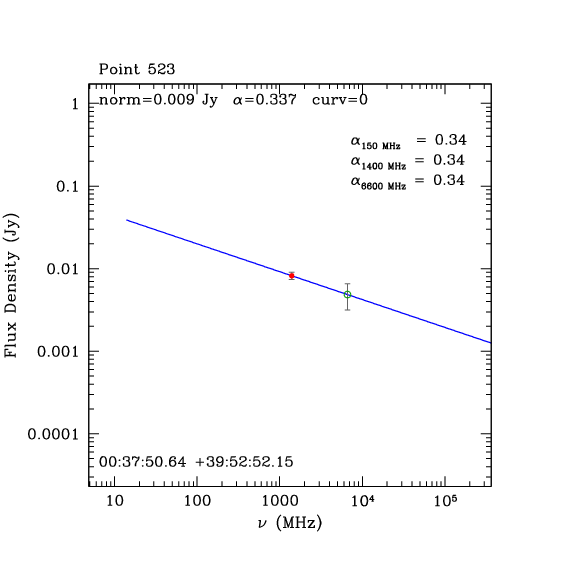} }
\subfigure{\includegraphics[scale=0.23]{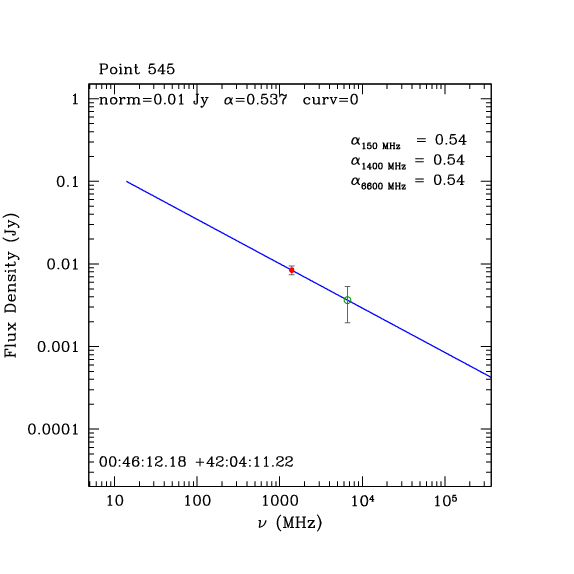} }
\subfigure{\includegraphics[scale=0.23]{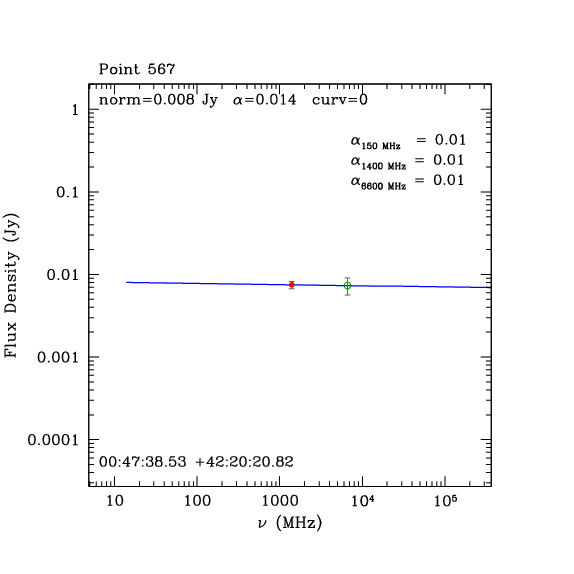} }
\subfigure{\includegraphics[scale=0.23]{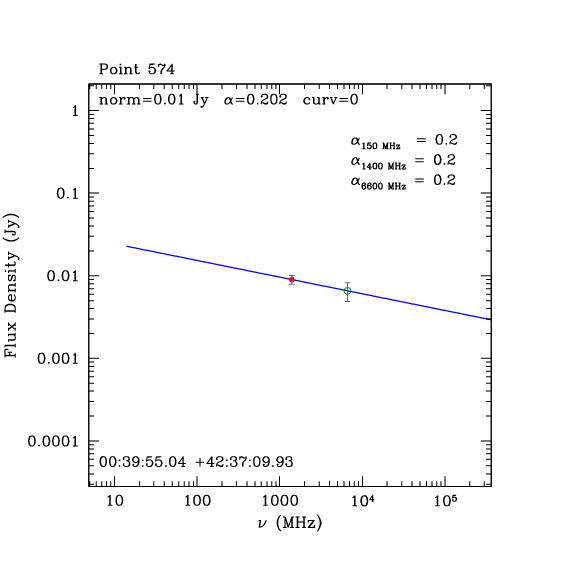} }
\subfigure{\includegraphics[scale=0.23]{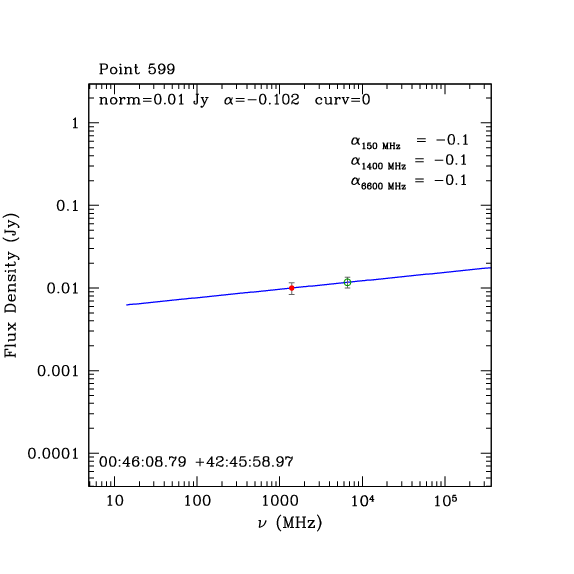} }
\subfigure{\includegraphics[scale=0.23]{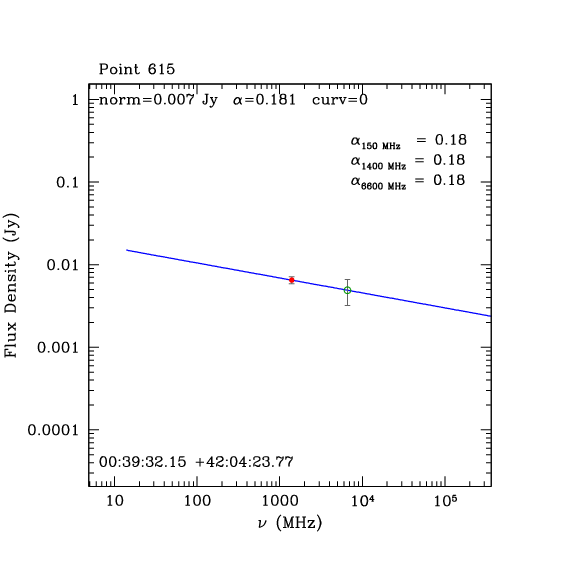} }
\subfigure{\includegraphics[scale=0.23]{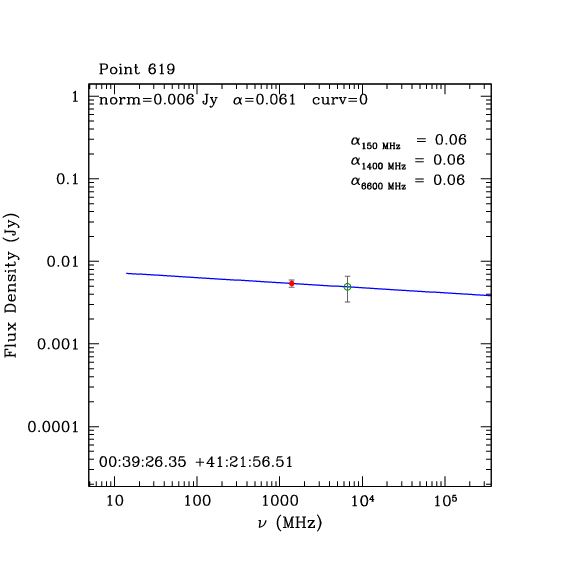} }
\subfigure{\includegraphics[scale=0.23]{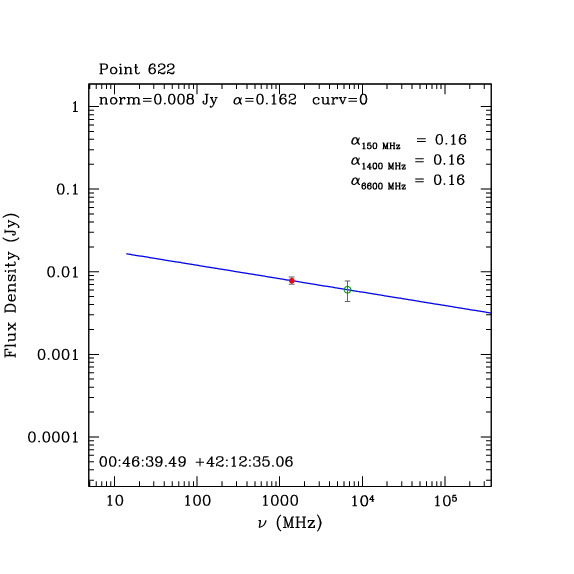} }
\subfigure{\includegraphics[scale=0.23]{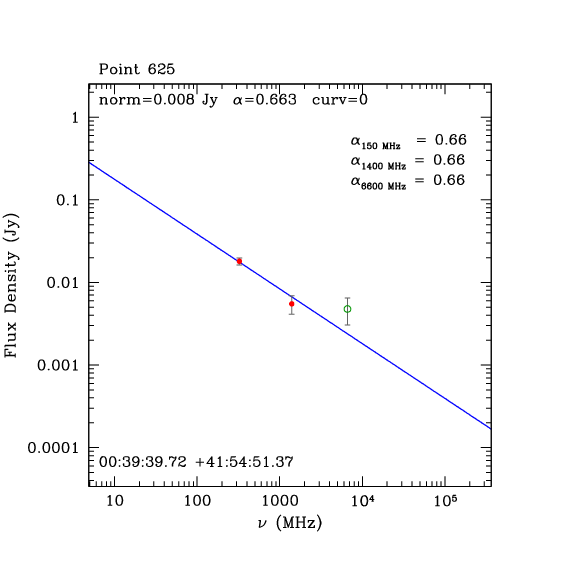} }
\end{figure*}

\begin{figure*}[h!]
\centering
\subfigure{\includegraphics[scale=0.23]{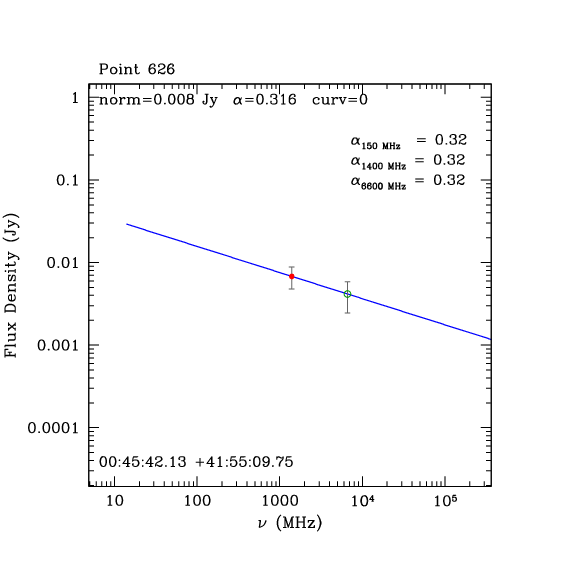} }
\subfigure{\includegraphics[scale=0.23]{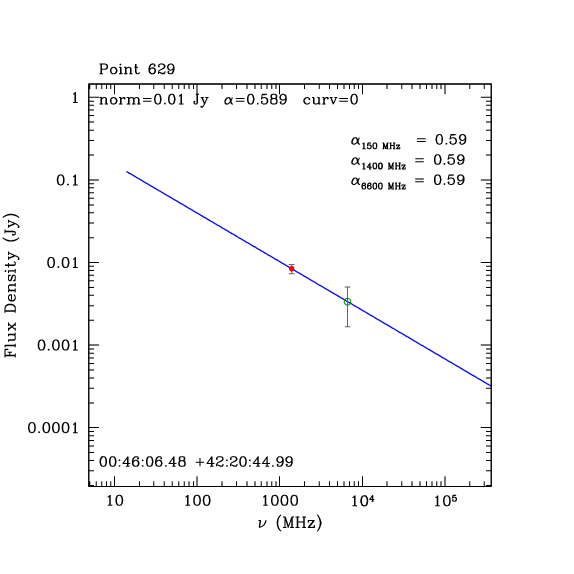} }
\subfigure{\includegraphics[scale=0.23]{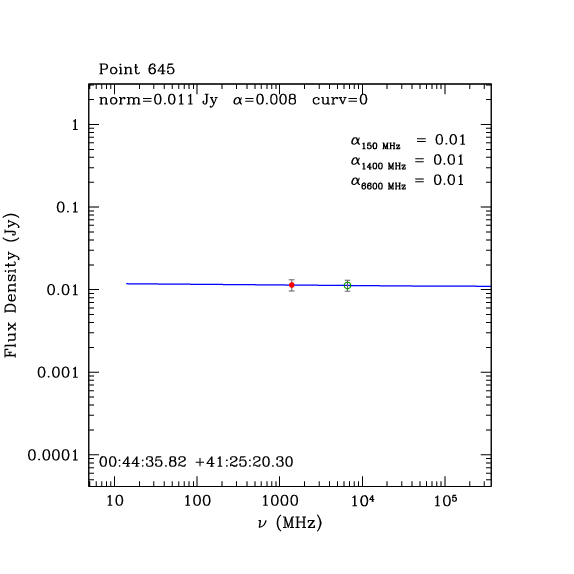} }
\caption{ Spectral fit of the 93 compact sources detected in our C-band map. Green points are the new flux densities extracted in this work. Red points are the ancillary data. Each figure shows the compact source RA-DEC coordinates (bottom left), the fit parameters (upper side), and the model spectral indices at 150, 1400, and 6600 MHz (right side). }
\label{fig:all_fit}
\end{figure*}

\clearpage
\onecolumn

\begin{longtable}{ c|c|c|c|c|c|c|c|c }
\hline
 N. &  Nearest Obj. &  RA($^{\circ}$) &  DEC($^{\circ}$) &  Flux density(Jy) &  $A$(Jy) &  $ \alpha $ &  $k$ & $\alpha _{1.4\,GHz} ^{6.6\,GHz}$ \\
\hline
\hline
         3 &           4C +39.03 [RadioS] &    10.57 &     40.16 &  $0.153 \pm 0.015$ &  2.397 &   1.26 & -0.40 &    1.01 \\
         4 &           B3 0035+413 [QSO] &     9.60 &     41.62 &  $0.388 \pm 0.039$ &  0.559 &   0.08 &  0.00 &    0.37 \\
         5 &           4C +42.03 [RadioS] &    11.16 &     42.80 &  $0.094 \pm 0.009$ &  1.067 &   1.16 & -0.25 &    1.02 \\
         6 &           4C +40.03 [G] &    12.05 &     40.36 &  $0.117 \pm 0.012$ &  1.223 &   1.15 & -0.20 &    1.06 \\
         7 &           B3 0039+412 [G] &    10.58 &     41.49 &   $0.096 \pm 0.01$ &  0.682 &   0.96 & -0.15 &    0.86 \\
         23 &           B3 0041+405 [RadioS] &    10.98 &     40.78 &  $0.062 \pm 0.006$ &  0.226 &   0.75 & -0.01 &    0.64 \\
         27 &           B3 0045+404 RadioS] &    12.18 &     40.75 &  $0.056 \pm 0.006$ &  0.343 &   0.98 & -0.11 &    0.89 \\
         33 &           B3 0044+418 [G] &    11.70 &     42.15 &  $0.032 \pm 0.003$ &  0.249 &   1.01 & -0.12 &    0.95 \\
         39 &           B2 0046+39 [G] &    12.19 &     39.89 &  $0.039 \pm 0.004$ &  0.218 &   0.91 & -0.12 &    0.81 \\
        41 &           B3 0036+409 [RadioS] &     9.70 &     41.27 &   $0.03 \pm 0.003$ &  1.171 &   1.44 & -0.96 &    0.86 \\
        48 &           B3 0041+416 [G] &    11.00 &     41.95 &  $0.019 \pm 0.002$ &  0.227 &   1.20 & -0.16 &    1.14 \\
        49 &           GALEXMSC J004755.61+394903.6 [UvS] &    11.98 &     39.82 &  $0.036 \pm 0.004$ &  0.060 &   0.25 &  0.20 &    0.62 \\
        52 &           B3 0043+405 [RadioS] &    11.51 &     40.80 &   $0.01 \pm 0.002$ &  0.403 &   1.58 & -0.56 &    1.34 \\
        56 &           B3 0039+416 [*] &    10.65 &     41.96 &  $0.018 \pm 0.002$ &  0.095 &   0.83 & -0.02 &    0.94 \\
        58 &           B3 0036+415 [RadioS] &     9.87 &     41.77 &  $0.016 \pm 0.002$ &  0.121 &   1.01 & -0.06 &    1.05 \\
        63 &           B3 0037+409 [RadioS] &     9.98 &     41.19 &  $0.019 \pm 0.002$ &  0.128 &   0.92 & -0.11 &    0.93 \\
        74 &           B3 0040+406 [RadioS] &    10.92 &     40.91 &  $0.016 \pm 0.002$ &  0.099 &   0.98 & -0.04 &    0.94 \\
        79 &           B3 0035+401 [RadioS] &     9.44 &     40.42 &  $0.011 \pm 0.002$ &  0.183 &   1.39 & -0.27 &    1.14 \\
        81 &           B3 0042+406 [RadioS] &    11.27 &     40.90 &  $0.011 \pm 0.002$ &  0.110 &   1.15 & -0.11 &    1.11 \\
        85 &           B3 0037+396 [RadioS] &    10.07 &     39.92 &  $0.011 \pm 0.002$ &  0.189 &   1.34 & -0.40 &    1.07 \\
        87 &           NVSS J004653+403219 [RadioS] &    11.72 &     40.54 &  $0.011 \pm 0.002$ &  0.080 &   1.00 & -0.03 &    1.09 \\
        88 &           B3 0034+413 [RadioS] &     9.27 &     41.60 &  $0.018 \pm 0.002$ &  0.062 &   0.74 &  0.15 &    0.82 \\
        93 &           B3 0043+422 [RadioS] &    11.65 &     42.53 &  $0.008 \pm 0.002$ &  0.138 &   1.25 & -0.34 &    1.17 \\
        97 &           B3 0036+398 [G] &     9.89 &     40.14 &  $0.014 \pm 0.002$ &  0.065 &   0.82 & -0.01 &    0.82 \\
        101 &           NVSS J004141+410333 [RadioS] &    10.42 &     41.06 &  $0.026 \pm 0.003$ &  0.031 &   0.27 &  0.31 &    0.40 \\
        105 &           NVSS J004147+411848 [RadioS] &    10.45 &     41.31 &  $0.008 \pm 0.002$ &  0.128 &   1.32 & -0.23 &    1.15 \\
        106 &           B3 0043+410 [RadioS] &    11.61 &     41.33 &  $0.013 \pm 0.002$ &  0.058 &   0.82 &  0.09 &    0.90 \\
        110 &           WISEA J004848.46+424112.5 [RadioS] &    12.20 &     42.69 &  $0.017 \pm 0.002$ &  0.102 &   0.84 & -0.17 &    0.78 \\
        111 &           B3 0035+416 [RadioS] &     9.63 &     41.88 &   $0.01 \pm 0.002$ &  0.095 &   1.04 & -0.18 &    1.05 \\
        112 &           B3 0039+420 [RadioS] &    10.55 &     42.28 &  $0.013 \pm 0.002$ &  0.135 &   1.08 & -0.34 &    0.83 \\
        113 &           B3 0044+404 [RadioS] &    11.85 &     40.74 &  $0.034 \pm 0.003$ &  0.025 &   0.06 &  0.39 &    0.20 \\
        116 &           B3 0038+399 [G] &    10.35 &     40.21 &  $0.021 \pm 0.002$ &  0.245 &   1.06 & -0.37 &    0.89 \\
        119 &           B3 0042+411 [RadioS] &    11.26 &     41.41 &   $0.01 \pm 0.002$ &  0.073 &   0.94 & -0.11 &    0.98 \\
        120 &           NVSS J004013+410845 [RadioS] &    10.06 &     41.15 &  $0.014 \pm 0.002$ &  0.068 &   0.80 & -0.03 &    0.87 \\
        126 &           B3 0037+405 [QSO] &    10.06 &     40.83 &  $0.018 \pm 0.002$ &  0.047 &   0.57 &  0.10 &    0.64 \\
        136 &           NVSS J003909+422736 [RadioS] &     9.79 &     42.46 &  $0.008 \pm 0.002$ &  0.064 &   0.96 & -0.15 &    0.99 \\
        139 &           B3 0034+417 [RadioS] &     9.34 &     42.02 &  $0.006 \pm 0.002$ &  0.048 &   0.96 & -0.01 &    1.12 \\
        145 &           WISEA J004139.28+413034.9 [G] &    10.41 &     41.51 &  $0.012 \pm 0.002$ &  0.059 &   0.77 & -0.13 &    0.66 \\
        146 &           B3 0042+409A [RadioS] &    11.30 &     41.19 &  $0.008 \pm 0.002$ &  0.046 &   0.86 &  0.01 &    0.95 \\
        154 &           B3 0042+409B [RadioS] &    11.42 &     41.20 &  $0.007 \pm 0.002$ &  0.062 &   1.08 & -0.10 &    1.08 \\
        160 &           WISEA J004638.18+405425.1 [RadioS] &    11.66 &     40.91 &  $0.008 \pm 0.002$ &  0.039 &   0.78 &  0.01 &    0.88 \\
        177 &           B3 0038+406 [RadioS] &    10.23 &     40.96 &  $0.007 \pm 0.002$ &  0.042 &   1.04 &  0.01 &    0.90 \\
        184 &           WISEA J004536.73+403003.6 [G] &    11.40 &     40.50 &  $0.009 \pm 0.002$ &  0.012 &   0.23 &  0.32 &    0.68 \\
        185 &           SSTSL2 J004150.78+411439.1 [IrS] &    10.46 &     41.24 &  $0.005 \pm 0.002$ &  0.024 &   0.84 &  0.19 &    1.03 \\
        196 &           NVSS J004624+402215 [RadioS] &    11.60 &     40.37 &  $0.006 \pm 0.002$ &  0.041 &   0.86 & -0.13 &    0.88 \\
        204 &           NVSS J004815+402341 [RadioS] &    12.07 &     40.39 &   $0.01 \pm 0.002$ &  0.029 &   0.55 &  0.00 &    0.54 \\
        208 &           WISEA J004344.69+412843.7 [*Cl] &    10.94 &     41.48 &  $0.004 \pm 0.002$ &  0.034 &   0.98 & -0.02 &    1.20 \\
        216 &           NVSS J004737+405522 [RadioS] &    11.91 &     40.92 &  $0.007 \pm 0.002$ &  0.020 &   0.66 &  0.20 &    0.81 \\
        224 &           WISEA J004410.47+422549.3 [RadioS] &    11.04 &     42.43 &  $0.009 \pm 0.002$ &  0.022 &   0.36 &  0.00 &    0.56 \\
        232 &           B3 0045+423 [RadioS] &    12.04 &     42.62 &  $0.009 \pm 0.002$ &  0.055 &   0.90 & -0.09 &    0.90 \\
        235 &           2MASS J00425141+4126344 [RadioS] &    10.71 &     41.44 &  $0.013 \pm 0.002$ &  0.023 &   0.25 & -0.05 &    0.42 \\
        241 &           WISEA J003959.01+401343.1 [RadioS] &    10.00 &     40.23 &  $0.007 \pm 0.002$ &  0.035 &   0.91 & -0.09 &    0.71 \\
        245 &           NVSS J003807+422413 [RadioS] &     9.53 &     42.40 &   $0.01 \pm 0.002$ &  0.055 &   0.81 & -0.08 &    0.95 \\
        248 &           B3 0043+398 [G] &    11.48 &     40.11 &  $0.003 \pm 0.002$ &  0.128 &   1.26 & -0.51 &    1.48 \\
        253 &           2MASX J00473217+4004212 [G] &    11.88 &     40.07 &  $0.008 \pm 0.002$ &  0.036 &   0.66 & -0.23 &    0.61 \\
        254 &           NVSS J004648+411408 [RadioS] &    11.70 &     41.24 &  $0.006 \pm 0.002$ &  0.178 &   1.49 & -0.73 &    0.96 \\
        255 &           NVSS J003643+404009 [RadioS] &     9.18 &     40.67 &  $0.006 \pm 0.002$ &  0.064 &   1.00 & -0.44 &    0.76 \\
        258 &           NVSS J003645+402700 [RadioS] &     9.19 &     40.45 &  $0.006 \pm 0.002$ &  0.051 &   0.94 & -0.13 &    1.09 \\
        266 &           NVSS J004119+411043 [RadioS] &    10.33 &     41.18 &  $0.008 \pm 0.002$ &  0.013 &   0.41 &  0.32 &    0.60 \\
        272 &           NVSS J003838+412838 [RadioS] &     9.66 &     41.48 &  $0.007 \pm 0.002$ &  0.016 &   0.51 &  0.14 &    0.64 \\
        273 &           SSTSL2 J004035.81+413510.7 [RadioS] &    10.15 &     41.59 &  $0.012 \pm 0.002$ &  0.020 &   0.28 &  0.00 &    0.27 \\
        274 &           WISEA J004835.31+401435.8 [RadioS] &    12.15 &     40.24 &   $0.01 \pm 0.002$ &  0.017 &   0.34 &  0.10 &    0.37 \\
        284 &           WISEA J004054.80+395317.2 [RadioS] &    10.23 &     39.89 &  $0.008 \pm 0.002$ &  0.012 &   0.34 &  0.23 &    0.55 \\
        289 &           WISEA J004153.42+402117.6 [IrS] &    10.47 &     40.35 &  $0.006 \pm 0.002$ &  0.007 &   0.28 &  0.37 &    0.69 \\
        311 &           WISEA J004424.00+413042.9 [IrS] &    11.10 &     41.51 &   $0.01 \pm 0.002$ &  0.017 &   0.39 &  0.10 &    0.34 \\
        328 &           NVSS J004510+410915 [RadioS] &    11.30 &     41.15 &  $0.004 \pm 0.002$ &  0.152 &   1.65 & -0.72 &    1.08 \\
        345 &           NVSS J004809+403239 [RadioS] &    12.04 &     40.54 &  $0.008 \pm 0.002$ &  0.028 &   0.59 &  0.00 &    0.78 \\
        351 &           SSTSL2 J004044.37+404845.4 [RadioS] &    10.18 &     40.81 &  $0.006 \pm 0.002$ &  0.002 &  -0.25 &  0.72 &    0.49 \\
        359 &           WISEA J004222.07+401605.6 [RadioS] &    10.59 &     40.27 &  $0.006 \pm 0.002$ &  0.014 &   0.49 &  0.00 &    0.49 \\
        363 &           WISEA J003806.84+424410.2 [RadioS &     9.53 &     42.74 &  $0.003 \pm 0.002$ &  0.007 &   0.31 &  0.31 &    1.05 \\
        364 &           NVSS J004828+402048 [RadioS] &    12.12 &     40.35 &  $0.013 \pm 0.002$ &  0.022 &   0.27 &  0.00 &    0.27 \\
        394 &           SSTSL2 J004418.42+411337.9 [RadioS] &    11.08 &     41.23 &  $0.005 \pm 0.002$ &  0.014 &   0.43 & -0.03 &    0.56 \\
    397 &               WISEA J003924.78+423005.0 RadioS] &     9.85 &     42.50 &  $0.006 \pm 0.002$ &  0.076 &   1.14 & -0.38 &    0.93 \\
        399 &           B3 0040+410 [RadioS] &    10.90 &     41.34 &  $0.008 \pm 0.002$ &  0.018 &   0.60 &  0.19 &    0.54 \\
        410 &           WISEA J003917.87+410300.4 [RadioS] &     9.83 &     41.05 &  $0.018 \pm 0.002$ &  0.011 &  -0.31 &  0.00 &   -0.28 \\
        448 &           NVSS J004437+414511 [RadioS] &    11.16 &     41.75 &  $0.003 \pm 0.002$ &  0.006 &   0.21 &  0.30 &    0.85 \\
        449 &           NVSS J004437+414511 [RadioS] &    11.16 &     41.75 &  $0.003 \pm 0.002$ &  0.006 &   0.21 &  0.30 &    0.85 \\
        466 &           NVSS J003830+424218 [RadioS] &     9.63 &     42.70 &  $0.005 \pm 0.002$ &  0.019 &   0.73 &  0.00 &    0.73 \\
        468 &           WISEA J004444.96+424334.6 [RadioS] &    11.19 &     42.73 &  $0.008 \pm 0.002$ &  0.007 &   0.05 &  0.32 &    0.28 \\
        479 &           NVSS J004006+402146 [RadioS] &    10.03 &     40.36 &  $0.005 \pm 0.002$ &  0.010 &   0.38 &  0.00 &    0.38 \\
        518 &           NVSS J004424+415421 [RadioS] &    11.10 &     41.91 &  $0.006 \pm 0.002$ &  0.003 &  -0.17 &  0.58 &    0.30 \\
        523 &           WISEA J003750.66+395253.4 [RadioS] &     9.46 &     39.88 &  $0.005 \pm 0.002$ &  0.009 &   0.34 &  0.00 &    0.33 \\
        545 &           NVSS J004612+420410 [XrayS] &    11.55 &     42.07 &  $0.004 \pm 0.002$ &  0.010 &   0.54 &  0.00 &    0.53 \\
        567 &           NVSS J004738+422020 [RadioS] &    11.91 &     42.34 &  $0.007 \pm 0.002$ &  0.008 &   0.01 &  0.00 &    0.01 \\
        574 &           NVSS J003954+423708 [RadioS] &     9.98 &     42.62 &  $0.007 \pm 0.002$ &  0.010 &   0.20 &  0.00 &    0.20 \\
        599 &           NVSS J004608+424558 [RadioS] &    11.54 &     42.77 &  $0.012 \pm 0.002$ &  0.010 &  -0.10 &  0.00 &   -0.10 \\
        615 &        WISEA J003932.05+420424.2 [RadioS]    &     9.88 &     42.07 &  $0.005 \pm 0.002$ &  0.007 &   0.18 &  0.00 &    0.18 \\
        619 &        SSTSL2 J003926.19+412154.2 [RadioS]    &     9.86 &     41.37 &  $0.005 \pm 0.002$ &  0.006 &   0.06 &  0.00 &    0.06 \\
        622 &          NVSS J004639+421236 [RadioS]  &    11.66 &     42.21 &  $0.006 \pm 0.002$ &  0.008 &   0.16 &  0.00 &    0.16 \\
        625 &           SSTSL2 J003939.81+415450.5 [RadioS] &     9.92 &     41.91 &  $0.005 \pm 0.002$ &  0.008 &   0.66 &  0.00 &    0.09 \\
        626 &        NVSS J004542+415512 [RadioS]    &    11.43 &     41.92 &  $0.004 \pm 0.002$ &  0.008 &   0.32 &  0.00 &    0.31 \\
        629 &           SSTSL2 J004607.00+422045.2 [IrS] &    11.53 &     42.35 &  $0.003 \pm 0.002$ &  0.010 &   0.59 &  0.00 &    0.58 \\
        645 &           NVSS J004435+412521 [SNR] &    11.15 &     41.42 &  $0.011 \pm 0.002$ &  0.011 &   0.01 &  0.00 &    0.01 \\
\hline
\caption{The SRT C-band Compact Source catalog.}
\label{Tab:SRT_Sources_catalogue}
\end{longtable}

\end{appendix}

\end{document}